\documentclass{preprint}

\usepackage{hyperref}       
\usepackage{url}            
\usepackage{booktabs}       
\usepackage{amsfonts}       
\usepackage{nicefrac}       
\usepackage{microtype}      
\usepackage{xcolor}         
\usepackage{siunitx}
\usepackage{titletoc}
\usepackage{array}
\usepackage{colortbl}
\usepackage{arydshln}

\usepackage{amsmath}
\usepackage{xspace}
\usepackage{graphicx}

\usepackage{framed}
\usepackage{amssymb}
\usepackage{color}
\usepackage{soul}

\usepackage[most]{tcolorbox}

\usepackage{enumitem}
\usepackage{multirow}

\usepackage{algorithm}
\usepackage{algpseudocode}

\usepackage{siunitx}
\newcommand{\model}{Trading-R1\xspace}
\newcommand{\database}{Tauric-TR1-DB\xspace}

\newtcbox{\hlwhite}{on line, box align=base, colback=red!20,colframe=white,size=fbox,arc=2pt, before upper=\strut, top=-3pt, bottom=-4.5pt, left=-2pt, right=-2pt, boxrule=0pt}

\definecolor{evidenceblue}{RGB}{233,242,255}
\newcommand{\hlevidence}[1]{\sethlcolor{evidenceblue}\hl{#1}}

\definecolor{tagblue}{RGB}{100,150,200}

\definecolor{FBest}{RGB}{4,130,53}

\newcommand{\tagblock}[3]{%
  \textcolor{#3}{\texttt{<#1>}} #2 \textcolor{#3}{\texttt{</#1>}}%
}

\newcommand{\think}[1]{\tagblock{think}{#1}{tagblue}}

\title{\model: Financial Trading with LLM Reasoning via Reinforcement Learning}

\author
{Yijia Xiao~$^{1}$,  Edward Sun~$^{1}$, Tong Chen~$^{2}$, Fang Wu~$^{3}$, Di Luo~$^{1}$, Wei Wang~$^{1}$\\
\vspace{1em} 
\normalfont{\small $^{1}$University of California, Los Angeles}\\
\normalfont{\small $^{2}$University of Washington}\\
\normalfont{\small $^{3}$Stanford University}\\
\normalfont{\small $^{\textsuperscript{\S}}$Tauric Research\footnote{Tauric Research Institute: https://tauric.ai}} \vspace{1em}
}

\begin{document}

\maketitle
\thispagestyle{firstpagestyle} 

\begin{abstract}
Developing professional, structured reasoning on par with human financial analysts and traders remains a central challenge in AI for finance, where markets demand interpretability and trust. Traditional time-series models lack explainability, while LLMs face challenges in turning natural-language analysis into disciplined, executable trades. Although reasoning LLMs have advanced in step-by-step planning and verification, their application to risk-sensitive financial decisions is underexplored. We present \textsc{\model}, a financially-aware model that incorporates strategic thinking and planning for comprehensive thesis composition, facts-grounded analysis, and volatility-adjusted decision making. \textsc{\model} aligns reasoning with trading principles through supervised fine-tuning and reinforcement learning with a three-stage easy-to-hard curriculum. Training uses \database, a 100k-sample corpus spanning 18 months, 14 equities, and five heterogeneous financial data sources. Evaluated on six major equities and ETFs, \textsc{\model} demonstrates improved risk-adjusted returns and lower drawdowns compared to both open-source and proprietary instruction-following models as well as reasoning models. The system generates structured, evidence-based investment theses that support disciplined and interpretable trading decisions. \textsc{\model} Terminal will be released at \url{https://github.com/TauricResearch/Trading-R1}.
\end{abstract}

\section{Introduction}
Financial exchange predates written history, yet the founding of the Amsterdam Stock Exchange is commonly treated as the birth of the \emph{modern} securities market \cite{petram2014world}. Since then, scholars and practitioners have proposed a wide range of theories to explain price formation and guide trading, spanning sociological and psychological accounts, econometric models, and technical paradigms such as Elliott Wave, Dow Theory, and price action \cite{elliott1938wave, dow_theory, schwager2012market, lefevre2010reminiscences, brooks2009reading}. As time passes, the amount of available market data and computing power and technology have drastically changed and increased, quantitative methods have flourished, and advances in natural language processing have enabled large-scale analysis of unstructured sources of various modalities, including news, earnings disclosures, and macroeconomic reports, using tools such as sentiment analysis. Yet these signals are seldom integrated into a coherent decision framework but individually used as tools for analysts and firms. Instead, bespoke factors are typically engineered in isolation and left to human traders to interpret and combine.

Recent breakthroughs in large language models (LLMs) have transformed automated reasoning across domains. NLP has shifted from single-purpose models to promptable intelligence: general purpose systems augmented with chain-of-thought, self-verification, and reinforcement learning, that now can tackle complex reasoning tasks with increasing reliability and scope \cite{bai2023qwentechnicalreport, openai2025gptoss120bgptoss20bmodel, deepseekai2025deepseekr1incentivizingreasoningcapability, openai2024openaio1card}. Yet their application to finance remains nascent. Markets are dynamic, noisy, and multi-factor, demanding adaptive and interpretable reasoning under uncertainty, requirements that differ markedly from the math, coding, and science tasks that have dominated recent LLM optimizations \cite{lu2024mathvistaevaluatingmathematicalreasoning, hendrycks2021measuringmassivemultitasklanguage}. Alternatives exist, notably purely quantitative approaches \cite{ericson2024deepgenerativemodelingfinancial, fjellström2022longshorttermmemoryneural}, but they are often opaque and brittle across regimes; meanwhile, general-purpose reasoning LLMs struggle to ground their inferences in financial contexts, provide verifiable logic, and produce traceable decisions \cite{liu2025finr1largelanguagemodel, tatsat2025blackboxinterpretabilityllms, Lee_2025}. Progress is further hampered by sparse, fragmented public financial data, which complicates model training; moreover, there is a mismatch between open-ended financial QA benchmarks \cite{qian2025fino1transferabilityreasoningenhancedllms, liu2025finr1largelanguagemodel} and the structured, sequential reasoning that trading demands. Unlike QA, market decisions are inherently uncertain and path-dependent: even reasonable choices can yield divergent, unforeseen outcomes.

To address these gaps, we propose \textsc{\model}, a financial trading reasoning foundation model tailored for financial and specifically trading-oriented reasoning. We curate a high-quality dataset of over 100K publicly sourced financial reasoning samples and train \textsc{\model} via supervised fine-tuning followed by a curriculum of reinforcement learning from easy to hard. This design directly tackles the core limitations of existing reasoning LLMs in finance, advancing models that are both grounded in market complexity and practically usable for trading. Our key contributions are as follows:

\begin{itemize}[leftmargin=*, itemsep=2pt, topsep=3pt]
    \item \textbf{\database: A large-scale, diverse financial reasoning corpus.} We curate a comprehensive dataset spanning 18 months from January 1, 2024, to May 31, 2025, across 14 major tickers, integrating heterogeneous financial data used by real traders, such as technical market data, fundamentals, news, insider sentiment, and macroeconomic indicators. The final corpus contains 100k samples of filtered, high-quality financial information, paired with supervision via reverse reasoning distillation and volatility-aware reward labeling.

    \item \textbf{Supervised fine-tuning with reverse chain-of-thought distillation.} Since proprietary LLMs provide only final answers without intermediate reasoning, we reconstruct hidden reasoning traces from high-performing but opaque API models and use them as supervision for reasoning-oriented training. This approach enables our model to generate concise, interpretable, and high-quality investment theses distilled from the insights of state-of-the-art reasoning systems.
    
    \item \textbf{Reinforcement learning for execution-grade decisions.} Beyond thesis writing, we refine the model for actionable decision making. We cast trade recommendations as an RL problem, labeling assets on the standard five-tier investment scale (\emph{Strong Buy}, \emph{Buy}, \emph{Hold}, \emph{Sell}, \emph{Strong Sell}). Ratings are volatility-adjusted and used as rewards to align model outputs with realistic trading objectives.
    
    \item \textbf{\model: A financial reasoning LLM for trading.} We propose \model, a large-scale financial reasoning LLM trained across diverse assets and market conditions (bull and bear). The model demonstrates strong generalization in trading scenarios, producing both high-quality analyses and profitable trade recommendations.
\end{itemize}
\section{Related Work}
\subsection{Large Language Models in Finance}

Large Language Models (LLMs) have demonstrated impressive capabilities across many domains, including finance. To adapt them for financial tasks, researchers typically rely on three strategies: pretraining on domain-specific data, fine-tuning on task-specific datasets, and applying reinforcement learning to align model behavior with desired outcomes. Another line of work explores using pretrained models directly off the shelf as specialized experts within multi-agent systems. In these setups, LLMs are assigned distinct roles, and their coordinated interactions are designed to elicit more explicit financial reasoning and enhance the overall reasoning capabilities of the system.

\paragraph{Pretraining and Fine-tuning LLMs for Finance}

Domain adaptation for LLMs in finance follows two main approaches: pretraining from scratch on financial corpora and fine-tuning existing models on financial data.

Models like BloombergGPT \citep{wu2023bloomberggpt}, XuanYuan 2.0 \citep{zhang2023xuanyuan}, and Fin-T5 \citep{lu2023bbtfin} were trained on combined public and finance-specific datasets. BloombergGPT, leveraging proprietary Bloomberg data, outperforms general-purpose counterparts like BLOOM-176B in market sentiment classification and summarization tasks, while maintaining competitive general language understanding compared to similar-sized open-source models.

The fine-tuning approach is exemplified by models such as PIXIU (FinMA) \citep{xie2023pixiu}, which fine-tuned LLaMA on 136K finance-related instructions; FinGPT \citep{yang2023fingpt}, which used LoRA to adapt LLaMA and ChatGLM with approximately 50K finance-specific samples; and Instruct-FinGPT \citep{zhang2023instructfingpt}, fine-tuned on 10K instruction samples from financial sentiment analysis datasets. These models demonstrate stronger performance in finance classification tasks compared to their base versions and other open-source LLMs like BLOOM and OPT \citep{zhang2022opt}, sometimes even surpassing BloombergGPT. However, their performance in generative tasks still lags behind powerful general-purpose models like GPT-4, indicating the need for higher-quality, domain-specific datasets.

\paragraph{Reinforcement Learning for LLMs}

Reinforcement learning from human feedback \citep[RLHF]{ouyang2022training, kaufmann2023survey} has emerged as a cornerstone technique for aligning LLMs with human preferences \citep{lambert2024rewardbench}. This approach ranges from Proximal Policy Optimization \citep[PPO]{ppo} to Direct Preference Optimization \citep[DPO]{rafailov2023direct} and Simple Preference Optimization \citep[SimPO]{meng2024simpo}, which eliminate the need for explicit reward modeling and help stabilize training. Recent innovations such as Group Relative Policy Optimization \citep[GRPO]{shao2024deepseekmath} address computational challenges by optimizing group-wise comparisons and implementing batch-normalized rewards. Notable advancements include DeepSeek-R1's multi-stage RL training \citep{deepseekai2025deepseekr1incentivizingreasoningcapability} and personalized alignment via variational preference learning \citep{poddar2024personalizing}. Despite significant progress, fundamental limitations persist, including risks of reward hacking, off-policy instability, and the need for pluralistic alignment to accommodate diverse human preferences \citep{casper2023openproblemsfundamentallimitations, kaufmann2024surveyreinforcementlearninghuman}.

\paragraph{Multi-agent LLMs for Finance}
While training models on financial data can improve performance, limited resources and data availability often make off-the-shelf LLMs an attractive alternative. Although not specialized in finance, large general-purpose models excel at reasoning and instruction following, which has fueled the rise of agent systems, frameworks that equip LLMs with memory, tools, and role specialization to achieve complex goals. This paradigm has spread rapidly across domains from coding to AI4Science to computer use agents \cite{gottweis2025aicoscientist, hong2024metagptmetaprogrammingmultiagent, liu2025pcagenthierarchicalmultiagentcollaboration}. In finance, multi-agent systems are often designed to replicate real decision-making processes, such as hedge fund structures, by assigning agents distinct roles and tools (e.g., news retrieval, indicator calculation). Recent frameworks like \textit{TradingAgents} explicitly model financial institutions, combining structured communication and debate to produce detailed reasoning reports segmented by information sources \cite{xiao2025tradingagentsmultiagentsllmfinancial}.

\subsection{Large Language Models in Financial Trading}

LLMs are employed in financial trading across four primary application paradigms: information processing, reasoning-based decision making, reinforcement learning optimization, and alpha factor generation.

\paragraph{Information-Driven Trading}
Information-driven approaches process news and market data to generate trading signals. Studies evaluating both closed-source models (e.g., GPT-4.1, Claude 3.7) and open-source LLMs (e.g., Qwen \citep{bai2023qwentechnicalreport}) have demonstrated the effectiveness of simple long-short strategies based on sentiment scores \citep{lopezlira2023chatgptforecaststockprice}. Fine-tuned LLMs like FinGPT show improved performance through domain-specific alignment \citep{yang2023fingpt, unveiling, sentitrade}. Advanced methods involve summarizing financial news and reasoning about their relationship with stock prices \citep{beatunveiling, wang2024llmfactorextractingprofitablefactors}.

\paragraph{Reasoning-Enhanced Trading}
Reasoning-enhanced approaches leverage LLMs' analytical capabilities through reflection and multi-agent debate. Reflection-based systems, such as FinMem \citep{finmem} and FinAgent \citep{multimodalfinmem}, employ layered memorization and multimodal data to summarize inputs, inform decisions, and incorporate technical indicators, achieving better backtest performance while mitigating hallucinations \citep{ji2023mitigatinghallucinationlargelanguage}. Multi-agent frameworks \citep{xing2024designingheterogeneousllmagents, xiao2025tradingagentsmultiagentsllmfinancial} enhance reasoning and factual validity by employing LLM debates among specialized agents. Systems like TradingGPT \citep{li2023tradinggpt} demonstrate improved sentiment classification and increased robustness in trading decisions through this collaborative approach.

\paragraph{Reinforcement Learning Optimization} 
Reinforcement learning optimized trading systems use backtesting performance as rewards to refine decision-making processes. SEP \citep{Koa_2024} employs RL with memorization and reflection to refine LLM predictions based on market history. Classical RL methods are also integrated in frameworks that combine LLM-generated embeddings with stock features, trained via algorithms like Proximal Policy Optimization (PPO) \citep{ding2023integratingstockfeaturesglobal, ppo}. These approaches systematically improve LLM trading capabilities through iterative feedback loops.

\paragraph{Alpha Factor Generation}
Rather than directly making trading decisions, LLMs can generate alpha factors—signals that predict stock returns. QuantAgent \citep{wang2023alpha} employs a dual-loop architecture: an inner loop where a writer agent generates code from trading ideas with feedback from a judge agent, and an outer loop where the code is tested in real markets to enhance the judging agent. Similarly, AlphaGPT \citep{wang2023alpha} proposes a human-in-the-loop framework for alpha mining. These approaches leverage LLMs' capabilities to automate and accelerate trading strategy development through systematic generation and refinement of predictive signals.

\section{\model Methodology}

\subsection{Motivations}
Training a reasoning model for financial trading is uniquely challenging compared to other domains. Financial decisions are high-stakes, multifaceted, market-dependent, and highly sensitive to noise. Simply extending chains of thought through standard reasoning training does not necessarily improve model quality; instead, it can amplify hallucinations and degrade the reliability of generated trading decisions. Since language models are \emph{conditional autoregressive} generators, the quality of the final action depends on two coupled priors: (i) the \emph{external prior}, given by the input context that initiates generation, and (ii) the \emph{internal prior}, shaped by the model’s own previously generated tokens during roll-out. These dynamics lead to two practical imperatives:

\begin{itemize}[leftmargin=*]
\item \textbf{Input quality (external prior)} If the prompting context is noisy, misaligned, or low signal-to-noise, the model’s analysis is anchored to poor evidence, degrading downstream reasoning regardless of decoding or prompting.
\item \textbf{Reasoning scaffolding (internal prior)} During generation, poorly structured intermediate thoughts accumulate and result in brittle theses and unreliable decisions. Trading requires a disciplined investment thesis with clear structure, defensible claims, explicit evidence, and careful attention to risk. Providing such scaffolding ensures that the reasoning process remains coherent and that the final action is grounded in sound logic.
\end{itemize}

These challenges motivate our methodology for \model. We control both \emph{what} the model conditions on and \emph{how} it reasons toward the decision. Specifically, we (1) implement a rigorous data collection, cleaning, and assembly pipeline to provide high signal-to-noise, finance-grounded context at training time, and (2) employ a multi-stage, easy-to-hard curriculum for supervised fine-tuning and reinforcement learning that first teaches the model to structure an investment thesis, then to construct logical, evidence-based arguments, and finally to make decisions grounded in market dynamics. This design enables \model to reason like a professional trader, generating substantiated and transparent analyses that lead to coherent, actionable decisions rather than merely producing longer text.

\subsection{Input Data Collection at Scale}
To manage external priors and ensure high-quality training inputs, we implement a rigorous data collection process that spans diverse time periods, market conditions, sectors, and analytical modalities. In financial trading research, the main challenge is not gaining access to data, but selecting information that sharpens the signal-to-noise ratio and yields actionable insights. Our dataset is built from reliable sources capturing market dynamics, company fundamentals, and public sentiment. We define input data broadly, encompassing a holistic view of macroeconomic trends and firm-specific conditions—what companies do, how they perform, and how they are perceived. To build generalizable market intelligence and provide the model with the strongest possible priors for generating high-quality theses, three core objectives guide our collection process:

\paragraph{Breadth across companies}
We include data from a diverse set of stocks spanning sectors and market capitalizations. By focusing on more than a dozen widely followed firms (e.g., NVDA, AAPL, JNJ) over the 18 months period from January 1, 2024, to May 31, 2025, we capture a broad range of market conditions and corporate developments.

\paragraph{Depth of information}
For each day and for each given asset, we aggregate features spanning technical data, fundamentals, news, sentiment, and macroeconomic factors. Sources include Finnhub, SimFin, Google News scraping, and \texttt{stockstats}, yielding a dense, multi-perspective snapshot of each company.

\paragraph{Robustness to variation}
Real-world data is often incomplete or unbalanced. To enhance resilience, we vary input composition during label generation by randomly sampling from market data, news, sentiment, fundamentals, and macroeconomic factors. This approach trains the model to reason effectively even when information is limited. It also enables the model to detect complex patterns across contexts while remaining adaptable to real-world variability, which is essential for success in dynamic financial environments. Further methodological details are provided in Appendix \ref{app:data_collection}.

\subsection{\model Training Overview}
\begin{figure}[htbp]
  \centering
  \includegraphics[width=\linewidth]{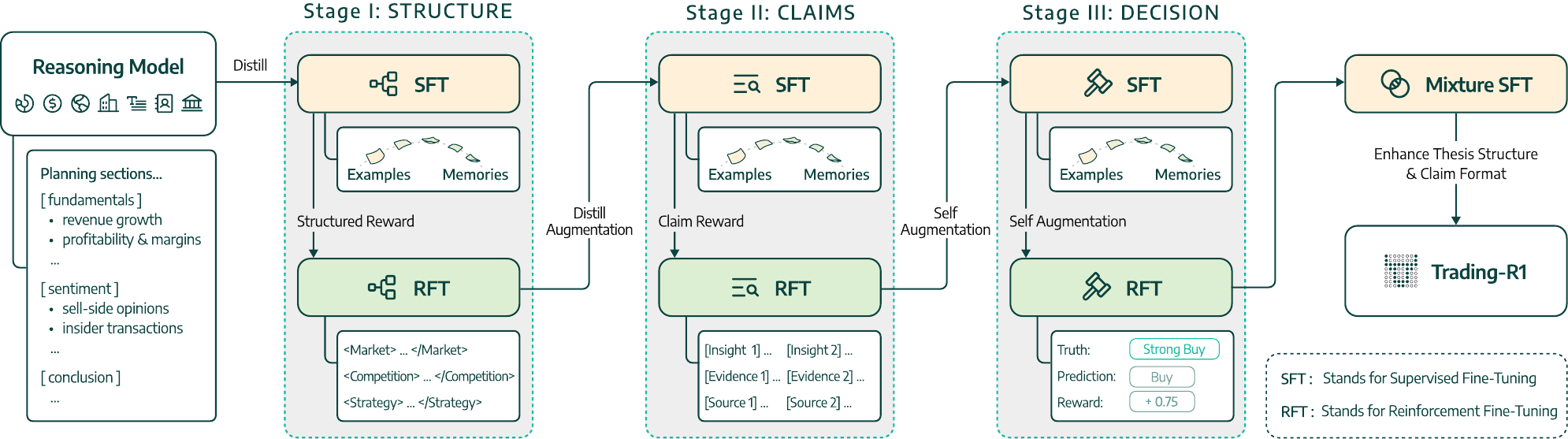}
  \caption{\model Training Schema}
  \label{fig:rl_curriculum}
\end{figure}

Managing the internal priors of an LLM’s generation is critical for trading. Without proper structure, intermediate reasoning steps can compound errors, producing brittle theses and unreliable final decisions. To address this, we design a multi-stage, easy-to-hard curriculum that interleaves supervised fine-tuning (SFT) with reinforcement learning fine-tuning (RFT) as seen in Figure~\ref{fig:rl_curriculum}. This curriculum progressively teaches the model to (i) structure its outputs like a professional investment thesis, (ii) construct logical and evidence-backed arguments, and (iii) make decisions grounded in real market dynamics. The curriculum unfolds across three stages, each warm-started with SFT (Figure~\ref{fig:sft}) to establish structural and stylistic priors, and refined with RFT to align behavior through task-specific rewards. This interleaving ensures that the model first acquires the general form of professional analysis before being guided toward evidence-grounded reasoning and, ultimately, market-aligned decision making. The staged progression stabilizes intermediate reasoning, mitigates error compounding, and builds the internal discipline required for coherent and actionable trading outputs.

\begin{table}[htbp]
\centering
\caption{Three-Stage Financial Trading Model Training Pipeline}
\label{tab:training_pipeline}
\resizebox{\linewidth}{!}{%
\begin{tabular}{@{}llll@{}}
\toprule
\textbf{Stage} & \textbf{Training Phase} & \textbf{Method Description} & \textbf{Objective} \\
\midrule
\multirow{6}{*}{\begin{tabular}[c]{@{}l@{}}\textbf{Stage I:}\\\textbf{STRUCTURE}\end{tabular}} 
 & SFT & Supervised Fine-Tuning on massive Qwen and & Initial structured thinking and \\
 & & OpenAI data (no reject sampling) & basic data organization \\
\cmidrule{2-4}
 & RFT & Reinforcement Fine-Tuning on sections & Enable systematic analysis and \\
 & & (introduction/claims/table/conclusion) & professional data categorization \\
\cmidrule{2-4}
 & Augmentation & Self-Distill with reject sampling cases & Reinforce structured reasoning \\
 & & with clear structure & patterns \\
\midrule
\multirow{6}{*}{\begin{tabular}[c]{@{}l@{}}\textbf{Stage II:}\\\textbf{CLAIMS}\end{tabular}} 
 & SFT & Supervised Fine-Tuning for evidence-based & Basic claim structure and \\
 & & reasoning foundation & evidence awareness \\
\cmidrule{2-4}
 & RFT & Reinforcement Fine-Tuning on opinion + & Ground claims with quotes and \\
 & & quote + source structure & sources, address hallucination \\
\cmidrule{2-4}
 & Augmentation & Self-Distill with reject sampling cases & Reinforce evidence-based \\
 & & with professional and faithful claims & professional reasoning \\
\midrule
\multirow{6}{*}{\begin{tabular}[c]{@{}l@{}}\textbf{Stage III:}\\\textbf{DECISION}\end{tabular}} 
 & SFT & Supervised Fine-Tuning for investment & Basic decision-making structure \\
 & & recommendation patterns & for investments \\
\cmidrule{2-4}
 & RFT & Reinforcement Fine-Tuning with Equity & Generate market-aware \\
 & & \& Volatility \& Smooth Adj & investment recommendations \\
\cmidrule{2-4}
 & Augmentation & Self-Distill with reject sampling prediction & Reinforce accurate directional \\
 & & (directional) correct cases & predictions \\
\bottomrule
\end{tabular}%
}
\end{table}

\begin{figure}[htbp]
  \centering
  \includegraphics[width=\linewidth]{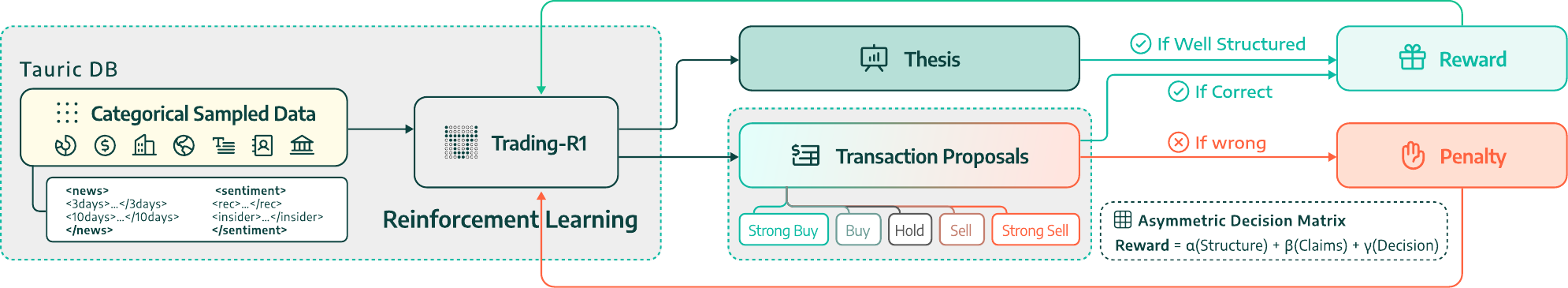}
  \caption{Reinforcement learning on Thesis Structure, Statement, and Decision}
  \label{fig:outcome_rl}
\end{figure}

In the \textbf{formatting stage}, the model is rewarded for following the professional structure of investment theses, systematically organizing technical, fundamental, and sentiment-based analyses. XML-tagged formatting is reinforced at this stage to promote consistent reasoning patterns and stable structured outputs. In the \textbf{evidence-grounding stage}, rewards encourage the model to support claims with direct citations and quotations from the input context, reducing hallucinations and fostering disciplined, evidence-based reasoning. Finally, in the \textbf{decision stage}, the model is trained with outcome-based rewards derived from the volatility-aware labels in Section~\ref{sec:label_generation}, penalizing poor predictions and incentivizing decisions that align with verifiable market outcomes, as illustrated in Figure~\ref{fig:outcome_rl}. Through this progression, \model learns first to produce the correct form of professional analysis, then to anchor its reasoning in evidence, and ultimately to generate coherent, market-driven trading decisions. 

\subsection{Supervised Investment Reasoning Distillation}

\begin{figure}[htbp]
  \centering
  \begin{subfigure}{\linewidth}
    \centering
    \includegraphics[width=\linewidth]{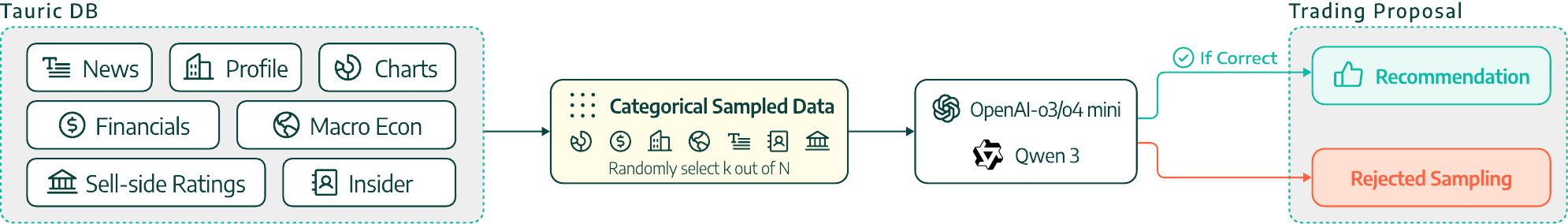}
    \caption{Investment Thesis Distillation from OpenAI Reasoning Models.}
    \label{fig:thesis}
  \end{subfigure}
  
  \vspace{0.8em} 
  
  \begin{subfigure}{\linewidth}
    \centering
    \includegraphics[width=\linewidth]{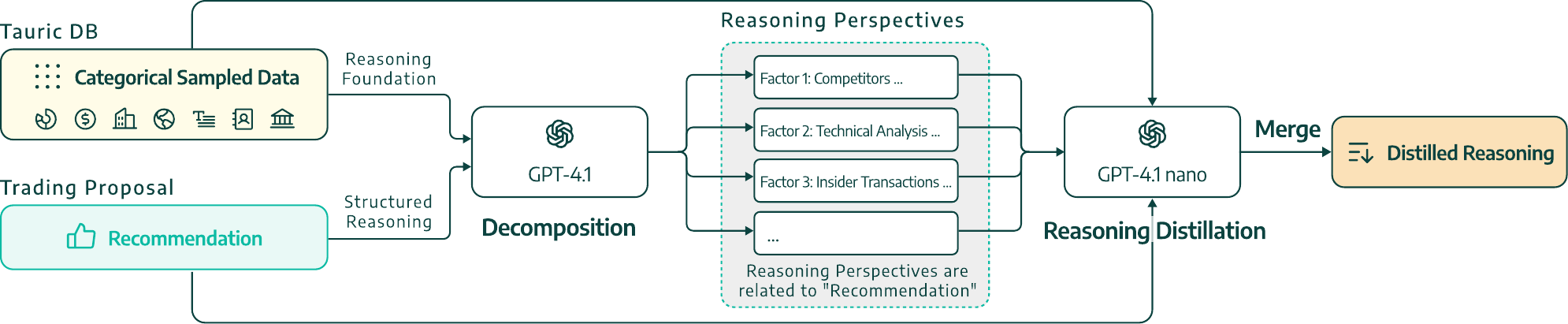}
    \caption{Reverse Reasoning Distillation.}
    \label{fig:rev_cot}
  \end{subfigure}
  
  \caption{Overview of \model distillation: (a) investment thesis distillation, (b) reverse reasoning distillation.}
  \label{fig:distillation_methods}
\end{figure}

To support the SFT warm-start stages in the interleaved easy-to-hard curriculum of \model (Figure~\ref{fig:rl_curriculum}), high-quality reasoning traces are required as supervised targets. Yet obtaining such labels for large language models (LLMs) is notoriously costly, and the difficulty is magnified in the financial domain, where ground truth is often ambiguous, unavailable, or prohibitively expensive to produce. To address this, we leverage the volatility-driven labeling method introduced in Section~\ref{sec:label_generation} to automatically construct \textit{input-to-investment-thesis pairs} for SFT. Each investment thesis provides a detailed reasoning trace that logically supports a trading decision consistent with the volatility-aware label assigned for that day.

\paragraph{Reverse Reasoning Distillation}
To overcome the difficulty of sourcing such detailed reasoning traces, we introduce a novel technique we call \textit{reverse reasoning distillation}. While commercial LLMs accessed via APIs (e.g., OpenAI's \texttt{o1}, \texttt{o3}) consistently outperform most open-source models in reasoning quality, they typically do not expose their full chain-of-thought (CoT) outputs for easy full distillation, returning only final conclusions without explanatory steps. To extract high-quality, long-form reasoning without hosting massive models ourselves, we propose a method for synthetically reconstructing reasoning paths from these black-box models.

As illustrated in Figure~\ref{fig:thesis}, we begin by inputting structured financial data for a specific ticker on a given date into proprietary reasoning models, such as \texttt{o3-mini} or \texttt{o4-mini}, and retrieve its final trading recommendation (i.e., a \textit{front-end response}). Next, as shown in Figure~\ref{fig:rev_cot}, we pass this final response, along with the original input, into a dedicated planner LLM tasked with inferring the key reasoning steps required to arrive at the given conclusion. To simulate the full reasoning process, we then use a lightweight LLM (e.g., \texttt{GPT-4.1-nano}) to elaborate how each data modality (e.g., market data, news, social media, fundamentals) contributes to the investment decision. These segments are then programmatically stitched into a coherent reasoning trace. The end result is a high-quality, synthetic dataset of structured financial inputs paired with plausible, step-by-step investment theses, suitable for use in SFT pipelines.

\subsection{Volatility-Driven Discretization for Label Generation}
\label{sec:label_generation}

Once a broad and diverse input corpus is assembled, the next step is to define a reliable target label. This serves as a clear indicator of the optimal trading action at a given point in time and supports market-verifiable, reward-driven reinforcement learning. Instead of attempting to predict exact future price movements, which are noisy, unstable, and especially difficult for language models to capture, we discretize the output space into five intuitive actions: \texttt{strong sell}, \texttt{sell}, \texttt{hold}, \texttt{buy}, and \texttt{strong buy}. This design serves two purposes. First, it mirrors real-world trading, where decisions are expressed as actions rather than precise price forecasts. Second, it provides a natural mapping from outputs to portfolio allocation weights that can be tailored to user-specific risk preferences.

Labels are generated using a principled, multi-horizon volatility-aware procedure. For each training instance, we sample inputs across all modalities (market, news, sentiment, fundamentals, and macroeconomic information) and construct a composite signal from multiple time horizons. Specifically, we compute exponential moving average (EMA) prices and calculate forward returns over 3, 7, and 15-day periods. Each return series is normalized by its rolling 20-period volatility to create Sharpe-like signals. These signals are then combined using empirically-determined weights (0.3, 0.5, 0.2 respectively) to form a composite weighted signal. Finally, labels are assigned based on percentile thresholds computed from the distribution of valid weighted signals, using asymmetric quantiles (85\%, 53\%, 15\%, 3\%) that reflect market dynamics. The full procedure is described in Algorithm~\ref{alg:vol_signals} and Appendix~\ref{app:signal_gen_alg}. This multi-horizon volatility-aware design provides four advantages. \textit{(i)}~Signals capture both short-term momentum and medium-term trends through multiple time horizons. \textit{(ii)}~Volatility normalization ensures consistent signal strength across different market regimes. \textit{(iii)}~The weighted combination balances immediate price action with broader trend information. \textit{(iv)}~Asymmetric quantile cutoffs preserve the market's long-term upward drift while maintaining class diversity for robust training.

The resulting proxy labels are highly valuable for downstream learning. They supply a natural reward signal for reinforcement learning (Section~\ref{sec:rl}) and enable the scalable creation of high-quality targets for supervised fine-tuning (Section~\ref{sec:sft}). This significantly lowers the cost of reasoning-based supervision, which would otherwise depend on manual or expert annotation.

\subsection{Supervised Fine-Tuning for Structured Analysis} \label{sec:sft}

\begin{figure}[htbp]
  \centering
  \includegraphics[width=\linewidth]{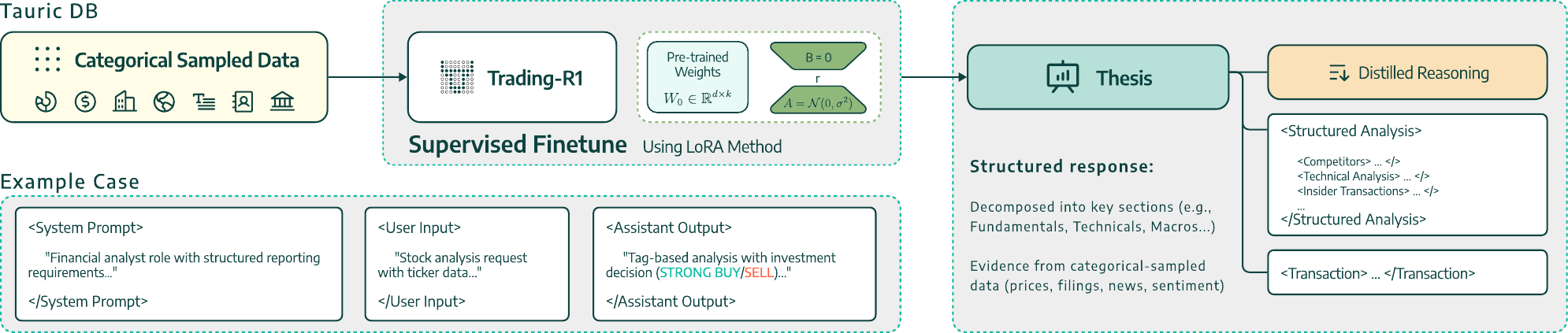}
  \caption{Supervised finetuning with reverse reasoning distilled data from \database}
  \label{fig:sft}
\end{figure}

\paragraph{SFT Warm-Start for Structured Reasoning}  
Using high-quality investment theses generated through reverse reasoning distillation, we perform SFT to warm-start each stage of the easy-to-hard curriculum (Figure~\ref{fig:sft}). Each training instance pairs structured market data with a detailed investment thesis, teaching the model to analyze, synthesize, and decide in ways that mirror expert financial workflows. Because the distillation process is controllable, stage-specific SFT targets can be designed to establish the correct reasoning priors before RFT refinement.  

\paragraph{Stage-Wise Targets}  
In \textbf{Stage I (Structure)}, SFT emphasizes the professional organization of theses, instilling structured thinking and systematic data organization. In \textbf{Stage II (Claims)}, SFT introduces evidence-based reasoning, guiding the model to build claims anchored in data. In \textbf{Stage III (Decision)}, SFT focuses on investment recommendation patterns, preparing the model to structure outputs around actionable decisions. This staged warm-start stabilizes intermediate reasoning, reduces compounding errors, and ensures that RFT operates on strong structural and evidential priors.  

\paragraph{Backbone Model and Stability}  
We adopt Qwen3-4B as the backbone model, since it is already optimized for reasoning tasks. This prior accelerates convergence during both SFT and RFT, while improving the model’s ability to generate structured, interpretable outputs. Without this warm-start, models tended to overfit to superficial heuristics, forget structures from earlier stages, and produce brittle, incoherent theses. Staged SFT initialization instead provides disciplined scaffolding that preserves prior knowledge and allows reinforcement learning to refine rather than overwrite the model’s analytical capabilities.

\subsection{Reinforcement Learning Fine-Tuning for Market-Aligned Decisions}
\label{sec:rl}

\paragraph{RFT Fine-Tuning after SFT Warm-Start}
While SFT equips the model with structured and interpretable reasoning, it often overfits to superficial patterns and falls short of producing decisions that are both robust and actionable. Directly transitioning from SFT to outcome-based RL is unstable, as the model lacks the discipline to balance structured reasoning with verifiable market performance. To address this, our interleaved easy-to-hard curriculum applies RFT after each SFT warm-start, reinforcing the stage-specific priors with outcome-based feedback. In this way, RFT refines the model’s reasoning so that quality analyses translate into coherent, market-aligned actions. Detailed reward specifications for each stage are provided in Appendix~\ref{app:reward_design}.

\paragraph{Action Space and Labeling}  
We define a five-class action space—\texttt{strong sell}, \texttt{sell}, \texttt{hold}, \texttt{buy}, \texttt{strong buy}—mapped to portfolio weights. This design reflects varying degrees of conviction and enables finer-grained position control compared to the traditional \texttt{buy/hold/sell} triad. Labels are asset-specific, constructed from the multi-horizon volatility-aware procedure described in Section~\ref{sec:label_generation}. To better capture market dynamics, we project returns onto asymmetric quantiles, producing a skewed distribution that reflects both the equity market's long-term upward drift and the growth-oriented characteristics of our blue-chip training universe, while maintaining sufficient class diversity for robust training:

\begin{table}[h]
\caption{Target class distribution for trading actions}
\label{tab:action_distribution}
\centering
\small
\begin{tabular}{*{5}{c}}
\toprule
\texttt{Strong Buy} & \texttt{Buy} & \texttt{Hold} & \texttt{Sell} & \texttt{Strong Sell} \\
\midrule
15\% & 32\% & 38\% & 12\% & 3\% \\
\bottomrule
\end{tabular}
\end{table}

This distribution encourages the model to learn realistic, market-consistent policies while preserving class diversity for robust training. The asymmetric allocation reflects both empirical equity market behavior and established analyst practices, with a bias toward positive actions that is particularly justified by our training universe composition. Our portfolio focuses on large-cap and mega-cap blue-chip companies—including technology leaders like NVIDIA, Microsoft, and Apple, established financials like Berkshire Hathaway and JPMorgan Chase, healthcare giants like Eli Lilly and Johnson \& Johnson, and broad market ETFs like SPY and QQQ. These companies collectively represent over \$11 trillion in market capitalization and are characterized by solid fundamentals, robust cash flows, dominant market positions, and strong competitive moats within their respective sectors. Given the inherent quality and growth orientation of this blue-chip universe, a structural bullish bias in the action distribution aligns with the long-term appreciation potential of these market-leading assets. Importantly, because \model trades with long–short strategies, a \texttt{sell} or \texttt{strong sell} signal implies initiating short positions rather than merely closing longs. While shorting introduces practical feasibility challenges, incorporating it during training provides a richer action space and sharper signal discrimination for tactical positioning around these fundamentally strong companies.  

\paragraph{Time Horizon}  
We target \textbf{medium-term strategies} with holding periods on the order of one week. This horizon balances actionability with feasibility: excluding high-frequency trading (limited by LLM inference latency) while avoiding long-horizon investing, which requires macroeconomic foresight beyond the current capabilities of language models. Medium-term trading provides a natural setting where structured reasoning, evidence grounding, and outcome alignment can be most effectively combined.

\paragraph{Policy Optimization}  
To optimize the policy during reinforcement learning, we adopt \textit{Group Relative Policy Optimization} (GRPO), a recent variant of Proximal Policy Optimization (PPO) that eliminates the need for a separate value model \cite{ppo, shao2024deepseekmathpushinglimitsmathematical}. Whereas PPO estimates per-token advantages with a learned value function, GRPO derives the baseline directly from a group of sampled trajectories for the same input. This relative scoring stabilizes training and reduces memory overhead. Concretely, for each input $q$, we sample $G$ candidate outputs $\{o_i\}_{i=1}^G$ from the old policy $\pi_{\theta_{\text{old}}}$ and assign each output a reward $r_i$ from the reward model. The group-relative advantage normalizes each candidate by its peers:
\[
\hat{A}_i = \frac{r_i - \mathrm{mean}(r)}{\mathrm{std}(r)}\!,
\qquad
r_t^{(i)}(\theta) \coloneqq 
\frac{\pi_\theta(o_{i,t}\mid q,o_{i,<t})}
     {\pi_{\theta_{\mathrm{old}}}(o_{i,t}\mid q,o_{i,<t})}.
\]

The GRPO objective is then:
\begin{align}
J_{\mathrm{GRPO}}(\theta)
&= \mathbb{E}_{q,\{o_i\}}\!\left[
\frac{1}{G}\sum_{i=1}^G \frac{1}{|o_i|}
\sum_{t=1}^{|o_i|}
\min\!\Big(
r_t^{(i)}(\theta)\,\hat{A}_{i,t},\,
\mathrm{clip}\!\big(r_t^{(i)}(\theta),\,1-\epsilon,\,1+\epsilon\big)\,\hat{A}_{i,t}
\Big)\right] \nonumber\\
&\quad - \beta\,\mathbb{E}_{q}\!\left[
D_{\mathrm{KL}}\!\big(\pi_\theta(\cdot\mid q)\,\|\,\pi_{\mathrm{ref}}(\cdot\mid q)\big)
\right],
\label{eq:grpo_obj}
\end{align}
where $\epsilon$ controls clipping, $\beta$ scales the KL penalty to the reference SFT model $\pi_{\text{ref}}$, and $\hat{A}_{i,t}$ is the normalized group-relative advantage.

For \model training, the reward $r_i$ integrates the \emph{structure}, \emph{evidence}, and \emph{decision} components (Section~\ref{app:reward_design}). Each sampled output is thus judged not only on the correctness of its trading decision, but also on the coherence of its thesis structure and the grounding of its claims. This holistic scoring aligns naturally with GRPO’s group-relative framework. Together, GRPO provides stable optimization without requiring a critic model, while our three-stage reward system supplies task-specific shaping signals that progressively refine structured reasoning, evidence-backed claims, and market-aligned trading decisions. Full reward formulations are deferred to Appendix~\ref{app:reward_design}.

\section{Experiments}
\label{sec:evaluation}

\subsection{Training Details}
Training \model involved processing multi-dimensional financial inputs (20--30k tokens) and generating comprehensive investment theses (6--8k tokens). The supervised fine-tuning stage with LoRA~\cite{hu2021loralowrankadaptationlarge} was conducted on one 8×H100 server (96GB), while the reinforcement learning stage utilized one 8×H200 server (141GB). This RL stage enhances the model's ability to transition from analytical reasoning to high-confidence trading decisions, completing the full pipeline from insight to action.

Our training portfolio encompasses a strategically selected universe of large-cap equity assets representing diverse market sectors and investment vehicles. The portfolio concentrates on mega-cap technology leaders including NVIDIA, Microsoft, and Apple, which collectively represent over \$11 trillion in market capitalization and serve as primary drivers of modern equity market dynamics. Beyond technology, the selection spans communication services (Meta), consumer discretionary (Amazon, Tesla), financials (Berkshire Hathaway, JPMorgan Chase), healthcare (Eli Lilly, Johnson \& Johnson), and energy (Exxon Mobil, Chevron). Additionally, two major ETFs (SPY and QQQ) provide exposure to broader market beta and technology sector concentration, respectively. This curated selection ensures \model encounters the full spectrum of market regimes, sector dynamics, and volatility patterns that characterize modern institutional trading environments. The complete portfolio breakdown by sector and market capitalization is detailed in Appendix~\ref{tab:portfolio_sectors}.

\subsection{Data, Prompts, and Reward Structure}
Inputs are standardized across models to ensure comparability. Each prompt provides a structured snapshot of market data, fundamentals, social sentiment, and recent news headlines for a given asset-day. Models are required to generate an investment thesis followed by a trading decision mapped to a five-class discrete action space (\texttt{Strong Sell, Sell, Hold, Buy, Strong Buy}).

Rewards are derived from the volatility-adjusted, percentile-based labeling scheme introduced in Section~\ref{sec:label_generation}. Labels are calibrated to each asset's return distribution, reflecting differences in volatility and drift. The resulting asymmetric target distribution (detailed in Table~\ref{tab:action_distribution}) mirrors both empirical equity market behavior and established analyst practices. By tailoring label distributions per asset, we ensure that both training rewards and evaluation outcomes are realistic, risk-aware, and aligned with professional financial analysis.

\subsection{Experimental Design and Evaluation Methodology}

We evaluate \textsc{Trading-R1} using a comprehensive historical backtesting framework on a curated set of high-volume equities, including Apple (AAPL), Google (GOOGL), and Amazon (AMZN), along with widely traded ETFs such as SPY. The backtest covers June 1 to August 31, 2024, a held-out period excluded from the training data that reflects diverse market conditions and provides a realistic benchmark for assessing generalization and robustness.

\paragraph{Baseline Models} 
We compare \model against a broad range of LLM-based analysis tools spanning small, medium, and large model classes. For small language models (SLMs), we evaluate \textsc{Qwen-4B}, \textsc{GPT-4.1-nano}, and \textsc{GPT-4.1-mini}.  For larger LLMs, we include \textsc{GPT-4.1}, \textsc{LLaMA-3.3}, \textsc{LLaMA-Scout}, and \textsc{Qwen3-32B}.  For reinforcement learning–enhanced models (RLMs), we consider \textsc{DeepSeek}, \textsc{O3-mini}, and \textsc{O4-mini}. In addition, we conduct an ablation study on \model variants: one initialized with SFT warm-start and another trained with RL only, to better understand the contributions of each training stage to overall performance.

\paragraph{Evaluation Metrics}
We evaluate model performance using standard finance metrics that capture both profitability and risk characteristics. Our evaluation framework includes Cumulative Return (CR) to measure total returns, Sharpe Ratio (SR) to assess risk-adjusted performance, Hit Rate (HR) to evaluate prediction accuracy, and Maximum Drawdown (MDD) to quantify downside risk. These metrics provide a comprehensive assessment of trading strategy effectiveness across different market conditions. Detailed mathematical definitions and calculation procedures for all metrics are provided in Appendix~\ref{app:signal_gen_alg}.

\paragraph{Backtesting Simulation}
We adopt a standard backtesting setup based on historical market data, collecting multi-dimensional inputs for \model such as daily news, price data, and derived indicators, like \database. Trades are executed using only information available up to each trading day, eliminating look-ahead bias and ensuring strictly causal evaluation in a fully out-of-sample setting. This controlled design isolates the effect of model quality on trading performance.

\section{Results}

Our experimental results demonstrate a clear hierarchy in trading performance across different model categories. Tables~\ref{tab:results_nvda_aapl_msft} and \ref{tab:results_amzn_meta_spy} present the performance metrics for \textsc{\model}.

\begin{table}[ht]
    \centering
    \caption{Model Performance on NVDA, AAPL, MSFT. Bolded green values indicate \textbf{\textcolor{FBest}{Best Performance}}, underlined green values indicate \underline{\textcolor{FBest}{Second Best Performance}}}

    \label{tab:results_nvda_aapl_msft}
    \resizebox{\textwidth}{!}{%
    \begin{tabular}{llcccccccccccc}
    \toprule
    & & \multicolumn{4}{c}{\textbf{NVDA}} & \multicolumn{4}{c}{\textbf{AAPL}} & \multicolumn{4}{c}{\textbf{MSFT}} \\
    \textbf{Category} & \textbf{Model} & \textbf{CR(\%)} & \textbf{SR} & \textbf{HR(\%)} & \textbf{MDD(\%)} & \textbf{CR(\%)} & \textbf{SR} & \textbf{HR(\%)} & \textbf{MDD(\%)} & \textbf{CR(\%)} & \textbf{SR} & \textbf{HR(\%)} & \textbf{MDD(\%)} \\
    \midrule
    \multirow{3}{*}{\textbf{SLM}} & Qwen-4B & -1.59 & -1.62 & 52.2 & 2.80 & -0.81 & -0.92 & 41.7 & 3.76 & -1.45 & -1.28 & 50.0 & 4.38 \\
     & GPT-4.1-nano & 0.76 & -0.09 & 56.0 & 3.82 & 0.44 & -0.31 & 51.9 & 3.52 & -0.01 & -0.95 & 39.3 & 1.60 \\
     & GPT-4.1-mini & 0.29 & -0.53 & 58.8 & \underline{\textcolor{FBest}{2.47}} & -2.14 & -1.92 & 40.0 & 3.69 & -2.34 & -1.74 & 27.3 & 4.00 \\
    \midrule
    \multirow{4}{*}{\textbf{LLM}} & GPT-4.1 & 3.15 & 0.85 & 65.5 & 2.81 & 4.02 & 1.24 & 50.0 & \underline{\textcolor{FBest}{2.89}} & \underline{\textcolor{FBest}{2.30}} & \underline{\textcolor{FBest}{0.97}} & \underline{\textcolor{FBest}{63.9}} & 1.92 \\
      & LLaMA-3.3 & 0.65 & -0.16 & 62.2 & 2.78 & \textbf{\textcolor{FBest}{6.73}} & \underline{\textcolor{FBest}{1.78}} & \textbf{\textcolor{FBest}{63.6}} & \textbf{\textcolor{FBest}{2.40}} & 1.58 & 0.54 & 58.1 & 1.59 \\
     & LLaMA-Scout & -1.96 & -1.64 & 31.8 & 2.90 & 2.03 & 0.58 & \underline{\textcolor{FBest}{59.4}} & 3.21 & -0.29 & -1.33 & 36.8 & 1.44 \\
     & Qwen3-32B & 1.74 & 0.27 & 64.5 & 2.80 & 0.62 & -0.12 & 33.3 & 3.39 & 2.14 & \textbf{\textcolor{FBest}{1.29}} & \textbf{\textcolor{FBest}{65.6}} & \textbf{\textcolor{FBest}{0.82}} \\
    \midrule
    \multirow{3}{*}{\textbf{RLM}} & DeepSeek & -0.79 & -0.66 & 50.0 & 3.66 & 0.68 & -0.13 & 55.3 & 4.78 & -0.38 & -1.01 & 33.3 & 2.06 \\
     & O3-mini & -2.97 & -1.48 & 46.9 & 5.33 & -1.89 & -1.13 & 50.0 & 3.72 & 1.19 & 0.15 & 47.4 & \underline{\textcolor{FBest}{1.19}} \\
     & O4-mini & -0.99 & -0.83 & 43.2 & 3.61 & -3.19 & -1.36 & 50.0 & 7.88 & -1.72 & -1.77 & 48.5 & 2.35 \\
    \midrule
    \multirow{3}{*}{\textbf{Ours}} & Supervise Finetune & \underline{\textcolor{FBest}{7.42}} & \textbf{\textcolor{FBest}{2.72}} & \textbf{\textcolor{FBest}{72.5}} & \textbf{\textcolor{FBest}{2.01}} & -2.37 & -1.27 & 45.2 & 5.20 & -0.24 & -0.64 & 56.1 & 3.87 \\
     & Reinforcement Learning & 3.27 & \underline{\textcolor{FBest}{1.25}} & 62.5 & 2.73 & 4.04 & 1.14 & 57.1 & 3.02 & -0.18 & -0.81 & 45.7 & 1.66 \\
     & \textsc{\model} & \textbf{\textcolor{FBest}{8.08}} & \textbf{\textcolor{FBest}{2.72}} & \underline{\textcolor{FBest}{70.0}} & 3.80 & \underline{\textcolor{FBest}{5.82}} & \textbf{\textcolor{FBest}{1.80}} & \textbf{\textcolor{FBest}{63.6}} & 3.68 & \textbf{\textcolor{FBest}{2.38}} & 0.87 & 60.4 & 1.90 \\
    \bottomrule
    \end{tabular}%
    }
\end{table}

Small Language Models (SLMs) perform the weakest, struggling with profitability due to their limited parameter capacity and shallow reasoning, which leads to unstable analyses, weak argumentation, and poor overall decision quality. Reasoning Language Models (RLMs) achieve modest improvements over SLMs but face significant challenges: their limited instruction-following ability sometimes prevents them from producing decisions in the required format, and their lengthy reasoning paths often drift away from market-relevant data. Large Language Models (LLMs) outperform both categories, demonstrating stronger consistency and decision quality even without domain-specific training.

Interestingly, despite their advanced reasoning capabilities, off-the-shelf RLMs often underperform LLMs on trading tasks. This underperformance stems from their unguided reasoning processes, which can drift away from financial analysis and result in unfocused outputs. In contrast, the \model series (SFT, RFT, and full \model) highlights the importance of specialized training: SFT enforces professional output formats and consistent decision-making patterns, while RFT progressively aligns reasoning with market outcomes.  

\begin{table}[ht]
    \centering
    \caption{Model Performance on AMZN, META, SPY. Bolded green values indicate \textbf{\textcolor{FBest}{Best Performance}}, underlined green values indicate \underline{\textcolor{FBest}{Second Best Performance}}}
    \label{tab:results_amzn_meta_spy}
    \resizebox{\textwidth}{!}{%
    \begin{tabular}{llcccccccccccc}
    \toprule
    & & \multicolumn{4}{c}{\textbf{AMZN}} & \multicolumn{4}{c}{\textbf{META}} & \multicolumn{4}{c}{\textbf{SPY}} \\
    \textbf{Category} & \textbf{Model} & \textbf{CR(\%)} & \textbf{SR} & \textbf{HR(\%)} & \textbf{MDD(\%)} & \textbf{CR(\%)} & \textbf{SR} & \textbf{HR(\%)} & \textbf{MDD(\%)} & \textbf{CR(\%)} & \textbf{SR} & \textbf{HR(\%)} & \textbf{MDD(\%)} \\
    \midrule
    \multirow{3}{*}{\textbf{SLM}} & Qwen-4B & -2.90 & -1.13 & 46.2 & 6.05 & 1.32 & 0.14 & 51.7 & 3.80 & -1.33 & -3.37 & 42.3 & 1.71 \\
     & GPT-4.1-nano & -4.88 & -2.34 & 40.7 & 6.20 & -3.07 & -1.69 & 47.8 & 5.19 & 0.04 & -1.23 & 47.6 & 1.38 \\
     & GPT-4.1-mini & 2.24 & 0.81 & 50.0 & \underline{\textcolor{FBest}{2.01}} & 1.21 & 0.16 & 56.5 & \textbf{\textcolor{FBest}{1.70}} & -1.03 & -2.47 & 43.5 & 1.44 \\
    \midrule
    \multirow{4}{*}{\textbf{LLM}} & GPT-4.1 & 3.80 & 1.15 & \textbf{\textcolor{FBest}{64.3}} & 2.44 & \textbf{\textcolor{FBest}{5.63}} & \textbf{\textcolor{FBest}{1.59}} & \underline{\textcolor{FBest}{68.8}} & \underline{\textcolor{FBest}{1.91}} & 0.35 & -0.74 & 43.3 & 1.21 \\
     & LLaMA-3.3 & -0.89 & -0.61 & 58.6 & 6.02 & 3.21 & \underline{\textcolor{FBest}{1.01}} & 62.5 & 2.55 & 1.27 & 0.27 & 64.7 & 1.35 \\
     & LLaMA-Scout & -3.47 & -1.48 & 35.7 & 5.95 & 3.51 & 0.92 & 53.1 & 2.78 & -1.34 & -3.36 & 36.0 & 1.65 \\
      & Qwen3-32B & \textbf{\textcolor{FBest}{5.61}} & \textbf{\textcolor{FBest}{2.12}} & \textbf{\textcolor{FBest}{64.3}} & \textbf{\textcolor{FBest}{1.89}} & -1.23 & -0.58 & 46.2 & 6.61 & \underline{\textcolor{FBest}{2.32}} & \textbf{\textcolor{FBest}{1.87}} & \textbf{\textcolor{FBest}{70.4}} & \underline{\textcolor{FBest}{0.65}} \\
    \midrule
    \multirow{3}{*}{\textbf{RLM}} & DeepSeek & -1.15 & -1.14 & 50.0 & 3.00 & 1.26 & 0.12 & 40.5 & 2.80 & -1.15 & -1.82 & 36.4 & 2.00 \\
     & O3-mini & -3.15 & -1.37 & 38.2 & 5.50 & 2.05 & 0.53 & \textbf{\textcolor{FBest}{73.1}} & 2.64 & 0.80 & -0.25 & 57.6 & \textbf{\textcolor{FBest}{0.62}} \\
     & O4-mini & -2.48 & -1.28 & 51.6 & 4.83 & -0.45 & -0.80 & 53.6 & 2.68 & -0.30 & -1.34 & 36.8 & 1.72 \\
    \midrule
    \multirow{3}{*}{\textbf{Ours}} & Supervise Finetune & 1.93 & 0.36 & 60.6 & 4.28 & 2.52 & 0.54 & 55.9 & 2.93 & 1.78 & 0.86 & 58.1 & 1.15 \\
      & Reinforcement Learning & -0.05 & -0.29 & 52.5 & 4.84 & -0.18 & -0.36 & 44.4 & 5.11 & 1.85 & 1.00 & \underline{\textcolor{FBest}{67.6}} & 0.69 \\
     & \textsc{\model} & \underline{\textcolor{FBest}{5.39}} & \underline{\textcolor{FBest}{1.72}} & \underline{\textcolor{FBest}{63.0}} & 3.20 & \underline{\textcolor{FBest}{5.12}} & 0.86 & 50.0 & 4.65 & \textbf{\textcolor{FBest}{3.34}} & \underline{\textcolor{FBest}{1.60}} & 64.0 & 1.52 \\
    \bottomrule
    \end{tabular}%
    }
\end{table}

We believe this trend reflects differences in training focus. General-purpose LLMs, exposed to a massive variety of user instructions during instruction tuning, remain flexible and open-ended in how they approach problems. RLMs, by contrast, have recently been optimized for narrow domains such as coding, mathematics, and scientific reasoning. This specialization yields strong performance in those fields but limits generalization, making them less effective for financial reasoning tasks. Although finance overlaps with mathematics and science, financial data differs in key ways: it is noisy, ambiguous, and filled with mixed signals, which makes step-by-step, verifiable rewards difficult to define. As a result, purely RFT-based approaches are not feasible off the shelf for financial LLM training. Our \model addresses these challenges by combining the strengths of both paradigms. Through SFT, it integrates structured thesis writing and consistent decision-making patterns, while RFT progressively reinforces stage-specific behaviors. This design stabilizes reasoning, prevents drift, and enables coherent, market-aligned trading decisions.

\begin{figure}[ht]
  \centering
  \includegraphics[width=0.55\linewidth]{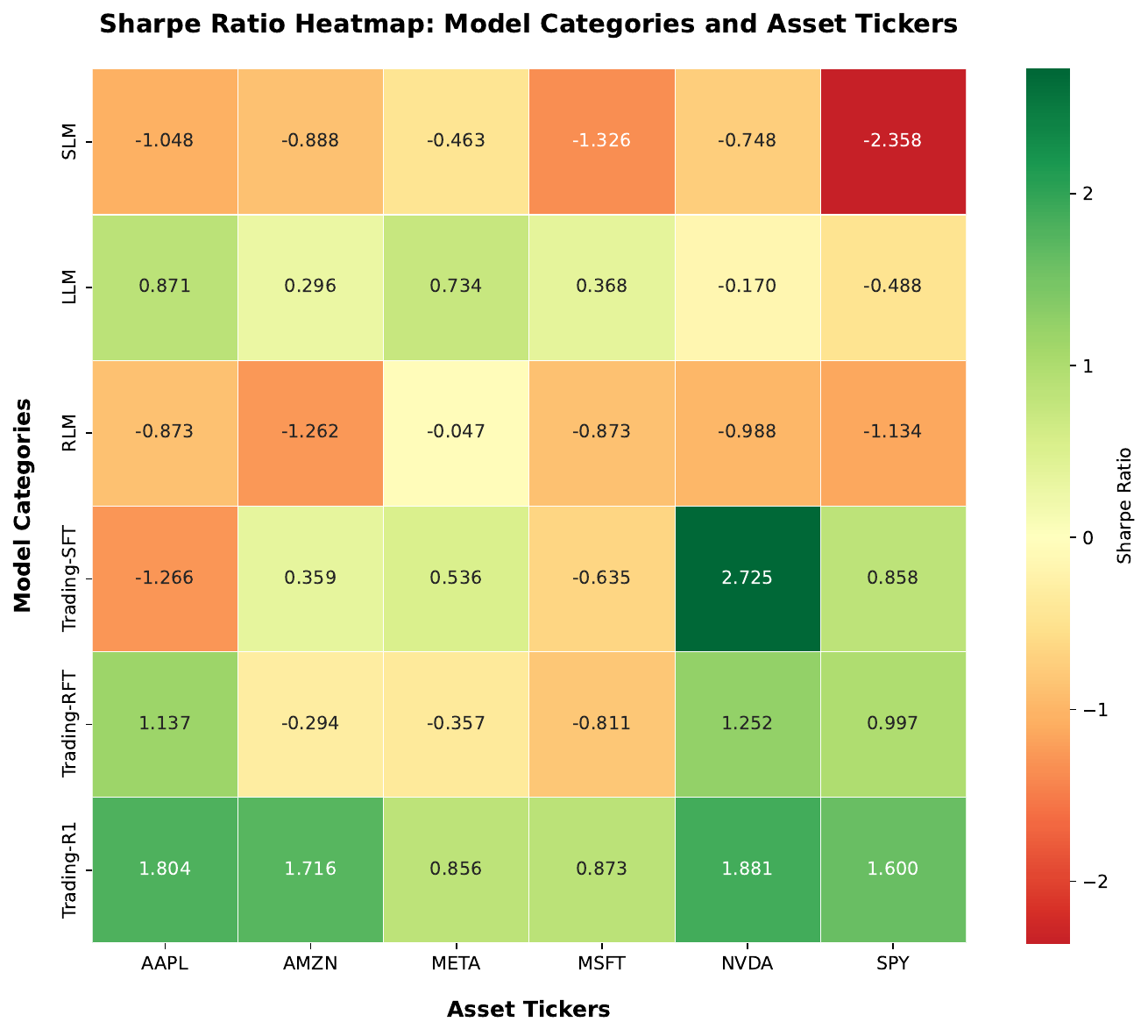}
  \caption{Sharpe Ratio Heatmap}
  \label{fig:sr_heatmap}
\end{figure}

Our Trading-R1 approach achieves the strongest overall performance by combining SFT and RFT to capture market dynamics effectively. Across all evaluated assets, \textsc{\model} delivers improvements over baseline models. It achieves a Sharpe ratio of $1.88$ with $8.08\%$ returns on NVDA, and outperforms GPT-4.1 on AAPL with a Sharpe of $1.80$ versus $1.24$, while maintaining lower drawdowns ($3.68\%$ vs.\ RLMs’ $7.88\%$). The model also attains leading hit ratios, including $70.0\%$ on NVDA and $64.0\%$ on SPY. By contrast, small LLMs such as Qwen-4B and GPT-4.1-nano often yield negative Sharpe ratios, and RLMs like O3-mini and O4-mini incur significant losses due to unguided reasoning processes. Overall, the performance hierarchy (SLM \(<\) RLM \(<\) LLM \(<\) Trading-SFT \(\approx\) Trading-RFT \(<\) \model) underscores the importance of both model scale and specialized reasoning in algorithmic trading, with our approach achieving the best balance between profitability and risk management. Figure~\ref{fig:sr_heatmap} shows this performance hierarchy across all evaluated assets, demonstrating the consistent improvements achieved by our \model series over baseline approaches.  

These results suggest that RFT-enhanced reasoning allows the model to adapt quickly to market fluctuations. The planning capabilities enable \model to compose structured, professional theses that evaluate assets from comprehensive aspects, providing high-quality context for final trading recommendations. Trading-R1 not only generates more profitable trading signals but also sustains lower maximum drawdowns, demonstrating strong potential for real-world trading applications.

\paragraph{Observations}
\model consistently ranks among the top models across all evaluated assets, demonstrating a good balance between performance and reliability. The stability and interpretability of its outputs distinguish it from baseline models, many of which either collapse into poor profitability or produce erratic, unreadable theses. Beyond competitive metrics, \model generates structured, facts-grounded investment theses that provide clear analytical reasoning. This combination of consistent performance with professional thesis generation makes the model particularly valuable for practical applications where both accuracy and interpretability are essential for decision-making support.

This outcome reflects a deliberate design choice. Our model's training with both SFT and RFT produces more stable reasoning and coherent, professional investment theses. We deliberately prioritize enhanced readability, interpretability, and structured argumentation over marginal performance gains. This trade-off aligns with established precedents in the field: recent work on reasoning models has similarly documented that optimizing purely for performance metrics can compromise output coherence and practical utility. For instance, DeepSeek-R1's developers found that their reinforcement learning approach led to language mixing issues, ultimately requiring additional alignment mechanisms that reduced raw performance to restore usability. Our design philosophy embraces this fundamental trade-off from the outset. In practical applications, investors and practitioners require not only accurate predictions but also clear, well-structured rationales that support decision-making. A model that consistently delivers professional, market-aligned analyses provides greater real-world value than one that maximizes performance metrics at the expense of interpretability and user trust.
\section{Conclusion}

We introduced \model, a framework for aligning large language models with financial decision-making through supervised fine-tuning and reinforcement learning. By integrating volatility-aware label generation and reasoning-based supervision, \model generates actionable, risk-aware trading decisions while maintaining analytical rigor and interpretability.

\model offers key advantages over existing approaches through its transparent, scalable pipeline that can be applied to proprietary data and research generation. Unlike general-purpose LLMs that rely on zero-shot prompting, our approach produces consistent, structured analytical reports and enables local deployment. Evaluation on historical market data shows \model achieves the best balance of risk-adjusted returns and drawdowns compared to instruction-following and reasoning models, while generating structured, facts-grounded investment theses that support interpretable decision-making.

\model is best suited for research support and structured analysis generation for financial professionals. The framework shows promise for institutional applications including data vendors, sell-side research generation, and buy-side decision support with customizable policies. We recommend its use as a tool to augment human decision-making in high-throughput scenarios where structured reasoning and interpretability are valued. Future work will focus on real-time deployment capabilities, scalable offline RL variants for improved sample efficiency, and integration of additional data modalities to enhance robustness and domain adaptability.

\clearpage
\newpage
\section*{Industrial Applications and Future Work}
Here, we outline the strengths and limitations of \model and highlight potential applications for the financial industry.

\paragraph{Strengths and Contributions} 
Our \model, trained entirely on publicly available data curated as \database, offers a transparent, modular, and easily modifiable pipeline for training, observation, and processing. This makes it accessible to researchers who wish to develop their own financial trading reasoning models. The current version shows strong potential in thesis drafting and raw data processing, producing high-quality analytical reports that support trade decisions with reasonable returns and favorable Sharpe ratios. These results indicate that by leveraging open and public data, it is possible to train a market-grounded RL-LLM capable of generating professional-quality financial analyses. The model provides interpretability, robust backtested returns, and strong potential for practical downstream applications. Its key strengths include structured reasoning chains for investment analysis, the ability to denoise diverse data sources effectively, and the capacity to mine actionable insights from textual information.

General-purpose LLMs can process large amounts of financial data through prompting, but relying solely on zero-shot performance is often inefficient. Outputs from such methods do not consistently guarantee fidelity, preferred formatting, or comprehensive coverage of relevant viewpoints. Our approach, by contrast, scales effectively with larger volumes of high-quality input data, combining open datasets with curated internal thesis reports that maintain stronger analytical standards.

All of this is packaged into a 4B-parameter form factor that can run on standard commercial GPUs without the need for massive or expensive servers. This makes deployment more affordable, inference faster on large datasets, and ensures the system can run locally and privately. As a result, sensitive information is protected while maintaining customizability and independence from the internet, which is essential for secure large-scale processing.

\paragraph{Limitations and Challenges}
Despite promising results, \model faces several important limitations. Trading decision-making itself remains highly challenging, as financial markets are inherently difficult to predict in terms of direction, timing, and portfolio management. Data quality further complicates this task: even after filtering and cleaning, noisy and conflicting information persists, and higher-quality sources remain costly and difficult to obtain. A significant thesis-to-decision gap also exists, since companies with strong fundamentals may still be poor short-term trades, while overvalued firms can continue to rise despite weak financials. Moreover, \model is best used as a research and thesis-generation tool, not as a substitute for independent due diligence, since traders with different risk tolerances and strategies can hold opposite yet profitable positions.

From a technical standpoint, our hybrid reward function can introduce training instability, while excessive reinforcement learning may erode the structured reasoning format established during supervised fine-tuning. Hallucinations also persist, particularly in smaller models handling long and noisy contexts, where outputs averaging 32K tokens make token-by-token dependencies fragile. To mitigate this, we penalize overly long generations to preserve conciseness and clarity. Another limitation lies in the training universe, which has been biased toward blue-chip and large-cap companies, especially in AI-related sectors during the bullish 2024–2025 cycle. This introduces a structural long bias. Customization would require incorporating longer historical data, expanding coverage to small- and mid-cap companies, and enabling flexible trading intervals aligned with client needs. Despite these constraints, our approach demonstrates how large language models can improve the consistency, quality, and throughput of financial data processing, offering particular value for data vendors and analysts who must manage large volumes of market information in standardized formats.

\paragraph{Recommended Applications}
Given the strengths and limitations of \model, we recommend its use primarily as \textit{a tool for data cleaning, data generation, and research support for human analysts and traders}. The model is particularly effective for producing daily key points, structured research summaries, and large-scale data processing that can assist decision-making. Despite encouraging backtesting results, \model is not suited for high-stakes scenarios, since even advanced LLMs remain prone to hallucinations and nondeterministic behavior. Backtesting performance should therefore be viewed as evidence of the model's ability to generate well-grounded reports, rather than as a guarantee of trading success. Reward design and citation mechanisms can improve faithfulness and grounding, but users must still exercise caution. The most appropriate applications are high-throughput tasks where efficiency gains outweigh the need for perfect accuracy. The best users are professionals with sufficient domain expertise to recognize and correct potential errors. For instance, we observed one data source incorrectly reporting the P/E ratio of the S\&P 500 as 38, illustrating the importance of pre-filtering input data rather than expecting the model to self-correct without internet access.

Key advantages of \model include faster and effective processing, local and private deployment, customization flexibility, and the ability to perform reasoning and thesis generation locally without relying on proprietary external models. This makes it particularly valuable for three key institutional applications:

\begin{itemize}[leftmargin=*, itemsep=2pt, topsep=3pt]
    \item \textbf{Data vendors} can process large amounts of raw financial data into structured feeds and standardized formats more cost-effectively than using proprietary models, while maintaining full control over output structure and thesis granularity through in-house deployment.
    \item \textbf{Sell-side institutions} can train models that understand their investment thesis frameworks and consistently produce research reports in their preferred analytical formats.
    \item \textbf{Buy-side institutions} can fine-tune the model's decision-making policy to align with firm-specific investment preferences. For instance, adjusting RFT label ratios to increase long exposure to preferred sectors like technology, or modifying hold ratios during training to reduce trading frequency and match the firm's investment horizon.
\end{itemize}

This customization capability enables \model to adapt to diverse institutional requirements while ensuring data privacy and analytical consistency across all deployment scenarios.

\clearpage
\newpage

\bibliographystyle{abbrvnat}
\bibliography{references}

\begin{thebibliography}{54}
\providecommand{\natexlab}[1]{#1}
\providecommand{\url}[1]{\texttt{#1}}
\expandafter\ifx\csname urlstyle\endcsname\relax
  \providecommand{\doi}[1]{doi: #1}\else
  \providecommand{\doi}{doi: \begingroup \urlstyle{rm}\Url}\fi

\bibitem[Bai et~al.(2023)Bai, Bai, Chu, Cui, Dang, Deng, Fan, Ge, Han, Huang, Hui, Ji, Li, Lin, Lin, Liu, Liu, Lu, Lu, Ma, Men, Ren, Ren, Tan, Tan, Tu, Wang, Wang, Wang, Wu, Xu, Xu, Yang, Yang, Yang, Yang, Yao, Yu, Yuan, Yuan, Zhang, Zhang, Zhang, Zhang, Zhou, Zhou, Zhou, and Zhu]{bai2023qwentechnicalreport}
J.~Bai, S.~Bai, Y.~Chu, Z.~Cui, K.~Dang, X.~Deng, Y.~Fan, W.~Ge, Y.~Han, F.~Huang, B.~Hui, L.~Ji, M.~Li, J.~Lin, R.~Lin, D.~Liu, G.~Liu, C.~Lu, K.~Lu, J.~Ma, R.~Men, X.~Ren, X.~Ren, C.~Tan, S.~Tan, J.~Tu, P.~Wang, S.~Wang, W.~Wang, S.~Wu, B.~Xu, J.~Xu, A.~Yang, H.~Yang, J.~Yang, S.~Yang, Y.~Yao, B.~Yu, H.~Yuan, Z.~Yuan, J.~Zhang, X.~Zhang, Y.~Zhang, Z.~Zhang, C.~Zhou, J.~Zhou, X.~Zhou, and T.~Zhu.
\newblock Qwen technical report, 2023.
\newblock URL \url{https://arxiv.org/abs/2309.16609}.

\bibitem[Brooks(2009)]{brooks2009reading}
A.~Brooks.
\newblock \emph{Reading Price Charts Bar by Bar: the Technical Analysis of Price Action for the Serious Trader}.
\newblock John Wiley \& Sons, Inc., Hoboken, New Jersey, USA, 2009.
\newblock ISBN 978-0-470-44395-8.

\bibitem[Casper et~al.(2023)Casper, Davies, Shi, Gilbert, Scheurer, Rando, Freedman, Korbak, Lindner, Freire, Wang, Marks, Segerie, Carroll, Peng, Christoffersen, Damani, Slocum, Anwar, Siththaranjan, Nadeau, Michaud, Pfau, Krasheninnikov, Chen, Langosco, Hase, Bıyık, Dragan, Krueger, Sadigh, and Hadfield-Menell]{casper2023openproblemsfundamentallimitations}
S.~Casper, X.~Davies, C.~Shi, T.~K. Gilbert, J.~Scheurer, J.~Rando, R.~Freedman, T.~Korbak, D.~Lindner, P.~Freire, T.~Wang, S.~Marks, C.-R. Segerie, M.~Carroll, A.~Peng, P.~Christoffersen, M.~Damani, S.~Slocum, U.~Anwar, A.~Siththaranjan, M.~Nadeau, E.~J. Michaud, J.~Pfau, D.~Krasheninnikov, X.~Chen, L.~Langosco, P.~Hase, E.~Bıyık, A.~Dragan, D.~Krueger, D.~Sadigh, and D.~Hadfield-Menell.
\newblock Open problems and fundamental limitations of reinforcement learning from human feedback, 2023.
\newblock URL \url{https://arxiv.org/abs/2307.15217}.

\bibitem[DeepSeek-AI et~al.(2025)DeepSeek-AI, Guo, Yang, Zhang, Song, Zhang, Xu, Zhu, Ma, Wang, Bi, Zhang, Yu, Wu, Wu, Gou, Shao, Li, Gao, Liu, Xue, Wang, Wu, Feng, Lu, Zhao, Deng, Zhang, Ruan, Dai, Chen, Ji, Li, Lin, Dai, Luo, Hao, Chen, Li, Zhang, Bao, Xu, Wang, Ding, Xin, Gao, Qu, Li, Guo, Li, Wang, Chen, Yuan, Qiu, Li, Cai, Ni, Liang, Chen, Dong, Hu, Gao, Guan, Huang, Yu, Wang, Zhang, Zhao, Wang, Zhang, Xu, Xia, Zhang, Zhang, Tang, Li, Wang, Li, Tian, Huang, Zhang, Wang, Chen, Du, Ge, Zhang, Pan, Wang, Chen, Jin, Chen, Lu, Zhou, Chen, Ye, Wang, Yu, Zhou, Pan, Li, Zhou, Wu, Ye, Yun, Pei, Sun, Wang, Zeng, Zhao, Liu, Liang, Gao, Yu, Zhang, Xiao, An, Liu, Wang, Chen, Nie, Cheng, Liu, Xie, Liu, Yang, Li, Su, Lin, Li, Jin, Shen, Chen, Sun, Wang, Song, Zhou, Wang, Shan, Li, Wang, Wei, Zhang, Xu, Li, Zhao, Sun, Wang, Yu, Zhang, Shi, Xiong, He, Piao, Wang, Tan, Ma, Liu, Guo, Ou, Wang, Gong, Zou, He, Xiong, Luo, You, Liu, Zhou, Zhu, Xu, Huang, Li, Zheng, Zhu, Ma, Tang, Zha, Yan, Ren, Ren, Sha, Fu, Xu, Xie, Zhang,
  Hao, Ma, Yan, Wu, Gu, Zhu, Liu, Li, Xie, Song, Pan, Huang, Xu, Zhang, and Zhang]{deepseekai2025deepseekr1incentivizingreasoningcapability}
DeepSeek-AI, D.~Guo, D.~Yang, H.~Zhang, J.~Song, R.~Zhang, R.~Xu, Q.~Zhu, S.~Ma, P.~Wang, X.~Bi, X.~Zhang, X.~Yu, Y.~Wu, Z.~F. Wu, Z.~Gou, Z.~Shao, Z.~Li, Z.~Gao, A.~Liu, B.~Xue, B.~Wang, B.~Wu, B.~Feng, C.~Lu, C.~Zhao, C.~Deng, C.~Zhang, C.~Ruan, D.~Dai, D.~Chen, D.~Ji, E.~Li, F.~Lin, F.~Dai, F.~Luo, G.~Hao, G.~Chen, G.~Li, H.~Zhang, H.~Bao, H.~Xu, H.~Wang, H.~Ding, H.~Xin, H.~Gao, H.~Qu, H.~Li, J.~Guo, J.~Li, J.~Wang, J.~Chen, J.~Yuan, J.~Qiu, J.~Li, J.~L. Cai, J.~Ni, J.~Liang, J.~Chen, K.~Dong, K.~Hu, K.~Gao, K.~Guan, K.~Huang, K.~Yu, L.~Wang, L.~Zhang, L.~Zhao, L.~Wang, L.~Zhang, L.~Xu, L.~Xia, M.~Zhang, M.~Zhang, M.~Tang, M.~Li, M.~Wang, M.~Li, N.~Tian, P.~Huang, P.~Zhang, Q.~Wang, Q.~Chen, Q.~Du, R.~Ge, R.~Zhang, R.~Pan, R.~Wang, R.~J. Chen, R.~L. Jin, R.~Chen, S.~Lu, S.~Zhou, S.~Chen, S.~Ye, S.~Wang, S.~Yu, S.~Zhou, S.~Pan, S.~S. Li, S.~Zhou, S.~Wu, S.~Ye, T.~Yun, T.~Pei, T.~Sun, T.~Wang, W.~Zeng, W.~Zhao, W.~Liu, W.~Liang, W.~Gao, W.~Yu, W.~Zhang, W.~L. Xiao, W.~An, X.~Liu, X.~Wang, X.~Chen, X.~Nie,
  X.~Cheng, X.~Liu, X.~Xie, X.~Liu, X.~Yang, X.~Li, X.~Su, X.~Lin, X.~Q. Li, X.~Jin, X.~Shen, X.~Chen, X.~Sun, X.~Wang, X.~Song, X.~Zhou, X.~Wang, X.~Shan, Y.~K. Li, Y.~Q. Wang, Y.~X. Wei, Y.~Zhang, Y.~Xu, Y.~Li, Y.~Zhao, Y.~Sun, Y.~Wang, Y.~Yu, Y.~Zhang, Y.~Shi, Y.~Xiong, Y.~He, Y.~Piao, Y.~Wang, Y.~Tan, Y.~Ma, Y.~Liu, Y.~Guo, Y.~Ou, Y.~Wang, Y.~Gong, Y.~Zou, Y.~He, Y.~Xiong, Y.~Luo, Y.~You, Y.~Liu, Y.~Zhou, Y.~X. Zhu, Y.~Xu, Y.~Huang, Y.~Li, Y.~Zheng, Y.~Zhu, Y.~Ma, Y.~Tang, Y.~Zha, Y.~Yan, Z.~Z. Ren, Z.~Ren, Z.~Sha, Z.~Fu, Z.~Xu, Z.~Xie, Z.~Zhang, Z.~Hao, Z.~Ma, Z.~Yan, Z.~Wu, Z.~Gu, Z.~Zhu, Z.~Liu, Z.~Li, Z.~Xie, Z.~Song, Z.~Pan, Z.~Huang, Z.~Xu, Z.~Zhang, and Z.~Zhang.
\newblock Deepseek-r1: Incentivizing reasoning capability in llms via reinforcement learning, 2025.
\newblock URL \url{https://arxiv.org/abs/2501.12948}.

\bibitem[Ding et~al.(2023)Ding, Jia, Ma, Mao, Zhou, Li, and Han]{ding2023integratingstockfeaturesglobal}
Y.~Ding, S.~Jia, T.~Ma, B.~Mao, X.~Zhou, L.~Li, and D.~Han.
\newblock Integrating stock features and global information via large language models for enhanced stock return prediction, 2023.
\newblock URL \url{https://arxiv.org/abs/2310.05627}.

\bibitem[Dow et~al.()Dow, Hamilton, Rhea, and Schaefer]{dow_theory}
C.~H. Dow, W.~P. Hamilton, R.~Rhea, and E.~G. Schaefer.
\newblock \emph{The Dow Theory on Stock Price Movement}.
\newblock Based on 255 editorials in The Wall Street Journal by Charles H. Dow.

\bibitem[Elliott(1938)]{elliott1938wave}
R.~N. Elliott.
\newblock \emph{The Wave Principle}.
\newblock 1938.

\bibitem[Ericson et~al.(2024)Ericson, Zhu, Han, Fu, Li, Guo, and Hu]{ericson2024deepgenerativemodelingfinancial}
L.~Ericson, X.~Zhu, X.~Han, R.~Fu, S.~Li, S.~Guo, and P.~Hu.
\newblock Deep generative modeling for financial time series with application in var: A comparative review, 2024.
\newblock URL \url{https://arxiv.org/abs/2401.10370}.

\bibitem[Fatouros et~al.(2024)Fatouros, Metaxas, Soldatos, and Kyriazis]{beatunveiling}
G.~Fatouros, K.~Metaxas, J.~Soldatos, and D.~Kyriazis.
\newblock Can large language models beat wall street? unveiling the potential of ai in stock selection, 2024.
\newblock URL \url{https://arxiv.org/abs/2401.03737}.

\bibitem[Fjellström(2022)]{fjellström2022longshorttermmemoryneural}
C.~Fjellström.
\newblock Long short-term memory neural network for financial time series, 2022.
\newblock URL \url{https://arxiv.org/abs/2201.08218}.

\bibitem[Gottweis et~al.(2025)Gottweis, Weng, Daryin, Tu, Palepu, Sirkovic, Myaskovsky, Weissenberger, Rong, Tanno, Saab, Popovici, Blum, Zhang, Chou, Hassidim, Gokturk, Vahdat, Kohli, Matias, Carroll, Kulkarni, Tomasev, Guan, Dhillon, Vaishnav, Lee, Costa, Penadés, Peltz, Xu, Pawlosky, Karthikesalingam, and Natarajan]{gottweis2025aicoscientist}
J.~Gottweis, W.-H. Weng, A.~Daryin, T.~Tu, A.~Palepu, P.~Sirkovic, A.~Myaskovsky, F.~Weissenberger, K.~Rong, R.~Tanno, K.~Saab, D.~Popovici, J.~Blum, F.~Zhang, K.~Chou, A.~Hassidim, B.~Gokturk, A.~Vahdat, P.~Kohli, Y.~Matias, A.~Carroll, K.~Kulkarni, N.~Tomasev, Y.~Guan, V.~Dhillon, E.~D. Vaishnav, B.~Lee, T.~R.~D. Costa, J.~R. Penadés, G.~Peltz, Y.~Xu, A.~Pawlosky, A.~Karthikesalingam, and V.~Natarajan.
\newblock Towards an ai co-scientist, 2025.
\newblock URL \url{https://arxiv.org/abs/2502.18864}.

\bibitem[Hendrycks et~al.(2021)Hendrycks, Burns, Basart, Zou, Mazeika, Song, and Steinhardt]{hendrycks2021measuringmassivemultitasklanguage}
D.~Hendrycks, C.~Burns, S.~Basart, A.~Zou, M.~Mazeika, D.~Song, and J.~Steinhardt.
\newblock Measuring massive multitask language understanding, 2021.
\newblock URL \url{https://arxiv.org/abs/2009.03300}.

\bibitem[Hong et~al.(2024)Hong, Zhuge, Chen, Zheng, Cheng, Zhang, Wang, Wang, Yau, Lin, Zhou, Ran, Xiao, Wu, and Schmidhuber]{hong2024metagptmetaprogrammingmultiagent}
S.~Hong, M.~Zhuge, J.~Chen, X.~Zheng, Y.~Cheng, C.~Zhang, J.~Wang, Z.~Wang, S.~K.~S. Yau, Z.~Lin, L.~Zhou, C.~Ran, L.~Xiao, C.~Wu, and J.~Schmidhuber.
\newblock Metagpt: Meta programming for a multi-agent collaborative framework, 2024.
\newblock URL \url{https://arxiv.org/abs/2308.00352}.

\bibitem[Hu et~al.(2021)Hu, Shen, Wallis, Allen-Zhu, Li, Wang, Wang, and Chen]{hu2021loralowrankadaptationlarge}
E.~J. Hu, Y.~Shen, P.~Wallis, Z.~Allen-Zhu, Y.~Li, S.~Wang, L.~Wang, and W.~Chen.
\newblock Lora: Low-rank adaptation of large language models, 2021.
\newblock URL \url{https://arxiv.org/abs/2106.09685}.

\bibitem[Ji et~al.(2023)Ji, Yu, Xu, Lee, Ishii, and Fung]{ji2023mitigatinghallucinationlargelanguage}
Z.~Ji, T.~Yu, Y.~Xu, N.~Lee, E.~Ishii, and P.~Fung.
\newblock Towards mitigating hallucination in large language models via self-reflection, 2023.
\newblock URL \url{https://arxiv.org/abs/2310.06271}.

\bibitem[Kaufmann et~al.(2023)Kaufmann, Weng, Bengs, and H{\"u}llermeier]{kaufmann2023survey}
T.~Kaufmann, P.~Weng, V.~Bengs, and E.~H{\"u}llermeier.
\newblock A survey of reinforcement learning from human feedback.
\newblock \emph{arXiv preprint arXiv:2312.14925}, 10, 2023.

\bibitem[Kaufmann et~al.(2024)Kaufmann, Weng, Bengs, and Hüllermeier]{kaufmann2024surveyreinforcementlearninghuman}
T.~Kaufmann, P.~Weng, V.~Bengs, and E.~Hüllermeier.
\newblock A survey of reinforcement learning from human feedback, 2024.
\newblock URL \url{https://arxiv.org/abs/2312.14925}.

\bibitem[Kirtac and Germano(2024)]{sentitrade}
K.~Kirtac and G.~Germano.
\newblock Sentiment trading with large language models.
\newblock \emph{Finance Research Letters}, 62:\penalty0 105227, 2024.
\newblock ISSN 1544-6123.
\newblock \doi{https://doi.org/10.1016/j.frl.2024.105227}.
\newblock URL \url{https://www.sciencedirect.com/science/article/pii/S1544612324002575}.

\bibitem[Koa et~al.(2024)Koa, Ma, Ng, and Chua]{Koa_2024}
K.~J. Koa, Y.~Ma, R.~Ng, and T.-S. Chua.
\newblock Learning to generate explainable stock predictions using self-reflective large language models, May 2024.
\newblock URL \url{http://dx.doi.org/10.1145/3589334.3645611}.

\bibitem[Lambert et~al.(2024)Lambert, Pyatkin, Morrison, Miranda, Lin, Chandu, Dziri, Kumar, Zick, Choi, et~al.]{lambert2024rewardbench}
N.~Lambert, V.~Pyatkin, J.~Morrison, L.~Miranda, B.~Y. Lin, K.~Chandu, N.~Dziri, S.~Kumar, T.~Zick, Y.~Choi, et~al.
\newblock Rewardbench: Evaluating reward models for language modeling.
\newblock \emph{arXiv preprint arXiv:2403.13787}, 2024.

\bibitem[Lee et~al.(2025)Lee, Stevens, and Han]{Lee_2025}
J.~Lee, N.~Stevens, and S.~C. Han.
\newblock Large language models in finance (finllms).
\newblock \emph{Neural Computing and Applications}, Jan. 2025.
\newblock ISSN 1433-3058.
\newblock \doi{10.1007/s00521-024-10495-6}.
\newblock URL \url{http://dx.doi.org/10.1007/s00521-024-10495-6}.

\bibitem[Lefevre and Markman(2010)]{lefevre2010reminiscences}
E.~Lefevre and J.~D. Markman.
\newblock \emph{Reminiscences of a stock operator: With new commentary and insights on the life and times of Jesse Livermore}.
\newblock John Wiley \& Sons, 2010.

\bibitem[Li et~al.(2023)Li, Yu, Li, Chen, and Khashanah]{li2023tradinggpt}
Y.~Li, Y.~Yu, H.~Li, Z.~Chen, and K.~Khashanah.
\newblock Tradinggpt: Multi-agent system with layered memory and distinct characters for enhanced financial trading performance.
\newblock \emph{arXiv preprint arXiv:2309.03736}, 2023.

\bibitem[Liu et~al.(2025{\natexlab{a}})Liu, Zhang, Xu, Wanyan, Wang, Yan, Zhang, Yuan, Xu, Hu, and Huang]{liu2025pcagenthierarchicalmultiagentcollaboration}
H.~Liu, X.~Zhang, H.~Xu, Y.~Wanyan, J.~Wang, M.~Yan, J.~Zhang, C.~Yuan, C.~Xu, W.~Hu, and F.~Huang.
\newblock Pc-agent: A hierarchical multi-agent collaboration framework for complex task automation on pc, 2025{\natexlab{a}}.
\newblock URL \url{https://arxiv.org/abs/2502.14282}.

\bibitem[Liu et~al.(2025{\natexlab{b}})Liu, Guo, Lou, Zeng, Niu, Wang, Xu, Cai, Yang, Zhao, Li, Xu, Chen, Chen, Bai, and Zhang]{liu2025finr1largelanguagemodel}
Z.~Liu, X.~Guo, F.~Lou, L.~Zeng, J.~Niu, Z.~Wang, J.~Xu, W.~Cai, Z.~Yang, X.~Zhao, C.~Li, S.~Xu, D.~Chen, Y.~Chen, Z.~Bai, and L.~Zhang.
\newblock Fin-r1: A large language model for financial reasoning through reinforcement learning, 2025{\natexlab{b}}.
\newblock URL \url{https://arxiv.org/abs/2503.16252}.

\bibitem[Lopez-Lira and Tang(2023)]{lopezlira2023chatgptforecaststockprice}
A.~Lopez-Lira and Y.~Tang.
\newblock Can chatgpt forecast stock price movements? return predictability and large language models, 2023.
\newblock URL \url{https://arxiv.org/abs/2304.07619}.

\bibitem[Lu et~al.(2023)Lu, Wu, Liang, Xu, He, Geng, Han, Xin, and Xiao]{lu2023bbtfin}
D.~Lu, H.~Wu, J.~Liang, Y.~Xu, Q.~He, Y.~Geng, M.~Han, Y.~Xin, and Y.~Xiao.
\newblock Bbt-fin: Comprehensive construction of chinese financial domain pre-trained language model, corpus and benchmark, 2023.
\newblock URL \url{https://arxiv.org/abs/2302.09432}.

\bibitem[Lu et~al.(2024)Lu, Bansal, Xia, Liu, Li, Hajishirzi, Cheng, Chang, Galley, and Gao]{lu2024mathvistaevaluatingmathematicalreasoning}
P.~Lu, H.~Bansal, T.~Xia, J.~Liu, C.~Li, H.~Hajishirzi, H.~Cheng, K.-W. Chang, M.~Galley, and J.~Gao.
\newblock Mathvista: Evaluating mathematical reasoning of foundation models in visual contexts, 2024.
\newblock URL \url{https://arxiv.org/abs/2310.02255}.

\bibitem[Meng et~al.(2024)Meng, Xia, and Chen]{meng2024simpo}
Y.~Meng, M.~Xia, and D.~Chen.
\newblock Simpo: Simple preference optimization with a reference-free reward.
\newblock \emph{Advances in Neural Information Processing Systems}, 37:\penalty0 124198--124235, 2024.

\bibitem[OpenAI et~al.(2024)OpenAI, :, Jaech, Kalai, Lerer, Richardson, El-Kishky, Low, Helyar, Madry, Beutel, Carney, Iftimie, Karpenko, Passos, Neitz, Prokofiev, Wei, Tam, Bennett, Kumar, Saraiva, Vallone, Duberstein, Kondrich, Mishchenko, Applebaum, Jiang, Nair, Zoph, Ghorbani, Rossen, Sokolowsky, Barak, McGrew, Minaiev, Hao, Baker, Houghton, McKinzie, Eastman, Lugaresi, Bassin, Hudson, Li, de~Bourcy, Voss, Shen, Zhang, Koch, Orsinger, Hesse, Fischer, Chan, Roberts, Kappler, Levy, Selsam, Dohan, Farhi, Mely, Robinson, Tsipras, Li, Oprica, Freeman, Zhang, Wong, Proehl, Cheung, Mitchell, Wallace, Ritter, Mays, Wang, Such, Raso, Leoni, Tsimpourlas, Song, von Lohmann, Sulit, Salmon, Parascandolo, Chabot, Zhao, Brockman, Leclerc, Salman, Bao, Sheng, Andrin, Bagherinezhad, Ren, Lightman, Chung, Kivlichan, O'Connell, Osband, Gilaberte, Akkaya, Kostrikov, Sutskever, Kofman, Pachocki, Lennon, Wei, Harb, Twore, Feng, Yu, Weng, Tang, Yu, Candela, Palermo, Parish, Heidecke, Hallman, Rizzo, Gordon, Uesato, Ward,
  Huizinga, Wang, Chen, Xiao, Singhal, Nguyen, Cobbe, Shi, Wood, Rimbach, Gu-Lemberg, Liu, Lu, Stone, Yu, Ahmad, Yang, Liu, Maksin, Ho, Fedus, Weng, Li, McCallum, Held, Kuhn, Kondraciuk, Kaiser, Metz, Boyd, Trebacz, Joglekar, Chen, Tintor, Meyer, Jones, Kaufer, Schwarzer, Shah, Yatbaz, Guan, Xu, Yan, Glaese, Chen, Lampe, Malek, Wang, Fradin, McClay, Pavlov, Wang, Wang, Murati, Bavarian, Rohaninejad, McAleese, Chowdhury, Chowdhury, Ryder, Tezak, Brown, Nachum, Boiko, Murk, Watkins, Chao, Ashbourne, Izmailov, Zhokhov, Dias, Arora, Lin, Lopes, Gaon, Miyara, Leike, Hwang, Garg, Brown, James, Shu, Cheu, Greene, Jain, Altman, Toizer, Toyer, Miserendino, Agarwal, Hernandez, Baker, McKinney, Yan, Zhao, Hu, Santurkar, Chaudhuri, Zhang, Fu, Papay, Lin, Balaji, Sanjeev, Sidor, Broda, Clark, Wang, Gordon, Sanders, Patwardhan, Sottiaux, Degry, Dimson, Zheng, Garipov, Stasi, Bansal, Creech, Peterson, Eloundou, Qi, Kosaraju, Monaco, Pong, Fomenko, Zheng, Zhou, McCabe, Zaremba, Dubois, Lu, Chen, Cha, Bai, He, Zhang, Wang,
  Shao, and Li]{openai2024openaio1card}
OpenAI, :, A.~Jaech, A.~Kalai, A.~Lerer, A.~Richardson, A.~El-Kishky, A.~Low, A.~Helyar, A.~Madry, A.~Beutel, A.~Carney, A.~Iftimie, A.~Karpenko, A.~T. Passos, A.~Neitz, A.~Prokofiev, A.~Wei, A.~Tam, A.~Bennett, A.~Kumar, A.~Saraiva, A.~Vallone, A.~Duberstein, A.~Kondrich, A.~Mishchenko, A.~Applebaum, A.~Jiang, A.~Nair, B.~Zoph, B.~Ghorbani, B.~Rossen, B.~Sokolowsky, B.~Barak, B.~McGrew, B.~Minaiev, B.~Hao, B.~Baker, B.~Houghton, B.~McKinzie, B.~Eastman, C.~Lugaresi, C.~Bassin, C.~Hudson, C.~M. Li, C.~de~Bourcy, C.~Voss, C.~Shen, C.~Zhang, C.~Koch, C.~Orsinger, C.~Hesse, C.~Fischer, C.~Chan, D.~Roberts, D.~Kappler, D.~Levy, D.~Selsam, D.~Dohan, D.~Farhi, D.~Mely, D.~Robinson, D.~Tsipras, D.~Li, D.~Oprica, E.~Freeman, E.~Zhang, E.~Wong, E.~Proehl, E.~Cheung, E.~Mitchell, E.~Wallace, E.~Ritter, E.~Mays, F.~Wang, F.~P. Such, F.~Raso, F.~Leoni, F.~Tsimpourlas, F.~Song, F.~von Lohmann, F.~Sulit, G.~Salmon, G.~Parascandolo, G.~Chabot, G.~Zhao, G.~Brockman, G.~Leclerc, H.~Salman, H.~Bao, H.~Sheng, H.~Andrin,
  H.~Bagherinezhad, H.~Ren, H.~Lightman, H.~W. Chung, I.~Kivlichan, I.~O'Connell, I.~Osband, I.~C. Gilaberte, I.~Akkaya, I.~Kostrikov, I.~Sutskever, I.~Kofman, J.~Pachocki, J.~Lennon, J.~Wei, J.~Harb, J.~Twore, J.~Feng, J.~Yu, J.~Weng, J.~Tang, J.~Yu, J.~Q. Candela, J.~Palermo, J.~Parish, J.~Heidecke, J.~Hallman, J.~Rizzo, J.~Gordon, J.~Uesato, J.~Ward, J.~Huizinga, J.~Wang, K.~Chen, K.~Xiao, K.~Singhal, K.~Nguyen, K.~Cobbe, K.~Shi, K.~Wood, K.~Rimbach, K.~Gu-Lemberg, K.~Liu, K.~Lu, K.~Stone, K.~Yu, L.~Ahmad, L.~Yang, L.~Liu, L.~Maksin, L.~Ho, L.~Fedus, L.~Weng, L.~Li, L.~McCallum, L.~Held, L.~Kuhn, L.~Kondraciuk, L.~Kaiser, L.~Metz, M.~Boyd, M.~Trebacz, M.~Joglekar, M.~Chen, M.~Tintor, M.~Meyer, M.~Jones, M.~Kaufer, M.~Schwarzer, M.~Shah, M.~Yatbaz, M.~Y. Guan, M.~Xu, M.~Yan, M.~Glaese, M.~Chen, M.~Lampe, M.~Malek, M.~Wang, M.~Fradin, M.~McClay, M.~Pavlov, M.~Wang, M.~Wang, M.~Murati, M.~Bavarian, M.~Rohaninejad, N.~McAleese, N.~Chowdhury, N.~Chowdhury, N.~Ryder, N.~Tezak, N.~Brown, O.~Nachum, O.~Boiko,
  O.~Murk, O.~Watkins, P.~Chao, P.~Ashbourne, P.~Izmailov, P.~Zhokhov, R.~Dias, R.~Arora, R.~Lin, R.~G. Lopes, R.~Gaon, R.~Miyara, R.~Leike, R.~Hwang, R.~Garg, R.~Brown, R.~James, R.~Shu, R.~Cheu, R.~Greene, S.~Jain, S.~Altman, S.~Toizer, S.~Toyer, S.~Miserendino, S.~Agarwal, S.~Hernandez, S.~Baker, S.~McKinney, S.~Yan, S.~Zhao, S.~Hu, S.~Santurkar, S.~R. Chaudhuri, S.~Zhang, S.~Fu, S.~Papay, S.~Lin, S.~Balaji, S.~Sanjeev, S.~Sidor, T.~Broda, A.~Clark, T.~Wang, T.~Gordon, T.~Sanders, T.~Patwardhan, T.~Sottiaux, T.~Degry, T.~Dimson, T.~Zheng, T.~Garipov, T.~Stasi, T.~Bansal, T.~Creech, T.~Peterson, T.~Eloundou, V.~Qi, V.~Kosaraju, V.~Monaco, V.~Pong, V.~Fomenko, W.~Zheng, W.~Zhou, W.~McCabe, W.~Zaremba, Y.~Dubois, Y.~Lu, Y.~Chen, Y.~Cha, Y.~Bai, Y.~He, Y.~Zhang, Y.~Wang, Z.~Shao, and Z.~Li.
\newblock Openai o1 system card, 2024.
\newblock URL \url{https://arxiv.org/abs/2412.16720}.

\bibitem[OpenAI et~al.(2025)OpenAI, :, Agarwal, Ahmad, Ai, Altman, Applebaum, Arbus, Arora, Bai, Baker, Bao, Barak, Bennett, Bertao, Brett, Brevdo, Brockman, Bubeck, Chang, Chen, Chen, Cheung, Clark, Cook, Dukhan, Dvorak, Fives, Fomenko, Garipov, Georgiev, Glaese, Gogineni, Goucher, Gross, Guzman, Hallman, Hehir, Heidecke, Helyar, Hu, Huet, Huh, Jain, Johnson, Koch, Kofman, Kundel, Kwon, Kyrylov, Le, Leclerc, Lennon, Lessans, Lezcano-Casado, Li, Li, Lin, Liss, Lily, Liu, Liu, Lu, Lu, Martinovic, McCallum, McGrath, McKinney, McLaughlin, Mei, Mostovoy, Mu, Myles, Neitz, Nichol, Pachocki, Paino, Palmie, Pantuliano, Parascandolo, Park, Pathak, Paz, Peran, Pimenov, Pokrass, Proehl, Qiu, Raila, Raso, Ren, Richardson, Robinson, Rotsted, Salman, Sanjeev, Schwarzer, Sculley, Sikchi, Simon, Singhal, Song, Stuckey, Sun, Tillet, Toizer, Tsimpourlas, Vyas, Wallace, Wang, Wang, Watkins, Weil, Wendling, Whinnery, Whitney, Wong, Yang, Yang, Yasunaga, Ying, Zaremba, Zhan, Zhang, Zhang, Zhang, and
  Zhao]{openai2025gptoss120bgptoss20bmodel}
OpenAI, :, S.~Agarwal, L.~Ahmad, J.~Ai, S.~Altman, A.~Applebaum, E.~Arbus, R.~K. Arora, Y.~Bai, B.~Baker, H.~Bao, B.~Barak, A.~Bennett, T.~Bertao, N.~Brett, E.~Brevdo, G.~Brockman, S.~Bubeck, C.~Chang, K.~Chen, M.~Chen, E.~Cheung, A.~Clark, D.~Cook, M.~Dukhan, C.~Dvorak, K.~Fives, V.~Fomenko, T.~Garipov, K.~Georgiev, M.~Glaese, T.~Gogineni, A.~Goucher, L.~Gross, K.~G. Guzman, J.~Hallman, J.~Hehir, J.~Heidecke, A.~Helyar, H.~Hu, R.~Huet, J.~Huh, S.~Jain, Z.~Johnson, C.~Koch, I.~Kofman, D.~Kundel, J.~Kwon, V.~Kyrylov, E.~Y. Le, G.~Leclerc, J.~P. Lennon, S.~Lessans, M.~Lezcano-Casado, Y.~Li, Z.~Li, J.~Lin, J.~Liss, Lily, Liu, J.~Liu, K.~Lu, C.~Lu, Z.~Martinovic, L.~McCallum, J.~McGrath, S.~McKinney, A.~McLaughlin, S.~Mei, S.~Mostovoy, T.~Mu, G.~Myles, A.~Neitz, A.~Nichol, J.~Pachocki, A.~Paino, D.~Palmie, A.~Pantuliano, G.~Parascandolo, J.~Park, L.~Pathak, C.~Paz, L.~Peran, D.~Pimenov, M.~Pokrass, E.~Proehl, H.~Qiu, G.~Raila, F.~Raso, H.~Ren, K.~Richardson, D.~Robinson, B.~Rotsted, H.~Salman, S.~Sanjeev,
  M.~Schwarzer, D.~Sculley, H.~Sikchi, K.~Simon, K.~Singhal, Y.~Song, D.~Stuckey, Z.~Sun, P.~Tillet, S.~Toizer, F.~Tsimpourlas, N.~Vyas, E.~Wallace, X.~Wang, M.~Wang, O.~Watkins, K.~Weil, A.~Wendling, K.~Whinnery, C.~Whitney, H.~Wong, L.~Yang, Y.~Yang, M.~Yasunaga, K.~Ying, W.~Zaremba, W.~Zhan, C.~Zhang, B.~Zhang, E.~Zhang, and S.~Zhao.
\newblock gpt-oss-120b and gpt-oss-20b model card, 2025.
\newblock URL \url{https://arxiv.org/abs/2508.10925}.

\bibitem[Ouyang et~al.(2022)Ouyang, Wu, Jiang, Almeida, Wainwright, Mishkin, Zhang, Agarwal, Slama, Ray, et~al.]{ouyang2022training}
L.~Ouyang, J.~Wu, X.~Jiang, D.~Almeida, C.~Wainwright, P.~Mishkin, C.~Zhang, S.~Agarwal, K.~Slama, A.~Ray, et~al.
\newblock Training language models to follow instructions with human feedback.
\newblock \emph{Advances in neural information processing systems}, 35:\penalty0 27730--27744, 2022.

\bibitem[Petram(2014)]{petram2014world}
L.~Petram.
\newblock \emph{The world's first stock exchange}.
\newblock Columbia University Press, 2014.

\bibitem[Poddar et~al.(2024)Poddar, Wan, Ivison, Gupta, and Jaques]{poddar2024personalizing}
S.~Poddar, Y.~Wan, H.~Ivison, A.~Gupta, and N.~Jaques.
\newblock Personalizing reinforcement learning from human feedback with variational preference learning.
\newblock In \emph{The Thirty-eighth Annual Conference on Neural Information Processing Systems}, 2024.
\newblock URL \url{https://openreview.net/forum?id=gRG6SzbW9p}.

\bibitem[Qian et~al.(2025)Qian, Zhou, Wang, Peng, Yi, Zhao, Huang, Xie, and yun Nie]{qian2025fino1transferabilityreasoningenhancedllms}
L.~Qian, W.~Zhou, Y.~Wang, X.~Peng, H.~Yi, Y.~Zhao, J.~Huang, Q.~Xie, and J.~yun Nie.
\newblock Fino1: On the transferability of reasoning-enhanced llms and reinforcement learning to finance, 2025.
\newblock URL \url{https://arxiv.org/abs/2502.08127}.

\bibitem[Rafailov et~al.(2023)Rafailov, Sharma, Mitchell, Manning, Ermon, and Finn]{rafailov2023direct}
R.~Rafailov, A.~Sharma, E.~Mitchell, C.~D. Manning, S.~Ermon, and C.~Finn.
\newblock Direct preference optimization: Your language model is secretly a reward model.
\newblock \emph{Advances in Neural Information Processing Systems}, 36:\penalty0 53728--53741, 2023.

\bibitem[Schulman et~al.(2017)Schulman, Wolski, Dhariwal, Radford, and Klimov]{ppo}
J.~Schulman, F.~Wolski, P.~Dhariwal, A.~Radford, and O.~Klimov.
\newblock Proximal policy optimization algorithms, 2017.
\newblock URL \url{https://arxiv.org/abs/1707.06347}.

\bibitem[Schwager(2012)]{schwager2012market}
J.~D. Schwager.
\newblock \emph{Market Wizards: Interviews with Top Traders}.
\newblock Wiley, updated edition, February 2012.
\newblock ISBN 978-1-118-27305-0.

\bibitem[Shao et~al.(2024{\natexlab{a}})Shao, Wang, Zhu, Xu, Song, Bi, Zhang, Zhang, Li, Wu, et~al.]{shao2024deepseekmath}
Z.~Shao, P.~Wang, Q.~Zhu, R.~Xu, J.~Song, X.~Bi, H.~Zhang, M.~Zhang, Y.~Li, Y.~Wu, et~al.
\newblock Deepseekmath: Pushing the limits of mathematical reasoning in open language models.
\newblock \emph{arXiv preprint arXiv:2402.03300}, 2024{\natexlab{a}}.

\bibitem[Shao et~al.(2024{\natexlab{b}})Shao, Wang, Zhu, Xu, Song, Bi, Zhang, Zhang, Li, Wu, and Guo]{shao2024deepseekmathpushinglimitsmathematical}
Z.~Shao, P.~Wang, Q.~Zhu, R.~Xu, J.~Song, X.~Bi, H.~Zhang, M.~Zhang, Y.~K. Li, Y.~Wu, and D.~Guo.
\newblock Deepseekmath: Pushing the limits of mathematical reasoning in open language models, 2024{\natexlab{b}}.
\newblock URL \url{https://arxiv.org/abs/2402.03300}.

\bibitem[Tatsat and Shater(2025)]{tatsat2025blackboxinterpretabilityllms}
H.~Tatsat and A.~Shater.
\newblock Beyond the black box: Interpretability of llms in finance, 2025.
\newblock URL \url{https://arxiv.org/abs/2505.24650}.

\bibitem[Wang et~al.(2024)Wang, Izumi, and Sakaji]{wang2024llmfactorextractingprofitablefactors}
M.~Wang, K.~Izumi, and H.~Sakaji.
\newblock Llmfactor: Extracting profitable factors through prompts for explainable stock movement prediction, 2024.
\newblock URL \url{https://arxiv.org/abs/2406.10811}.

\bibitem[Wang et~al.(2023)Wang, Yuan, Zhou, Ni, Shum, and Guo]{wang2023alpha}
S.~Wang, H.~Yuan, L.~Zhou, L.~M. Ni, H.-Y. Shum, and J.~Guo.
\newblock Alpha-gpt: Human-ai interactive alpha mining for quantitative investment.
\newblock \emph{arXiv preprint arXiv:2308.00016}, 2023.
\newblock URL \url{https://arxiv.org/abs/2308.00016}.

\bibitem[Wu et~al.(2023)Wu, Irsoy, Lu, Dabravolski, Dredze, Gehrmann, Kambadur, Rosenberg, and Mann]{wu2023bloomberggpt}
S.~Wu, O.~Irsoy, S.~Lu, V.~Dabravolski, M.~Dredze, S.~Gehrmann, P.~Kambadur, D.~Rosenberg, and G.~Mann.
\newblock Bloomberggpt: A large language model for finance, 2023.
\newblock URL \url{https://arxiv.org/abs/2303.17564}.

\bibitem[Xiao et~al.(2025)Xiao, Sun, Luo, and Wang]{xiao2025tradingagentsmultiagentsllmfinancial}
Y.~Xiao, E.~Sun, D.~Luo, and W.~Wang.
\newblock Tradingagents: Multi-agents llm financial trading framework, 2025.
\newblock URL \url{https://arxiv.org/abs/2412.20138}.

\bibitem[Xie et~al.(2023)Xie, Han, Zhang, Lai, Peng, Lopez-Lira, and Huang]{xie2023pixiu}
Q.~Xie, W.~Han, X.~Zhang, Y.~Lai, M.~Peng, A.~Lopez-Lira, and J.~Huang.
\newblock Pixiu: A large language model, instruction data and evaluation benchmark for finance, 2023.
\newblock URL \url{https://arxiv.org/abs/2306.05443}.

\bibitem[Xing(2024)]{xing2024designingheterogeneousllmagents}
F.~Xing.
\newblock Designing heterogeneous llm agents for financial sentiment analysis, 2024.
\newblock URL \url{https://arxiv.org/abs/2401.05799}.

\bibitem[Yang et~al.(2023)Yang, Liu, and Wang]{yang2023fingpt}
H.~Yang, X.-Y. Liu, and C.~D. Wang.
\newblock Fingpt: Open-source financial large language models, 2023.
\newblock URL \url{https://arxiv.org/abs/2306.06031}.

\bibitem[Yu et~al.(2023)Yu, Li, Chen, Jiang, Li, Zhang, Liu, Suchow, and Khashanah]{finmem}
Y.~Yu, H.~Li, Z.~Chen, Y.~Jiang, Y.~Li, D.~Zhang, R.~Liu, J.~W. Suchow, and K.~Khashanah.
\newblock Finmem: A performance-enhanced llm trading agent with layered memory and character design, 2023.
\newblock URL \url{https://arxiv.org/abs/2311.13743}.

\bibitem[Zhang et~al.(2023{\natexlab{a}})Zhang, Yang, and Liu]{zhang2023instructfingpt}
B.~Zhang, H.~Yang, and X.-Y. Liu.
\newblock Instruct-fingpt: Financial sentiment analysis by instruction tuning of general-purpose large language models, 2023{\natexlab{a}}.
\newblock URL \url{https://arxiv.org/abs/2306.12659}.

\bibitem[Zhang et~al.(2024{\natexlab{a}})Zhang, Hua, Xu, Kong, Zuo, and Guo]{unveiling}
H.~Zhang, F.~Hua, C.~Xu, H.~Kong, R.~Zuo, and J.~Guo.
\newblock Unveiling the potential of sentiment: Can large language models predict chinese stock price movements?, 2024{\natexlab{a}}.
\newblock URL \url{https://arxiv.org/abs/2306.14222}.

\bibitem[Zhang et~al.(2022)Zhang, Roller, Goyal, Artetxe, Chen, Chen, Dewan, Diab, Li, Lin, Mihaylov, Ott, Shleifer, Shuster, Simig, Koura, Sridhar, Wang, and Zettlemoyer]{zhang2022opt}
S.~Zhang, S.~Roller, N.~Goyal, M.~Artetxe, M.~Chen, S.~Chen, C.~Dewan, M.~Diab, X.~Li, X.~V. Lin, T.~Mihaylov, M.~Ott, S.~Shleifer, K.~Shuster, D.~Simig, P.~S. Koura, A.~Sridhar, T.~Wang, and L.~Zettlemoyer.
\newblock Opt: Open pre-trained transformer language models, 2022.
\newblock URL \url{https://arxiv.org/abs/2205.01068}.

\bibitem[Zhang et~al.(2024{\natexlab{b}})Zhang, Zhao, Xia, Sun, Sun, Qin, Li, Zhao, Zhao, Cai, Zheng, Wang, and An]{multimodalfinmem}
W.~Zhang, L.~Zhao, H.~Xia, S.~Sun, J.~Sun, M.~Qin, X.~Li, Y.~Zhao, Y.~Zhao, X.~Cai, L.~Zheng, X.~Wang, and B.~An.
\newblock A multimodal foundation agent for financial trading: Tool-augmented, diversified, and generalist, 2024{\natexlab{b}}.
\newblock URL \url{https://arxiv.org/abs/2402.18485}.

\bibitem[Zhang et~al.(2023{\natexlab{b}})Zhang, Yang, and Xu]{zhang2023xuanyuan}
X.~Zhang, Q.~Yang, and D.~Xu.
\newblock Xuanyuan 2.0: A large chinese financial chat model with hundreds of billions parameters, 2023{\natexlab{b}}.
\newblock URL \url{https://arxiv.org/abs/2305.12002}.

\end{thebibliography}

\clearpage
\newpage
\beginsupplement

\section{Data Collection} \label{app:data_collection}
Training a financial reasoning model that is both useful and auditable requires data that captures the diverse information channels traders and firms rely on in practice. To this end, we assemble a large, time-stamped corpus spanning assets, market regimes, and horizons, with multiple modalities and data dimensions. Each reasoning artifact (corporate filings, earnings commentary, reputable news, and technical signals) is aligned with price series under strict pre/post ordering to prevent lookahead bias.

This diversity and temporal discipline are essential for two reasons. First, the format objectives require high-quality exemplars of market analysis so the model can learn reasoning patterns that are coherent, verifiable, and well structured. Second, the outcome objectives require clean data of heterogeneous inputs from technical indicators, news, and sentiment to discrete trading actions and benchmarked excess returns over fixed horizons so that the model can learn meaningful connections of the data to trends in the market. This ensures that the learned reward captures directional accuracy, signal strength, and trading frictions. In short, the quality of the input data directly determines whether the model produces reasoning that is coherent, logical, and grounded in real market conditions.

The dataset is built in two stages. In the first stage, we collect large-scale raw inputs consisting of textual sources and numerical technical indicators. These inputs form the contextual foundation provided to the reasoning model. In the second stage, we integrate and structure these raw inputs into temporally grounded samples. For each trading day and ticker, we assemble the relevant documents and signals into a single input prompt that reflects the information available at that time. Each structured sample can then be paired with downstream labels for both supervised fine-tuning (SFT) and reinforcement learning (RL).

To ensure fairness and reproducibility, all data are sourced from publicly available providers with documented provenance and are collected through transparent, versioned pipelines.

\subsection{Large-Scale Raw Data Collection}

To ensure the training data captures a broad and meaningful diversity of inputs, we draw from five major categories of financial information: news related to the asset, technical indicators for both the asset and the broader market, fundamental financial data, sentiment surrounding the company or asset, and macroeconomic factors. Together, these categories provide comprehensive coverage of sources and data types, allowing the model to identify reliable patterns that reflect both market conditions and asset-specific dynamics. The collection process for each category is detailed below.

\subsubsection{News} 
\begin{table}[ht]
    \centering
    \caption{Temporal segmentation and sampling scheme for news data (Finnhub and Google News).}
    
    \label{tab:news_horizon}
    \begin{tabular}{lll}
    \toprule
    \textbf{Time Horizon} & \textbf{Date Range (relative to $t$)} & \textbf{Max Samples} \\
    \midrule
    Last 3 days & $t - 3$ days to $t$ & 10 \\
    Last 4--10 days & $t - 10$ days to $t - 4$ days & 20 \\
    Last 11--30 days & $t - 30$ days to $t - 11$ days & 20 \\
    \bottomrule
    \end{tabular}
\end{table}

To construct a robust news dataset, we implemented two complementary pipelines: structured financial APIs and broader web sources. This dual-source design balances timeliness with diversity, exposing the model to both precise market-moving signals and a wide range of narrative perspectives. All news was segmented into temporal buckets (Table~\ref{tab:news_horizon}), enabling the model to distinguish between recent developments, medium-term narratives, and older but still relevant context. This captures the natural decay of informational value over time, aligning the dataset with real-world trading dynamics. Finnhub was chosen for its structured, real-time financial coverage, while Google News was used to aggregate heterogeneous media perspectives, mitigating source bias and capturing broader narrative context. Integrating both sources preserves temporal precision while enriching coverage breadth, supporting downstream tasks that depend on both hard financial events and softer sentiment-driven dynamics.

\paragraph{Finnhub API}
We employed the Finnhub company news API to retrieve asset-specific news over a 30-day lookback horizon. Articles were filtered through a custom relevance function to remove unrelated items, then grouped by publication date. Following this, news items were segmented into three temporal buckets: \textit{last 3 days}, \textit{last 4--10 days}, and \textit{last 11--30 days}. Within each segment, articles were randomly sampled to a maximum of 10, 20, and 20 items, respectively. The retained items were chronologically sorted in descending order and formatted into standardized string outputs tagged by time horizon.

\paragraph{Google News Scraper}
To complement the structured Finnhub feed, we developed a custom scraper for Google News. This allowed us to capture a broader range of sources, including international and sector-specific outlets that may not be covered by financial APIs. For each article, we parsed the headline, snippet, and publication date, normalizing the timestamp into a consistent format. Articles were then segmented into the same three temporal buckets used for Finnhub (\textit{last 3 days}, \textit{last 4--10 days}, and \textit{last 11--30 days}), with random sampling caps of 10, 20, and 20 items, respectively. This ensured comparability across data sources. Final outputs were sorted in reverse chronological order and formatted into standardized, time-tagged strings.

\subsubsection{Technicals}

\begin{table}[ht]
    \centering
    \caption{Technical indicators collected in the dataset, grouped by category.}
    
    \label{tab:data_indicators}
    \resizebox{0.85\linewidth}{!}{%
    \begin{tabularx}{\linewidth}{@{}l l X@{}}
    \toprule
    \textbf{Category} & \textbf{Indicators} & \textbf{Purpose} \\
    \midrule
    Moving Averages & 50 SMA, 200 SMA & Medium/long-term trend; support/resistance. \\
                    & 50 EMA, 10 EMA  & Short/mid-term momentum, responsive to shifts. \\
    \midrule
    MACD Family     & MACD, Signal, Histogram & Momentum via EMA differentials; crossovers, divergence. \\
    \midrule
    Ichimoku System & Cloud, Conversion, Base, Span B & Trend + support/resistance framework, crossovers. \\
    \midrule
    Momentum        & RSI, KDJ (K/D/J), CCI, ROC & Overbought/oversold, stochastic turns, momentum shifts. \\
    \midrule
    Volatility      & ATR, ATR(5), Z-score(75) & Price volatility, dynamic stops, statistical extremes. \\
    \midrule
    Volume-Based    & PVO, MFI, ADX/ADXR, VWMA & Volume trends, buying/selling pressure, trend strength. \\
    \midrule
    Bollinger Bands & Middle, Upper, Lower & Mean reversion baseline; volatility expansion zones. \\
    \bottomrule
    \end{tabularx}%
    }
\end{table}

Price data and technical indicators are fundamental to trading decisions because they capture the market's internal dynamics in real time. We included raw price and volume series alongside widely used indicators to provide quantitative signals of how traders collectively respond to evolving conditions. Unlike news, which provides narrative context, technical data highlights patterns of momentum, volatility, and trend reversals that often anticipate or reinforce news-driven movements. Incorporating standardized indicators grounds the dataset in practitioner heuristics, while combining technicals with other sources balances external drivers of asset performance with endogenous market behavior. This integration enables the model to learn a more complete and realistic mapping of cause and effect in trading environments.

\paragraph{Price Data (Yahoo Finance)}  
We obtained historical price and volume data from the Yahoo Finance API. For each trading day, a 15-day rolling window was extracted containing open, high, low, close (OHLC) values and trading volume. This windowed design ensures that the model can learn short-term trading patterns and the immediate market context around a given date.

\paragraph{Technical Indicators (Yahoo Finance + Stockstats)}  
In addition to raw prices, we computed a suite of widely used technical indicators using the \texttt{stockstats} library. To ensure accurate calculation of long-horizon indicators such as the 200-day SMA or Ichimoku components, a two-year historical lookback was fetched from Yahoo Finance. Indicators included moving averages (SMA, EMA), momentum measures (MACD, RSI, ROC, KDJ), volatility metrics (ATR, Bollinger Bands, Z-scores), and volume-based indicators (MFI, PVO, VWMA, ADX). For each indicator, a 15-day output window was generated and tagged with its corresponding name (e.g., \texttt{<macd>}, \texttt{<rsi>}), alongside a brief description of its purpose and usage. This ensured that both the values and contextual interpretations were preserved for downstream reasoning tasks. A full list of indicators collected can be seen in Table~\ref{tab:data_indicators}

\subsubsection{Fundamentals}
While news and technical data capture short-term signals, \textit{fundamental information} anchors valuation to a firm's underlying financial health. Balance sheets, income statements, and cash flow statements provide structured views of profitability, leverage, liquidity, and growth prospects—factors that shape long-term price dynamics. Incorporating fundamentals grounds predictions in metrics widely used for valuation and risk assessment, complementing sentiment and market behavior with exogenous signals not visible in other information sources.

\paragraph{SimFin API}  
We collected structured balance sheet, income statement, and cash flow data from the SimFin API at both \textit{quarterly} and \textit{annual} frequencies (with trailing-twelve-month (TTM) variants included). For each ticker and frequency, we filtered reports to those published on or before the target trading date to ensure chronological validity. From each statement, we extracted key line items such as total assets, liabilities, equity, revenue, gross profit, operating expenses, net income, and cash flow components.

\paragraph{SEC Filings}  
To complement SimFin's structured data, we also integrated \textit{primary-source SEC filings}. Using a custom pipeline, we parsed quarterly filings (10-Q, 10-K) made available through the SEC EDGAR dataset. Each filing was aligned with reporting quarters and mapped to a ticker's CIK code. We extracted a curated set of key tags, including revenue, net income, operating income, earnings per share, total assets, liabilities, equity, and cash flow items. For each quarter, we associated the relevant SEC filing with all dates falling within that reporting window, ensuring temporal consistency. This design provided redundancy with SimFin while enriching coverage with directly reported financials.

\subsubsection{Sentiment}
Sentiment complements news and technicals by capturing how insiders and analysts respond to evolving fundamentals. Insider activity reflects management's private information and incentives, while analyst recommendations summarize institutional views and value revisions. These behaviorally grounded signals are exogenous to price action, often preceding disclosures or broader narratives. This integration helps the model learn how belief updates translate into order flow and price, strengthening the mapping from information to trading outcomes.

\paragraph{Insider Sentiment (Finnhub)}
We queried Finnhub's insider sentiment endpoint for each ticker using a 90-day lookback anchored to the target trading date. Reports were filtered to the \emph{current} month and the \emph{two} preceding months, deduplicated per month, sorted most-recent-first, and formatted with the two key fields: (i) \textit{change} (net insider buying/selling, humanized as K/M/B/T) and (ii) \textit{mspr} (monthly share purchase ratio). 

\paragraph{Insider Transactions (Finnhub)}
To obtain transaction-level detail, we iteratively fetched insider transactions in non-overlapping 30-day windows walking backward up to two years (or until at least 40 unique filings were found). Entries were deduplicated by record ID, then randomly subsampled to at most 25 transactions, and finally sorted by \textit{filingDate} (newest first).

\paragraph{Analyst Recommendations (Yahoo Finance)}
Professional sentiment was captured via Yahoo Finance's \texttt{upgrades\_downgrades} feed. For each ticker, we selected the latest recommendation date on or before the trading date and built a trailing 90-day window.

\subsubsection{Macros}
Macroeconomic signals form the backdrop for firm- and sector-level dynamics. Interest rates, inflation, employment, housing, and sentiment shape discount rates, risk premia, and growth expectations, providing essential context for asset-specific signals. By introducing these slow-moving, economy-wide constraints, macro data complements news, technicals, fundamentals, and sentiment. It allows the model to condition short-horizon signals on the broader regime, improving generalization across cycles and grounding rewards in economically meaningful state variables.

\paragraph{FRED API Collection}
We sourced U.S. macro indicators from the Federal Reserve Economic Data (FRED) API using an authenticated key. For each target trading date, we pulled a two-year history per series and retained first-of-month observations to ensure a consistent monthly frequency.

\subsection{Input Data Assembly}
\paragraph{Assembly Strategy}
After collecting the raw inputs, we construct each training sample by assembling data that describes an asset's financial situation on a given date. To improve flexibility, the model is trained on inputs with varying levels of information availability: subsets of technicals, sentiment, and fundamentals are randomly sampled and shuffled in order (e.g., news first, then fundamentals, or vice versa). This ensembling process generates diverse representations of the same underlying state, helping the model learn to reason under incomplete or differently structured inputs. For each date–ticker pair, we produce roughly 20 such variations, expanding the dataset and increasing diversity. These samples can then be paired with either SFT targets or RL labels during training.

\paragraph{Token Saving Strategies} High data diversity and volume mean that inputs, especially news and social media, can easily exceed 80K tokens if left unprocessed. To control this, we apply several preprocessing strategies that substantially reduce token usage. Numerical values are abbreviated (e.g., 1000 to 1k), saving space across large datasets. Overlong posts are truncated, irrelevant articles are filtered via regex, and markdown formatting is stripped from freeform text. For sources that cannot be statically checked, such as news or social media, we apply an LLM-based relevance filter to retain only high–information content. Together, these steps preserve core signals while minimizing wasted tokens.

\subsection{Dataset Statistics}

\begin{table}[ht]
    \centering
    \caption{Equity Portfolio by Sector: Market Capitalizations and ETF AUM as of September 2025}
    \label{tab:portfolio_sectors}
   \resizebox{0.8\textwidth}{!}{%
    \begin{tabular}{llc}
    \toprule
    \textbf{Sector} & \textbf{Company} & \textbf{Market Cap/AUM (USD)} \\
    \midrule
    \multirow{3}{*}{Information Technology (45)} & NVIDIA Corporation (NVDA) & \$4.18T \\
                                & Microsoft Corporation (MSFT) & \$3.78T \\
                                & Apple Inc. (AAPL) & \$3.56T \\
    \midrule
    Communication Services (50) & Meta Platforms Inc. (META) & \$1.89T \\
    \midrule
    \multirow{2}{*}{Consumer Discretionary (25)} & Amazon.com Inc. (AMZN) & \$2.48T \\
                                           & Tesla Inc. (TSLA) & \$1.13T \\
    \midrule
    \multirow{2}{*}{Financials (40)} & Berkshire Hathaway (BRK.B) & \$1.08T \\
                                & JPMorgan Chase \& Co. (JPM) & \$809B \\
    \midrule
    \multirow{2}{*}{Health Care (35)} & Eli Lilly and Company (LLY) & \$666B \\
                                & Johnson \& Johnson (JNJ) & \$430B \\
    \midrule
    \multirow{2}{*}{Energy (10)} & Exxon Mobil Corp. (XOM) & \$466B \\
                           & Chevron Corporation (CVX) & \$310B \\
    \midrule
    \multirow{2}{*}{ETFs} & SPDR S\&P 500 ETF Trust (SPY) & \$655B \\
                         & Invesco QQQ Trust (QQQ) & \$366B \\
    \bottomrule
    \end{tabular}%
    }
\end{table}

We construct our dataset from 14 large-cap tickers spanning diverse sectors: \texttt{BRK.B}, \texttt{JPM}, \texttt{LLY}, \texttt{JNJ}, \texttt{XOM}, \texttt{CVX}, \texttt{AAPL}, \texttt{NVDA}, \texttt{AMZN}, \texttt{META}, \texttt{MSFT}, \texttt{TSLA}, \texttt{QQQ}, and \texttt{SPY}. The collection covers January 1, 2024 through May 31, 2025 (roughly 354 trading days). For each day–ticker pair, we generate 20 sample variations, yielding approximately 100k training examples for downstream SFT and RFT experiments. A sectoral and market-capitalization breakdown of these assets is provided in Table~\ref{tab:portfolio_sectors}. 

\begin{table}[ht]
    \centering
    \caption{Token distribution statistics across tickers}
    
    \label{tab:token_dist}
    \resizebox{0.7\linewidth}{!}{%
    \begin{tabular}{lrrrrrrrr}
\toprule
\textbf{Ticker} & \textbf{Mean} & \textbf{Median} & \textbf{Min} & \textbf{Max} & \textbf{Std Dev} & \textbf{25th \%ile} & \textbf{75th \%ile} \\
\midrule
NVDA  & 18{,}169.4 & 18{,}285.0 & 15{,}044 & 20{,}791 & 1{,}103.4 & 17{,}351.0 & 19{,}131.2 \\
MSFT  & 22{,}684.2 & 22{,}478.0 & 19{,}101 & 27{,}841 & 1{,}605.0 & 21{,}658.5 & 23{,}096.2 \\
AAPL  & 20{,}196.9 & 20{,}590.5 & 16{,}855 & 23{,}810 & 1{,}360.3 & 19{,}181.8 & 21{,}353.2 \\
\midrule
META  & 19{,}030.1 & 19{,}216.0 & 15{,}029 & 22{,}180 &   869.9 & 18{,}372.0 & 19{,}651.0 \\
\midrule
AMZN  & 20{,}349.2 & 20{,}443.5 & 16{,}413 & 24{,}368 & 1{,}143.0 & 19{,}465.0 & 21{,}324.0 \\
TSLA  & 18{,}342.9 & 18{,}316.0 & 13{,}662 & 20{,}962 & 1{,}174.2 & 17{,}575.0 & 19{,}405.2 \\
\midrule
BRK.B & 15{,}399.5 & 15{,}375.0 & 12{,}243 & 16{,}955 &   792.6 & 15{,}023.0 & 16{,}015.5 \\
JPM   & 15{,}600.6 & 15{,}576.0 & 12{,}114 & 18{,}207 & 1{,}021.0 & 14{,}788.0 & 16{,}568.2 \\
\midrule
LLY   & 16{,}134.0 & 16{,}203.5 & 13{,}487 & 18{,}260 &   595.7 & 15{,}862.0 & 16{,}491.0 \\
JNJ   & 34{,}623.7 & 34{,}591.0 & 26{,}098 & 43{,}699 & 5{,}194.8 & 29{,}130.5 & 39{,}740.0 \\
\midrule
XOM   & 16{,}672.7 & 16{,}621.0 & 13{,}410 & 18{,}527 &   651.2 & 16{,}210.0 & 17{,}125.0 \\
CVX   & 24{,}799.4 & 21{,}374.0 & 17{,}299 & 34{,}750 & 6{,}089.4 & 20{,}247.0 & 32{,}249.0 \\
\midrule
SPY   &  4{,}829.9 &  4{,}469.0 &  3{,}808 &  7{,}172 &   813.7 &  4{,}277.0 &  5{,}217.5 \\
QQQ   &  5{,}285.4 &  5{,}026.0 &  3{,}983 &  7{,}877 &   838.2 &  4{,}738.8 &  5{,}571.0 \\
    \bottomrule
    \end{tabular}%
    }
\end{table}

Table~\ref{tab:token_dist} further summarizes the token distributions for each ticker. We observe that token lengths vary between 3.8k (minimum for \texttt{SPY}) and 43k (maximum for \texttt{JNJ}). Blue-chip healthcare names such as \texttt{JNJ} and \texttt{LLY} exhibit the longest sequences (means of 34.6k and 16.1k, respectively), while index trackers (\texttt{QQQ}, \texttt{SPY}) have the shortest (means of 5.3k and 4.8k). This reduction is expected since exchange-traded funds lack firm-level fundamentals (e.g., earnings reports, balance sheets, product pipelines) that contribute heavily to the token length of individual equities. Most other tickers lie in the 15k–23k range, with relatively narrow interquartile spreads (e.g., \texttt{AAPL}, \texttt{AMZN}, \texttt{META}). The consistency of mean and median values within each ticker indicates stable sequence lengths, while standard deviations remain modest relative to means (with the exception of \texttt{CVX}, where energy-sector reporting introduces occasional spikes). Overall, the dataset provides balanced input lengths across tickers, mitigating the risk of bias from systematically longer or shorter token sequences.

\section{Volatility-Adjusted Label Generation and Evaluation Metrics} \label{app:signal_gen_alg}

\subsection{Volatility-Based Label Generation}

Algorithm~\ref{alg:vol_signals} outlines the procedure used to generate multi-horizon volatility-informed labels, which serve as verifiable outcome rewards for reinforcement learning. The procedure constructs composite signals by computing exponential moving average prices, calculating forward returns over multiple time horizons (3, 7, 15 days), and normalizing each return series by its rolling volatility. These normalized signals are weighted and combined to form a composite signal. Finally, percentile thresholds from the composite signal distribution assign observations to one of five trading actions (\texttt{strong sell}, \texttt{sell}, \texttt{hold}, \texttt{buy}, \texttt{strong buy}). 
This design captures multi-horizon market dynamics, ensures signals are normalized by time-varying volatility, and provides suitable outcome rewards for reinforcement learning.

\begin{algorithm}
\caption{\model Multi-Horizon Volatility-Based Trading Signal Generation}
\label{alg:vol_signals}
\begin{algorithmic}[1]
\Require Asset price time series $P = \{P_1, P_2, \ldots, P_T\}$
\Require Horizon set $H = \{3, 7, 15\}$ days
\Require Signal weights $w = \{0.3, 0.5, 0.2\}$
\Require Percentile thresholds $q = \{0.03, 0.15, 0.53, 0.85\}$
\Ensure Trading signals $L = \{L_t\}$ for each time $t$
\State $\text{EMA} \gets P.\text{ewm}(\text{span}=3).\text{mean}()$
\For{$\tau \in H$}
    \State $R_\tau \gets (\text{EMA} - \text{EMA}.\text{shift}(\tau)) / \text{EMA}.\text{shift}(\tau)$
    \State $V_\tau \gets R_\tau.\text{rolling}(20).\text{std}()$
    \State $S_\tau \gets R_\tau / V_\tau$
\EndFor
\State $\text{WeightedSignal} \gets \sum_{\tau \in H} w_\tau \cdot S_\tau$
\State $\text{Valid} \gets \text{WeightedSignal}.\text{notna}()$
\State $\text{ValidSignals} \gets \text{WeightedSignal}[\text{Valid}]$
\If{ValidSignals is empty}
    \State $L \gets \text{NaN}$
    \State \Return $L$
\EndIf
\For{$i \in \{1,2,3,4\}$}
    \State $\text{threshold}_i \gets \text{ValidSignals}.\text{quantile}(q_i)$
\EndFor
\For{each time $t$ in the dataset}
    \If{$\text{Valid}[t]$}
        \State $x \gets \text{WeightedSignal}[t]$
        \If{$x \geq \text{threshold}_4$}
            \State $L_t \gets \text{STRONG BUY}$
        \ElsIf{$x \geq \text{threshold}_3$}
            \State $L_t \gets \text{BUY}$
        \ElsIf{$x \geq \text{threshold}_2$}
            \State $L_t \gets \text{HOLD}$
        \ElsIf{$x \geq \text{threshold}_1$}
            \State $L_t \gets \text{SELL}$
        \Else
            \State $L_t \gets \text{STRONG SELL}$
        \EndIf
    \Else
        \State $L_t \gets \text{NaN}$
    \EndIf
\EndFor
\State \Return Trading signals $L = \{L_t\}$
\end{algorithmic}
\end{algorithm}

\subsection{Evaluation Metrics}

We employ standard finance metrics that capture both profitability and risk characteristics of trading strategies. The metrics are defined as follows:

\begin{itemize}
  \item \textbf{Cumulative Return (CR).} For per-period returns $r_t$ ($t=1,\dots,N$), define the equity curve
  $V_t = V_0 \prod_{t=1}^{N} (1 + r_t)$. Then
  \[
    \mathrm{CR} \;=\; \frac{V_N}{V_0} - 1 \;=\; \prod_{t=1}^{N} (1+r_t) - 1.
  \]

  \item \textbf{Sharpe Ratio (SR).} We use the 10-year US Treasury yield (US10Y) as the risk-free rate benchmark, setting $r_f = 4\%$ annually. With per-period risk-free rate $r_f$, excess returns are $x_t = r_t - r_f$. Let
  $\bar{x} = \frac{1}{N}\sum_{t=1}^{N} x_t$ and
  $s_x = \sqrt{\frac{1}{N-1}\sum_{t=1}^{N} (x_t - \bar{x})^2}$. The (per-period) Sharpe ratio is
  \[
    \mathrm{SR}_{\text{per}} \;=\; \frac{\bar{x}}{s_x}.
  \]
  If using daily (weekly, monthly) data, annualize via
  \[
    \mathrm{SR}_{\text{ann}} \;=\; \sqrt{K}\,\mathrm{SR}_{\text{per}},
  \]
  where $K$ is the number of periods per year (e.g., $K{=}252$ for daily).

  \item \textbf{Hit Rate (HR).} The fraction of trading recommendations that correctly predict the direction of price movement upon position closure. For each recommendation $a_t$ and corresponding return $r_t$:
  \[
    \mathrm{HR} \;=\; \frac{1}{N}\sum_{t=1}^{N} \mathbf{1}\{\text{sign}(a_t) = \text{sign}(r_t)\}
  \]
  where $\text{sign}(a_t)$ represents the direction of the trading recommendation (buy/sell) and $\text{sign}(r_t)$ represents the direction of the actual price movement.

  \item \textbf{Maximum Drawdown (MDD).} Using the equity curve $\{V_t\}$, define the running peak
  $P_t = \max_{1 \le u \le t} V_u$ and drawdown $D_t = 1 - \frac{V_t}{P_t}$. Then
  \[
    \mathrm{MDD} \;=\; \max_{1 \le t \le N} D_t \;=\; \max_{1 \le t \le N} \left( 1 - \frac{V_t}{\max_{1 \le u \le t} V_u} \right).
  \]
\end{itemize}

\section{Trading Proposal Specifications}

We implement a three-stage reward system progressing from structure to evidence to decision. These tasks are designed to incentivize the model to develop financial trading reasoning capabilities, advancing from superficial task aspects to fundamental decision-making. The model progresses through these stages sequentially, with tasks increasing in difficulty. Each stage features a smoothed reward landscape to provide a gradual learning curve, enabling the model to acquire structured thinking, professional investment claim composition, and sound investment decision-making capabilities.

\subsection{Structure Reward}

The structure reward incentivizes the model to think systematically and adopt professional perspectives when analyzing input data. Our input data is extensive, averaging 20,000 to 30,000 tokens, and given the use of open public data, the information can be noisy and chaotic. Through experimentation, we determined that social media data should be excluded. While social media data can collectively provide valuable signals, the observable fraction is highly biased and therefore uninformative in our context.

Given this massive dataset, we want the model to think from a feasible number of categories, such as fundamentals, news, and sell-side analyst ratings. This task is relatively straightforward because we have structured the input to facilitate the model's identification of different data sources. We initially experimented with unstructured input data without clear sectioning, but found that the model struggled to identify sources effectively. To strike an appropriate balance, while we had initially intended for the model to identify hidden structures in the data, we ultimately decided that providing more structured input would be more effective than feeding massive amounts of unstructured data, particularly given that the model is small while the input data exists in long context.

The structure training aims to: (1) enable the model to think systematically and analyze data professionally, mirroring how investment theses are typically constructed, and (2) better prepare the model for subsequent training stages. An additional benefit of this approach is its applicability to proprietary training data. Proprietary data is of higher quality, and investment theses may cover various aspects that are not uniformly positive or negative. With manual annotation, fine-grained labeling (e.g., fundamentals as positive, insider transactions as negative) enables section-by-section supervision of the investment thesis.

\subsection{Evidence Reward}

The evidence reward addresses hallucination issues observed in earlier internal model versions. We incentivize the model to engage in longer reasoning traces and expand the variability space of final decisions. We observed that smaller models tend to be more stubborn, and without sufficient sampling or increased reasoning length to expand the possibility space, generating valid reward signals becomes difficult. However, one drawback of expanded thinking is that given the small model size and long context, the model can become less faithful to the input data.

To address this faithfulness issue, we introduce evidence requirements, including quotes and sources, to make claims more grounded. For this reward stage, we seek claims that follow the opinion-quote-source structure. We control the length of opinions, quotes, and sources to prevent excessive verbosity. Additionally, we regulate the order and separating symbols using regular text, italic, and inline code formatting. This stage serves two purposes: (1) increasing model faithfulness to input data, and (2) better aligning the investment thesis with professional investment thesis standards.

\subsection{Decision Reward}

The decision reward represents the ultimate objective: given the reasoning and investment thesis, what final recommendation the model should provide. Considering the multiple factors involved in financial markets, we standardized and denoised stock returns, categorizing them by percentile and using these as decision labels for reinforcement learning training.

We incorporated market characteristics into our reward matrix design and asset selection. Since we are testing blue-chip companies (large-cap stocks with solid fundamentals), we standardized the signal to be skewed towards bullish outcomes. We also incorporated the prior knowledge that, historically, the US equity market exhibits the characteristic pattern where bear markets tend to be short and sharp, while bull markets are typically longer-lasting and more gradual.

\section{Three-Stage Investment Analysis Reward} \label{app:reward_design}

Beyond the general format and outcome rewards, we implement a specialized three-stage reward system designed specifically for structured investment analysis. This system evaluates completions that follow an XML-structured format with distinct analysis categories followed by a conclusion section.

\paragraph{Stage I: Structure Reward: XML Section Organization.}
The structure reward evaluates the organization and content quality of XML-tagged analysis sections. For a completion $x$ with XML sections $\mathcal{S}(x) = \{(t_i, c_i)\}_{i=1}^{N}$ where $t_i$ is the tag name and $c_i$ is the section content, we exclude the \texttt{think} tag and require a mandatory \texttt{conclusion} section.

Let $S = |\mathcal{S}(x)| - 1$ be the number of analysis sections (excluding conclusion). The section count reward targets 5-7 analysis sections:
\[
R_{\text{count}}(S) = \begin{cases}
1, & \text{if } 5 \leq S \leq 7, \\
\max(0.3, S/5 \times 0.7), & \text{if } S < 5, \\
\max(0.3, 1 - 0.15(S-7)), & \text{if } S > 7.
\end{cases}
\]

For each section $s$ with content $c$, we evaluate structural elements:
\[
R_{\text{struct}}(s) = 0.3 \cdot \mathbf{1}_{\text{headers}}(c) + 0.4 \cdot \mathbf{1}_{\text{bullets}}(c) + 0.2 \cdot \mathbf{1}_{\text{bold}}(c) + 0.1 \cdot \mathbf{1}_{\text{tables}}(c),
\]
where indicators check for markdown headers ($\mathbf{1}_{\text{headers}}$), bullet points ($\mathbf{1}_{\text{bullets}}$), bold text ($\mathbf{1}_{\text{bold}}$), and tables ($\mathbf{1}_{\text{tables}}$). Sections below a minimum word threshold $w_{\min} = 50$ receive $R_{\text{struct}}(s) = 0.2$.

The overall structure reward combines section count and average structural quality:
\[
R_{\text{structure}}(x) = 0.6 \cdot R_{\text{count}}(S) + 0.4 \cdot \frac{1}{N}\sum_{i=1}^{N} R_{\text{struct}}(s_i).
\]

\paragraph{Stage II: Evidence Reward: Opinion-Quote-Source Validation.}
The evidence reward evaluates the quality of evidential reasoning within analysis sections. For each non-conclusion XML section, we extract bullet points $\mathcal{B}(c)$ and analyze their opinion-evidence structure.

For bullet $b$, let $\mathcal{Q}(b)$ denote quoted text (italic markdown \texttt{*quote*}) and $\mathcal{S}(b)$ denote source citations (backtick markdown \texttt{`source`}). We identify the opinion as text preceding the first citation marker. Let $w_{\text{op}}$ be the opinion word count and define optimal ranges $w_{\min}^{\text{op}} = 15$, $w_{\max}^{\text{op}} = 90$. We define the citation indicator $C(b) = \mathbf{1}_{|\mathcal{Q}(b)| > 0 \text{ and } |\mathcal{S}(b)| > 0}$.

The opinion quality score is:
\[
R_{\text{opinion}}(b) = \begin{cases}
1, & \text{if } w_{\min}^{\text{op}} \leq w_{\text{op}} \leq w_{\max}^{\text{op}} \text{ and } C(b) = 1, \\
w_{\text{op}}/w_{\min}^{\text{op}}, & \text{if } w_{\text{op}} < w_{\min}^{\text{op}} \text{ and } C(b) = 1, \\
\max(0.5, 1 - 0.02(w_{\text{op}} - w_{\max}^{\text{op}})), & \text{if } w_{\text{op}} > w_{\max}^{\text{op}} \text{ and } C(b) = 1, \\
\min(0.3, w_{\text{op}}/w_{\min}^{\text{op}} \times 0.3), & \text{if } C(b) = 0.
\end{cases}
\]

The bullet-level evidence score combines opinion quality with citation presence:
\[
R_{\text{bullet}}(b) = 0.4 \cdot R_{\text{opinion}}(b) + 0.35 \cdot \mathbf{1}_{|\mathcal{Q}(b)| > 0} + 0.25 \cdot \mathbf{1}_{|\mathcal{S}(b)| > 0}.
\]

For section $c$ with bullets $\mathcal{B}(c)$, we first evaluate bullet count structure with optimal range $[4, 7]$:
\[
R_{\text{count}}^{\text{bullet}}(c) = \begin{cases}
1, & \text{if } 4 \leq |\mathcal{B}(c)| \leq 7, \\
|\mathcal{B}(c)|/4, & \text{if } |\mathcal{B}(c)| < 4, \\
\max(0.3, 1 - 0.1(|\mathcal{B}(c)| - 7)), & \text{if } |\mathcal{B}(c)| > 7.
\end{cases}
\]

The section evidence score uses harmonic mean to prevent extreme outliers:
\[
R_{\text{evidence}}^{\text{section}}(c) = 0.3 \cdot R_{\text{count}}^{\text{bullet}}(c) + 0.7 \cdot \frac{|\mathcal{B}(c)|}{\sum_{b \in \mathcal{B}(c)} 1/\max(R_{\text{bullet}}(b), 0.01)}.
\]

The overall evidence reward averages across all analysis sections using harmonic mean:
\[
R_{\text{evidence}}(x) = \frac{|\mathcal{S}_{\text{analysis}}(x)|}{\sum_{c \in \mathcal{S}_{\text{analysis}}(x)} 1/\max(R_{\text{evidence}}^{\text{section}}(c), 0.01)}.
\]

\paragraph{Stage III: Decision Reward: Asymmetric Trading Performance.}
The decision reward implements an asymmetric penalty structure reflecting institutional risk management priorities. The asymmetric design stems from three core principles:

(1) \textbf{Market reality}: markets fall faster than they rise, making false bullish signals during downturns more devastating.

(2) \textbf{Capital preservation}: professional trading prioritizes downside protection over opportunity maximization, as losses compound differently than gains.

(3) \textbf{Volatility asymmetry}: timing is more critical during selloffs when volatility spikes occur.

\begin{figure}
    \centering
    \includegraphics[width=0.7\linewidth]{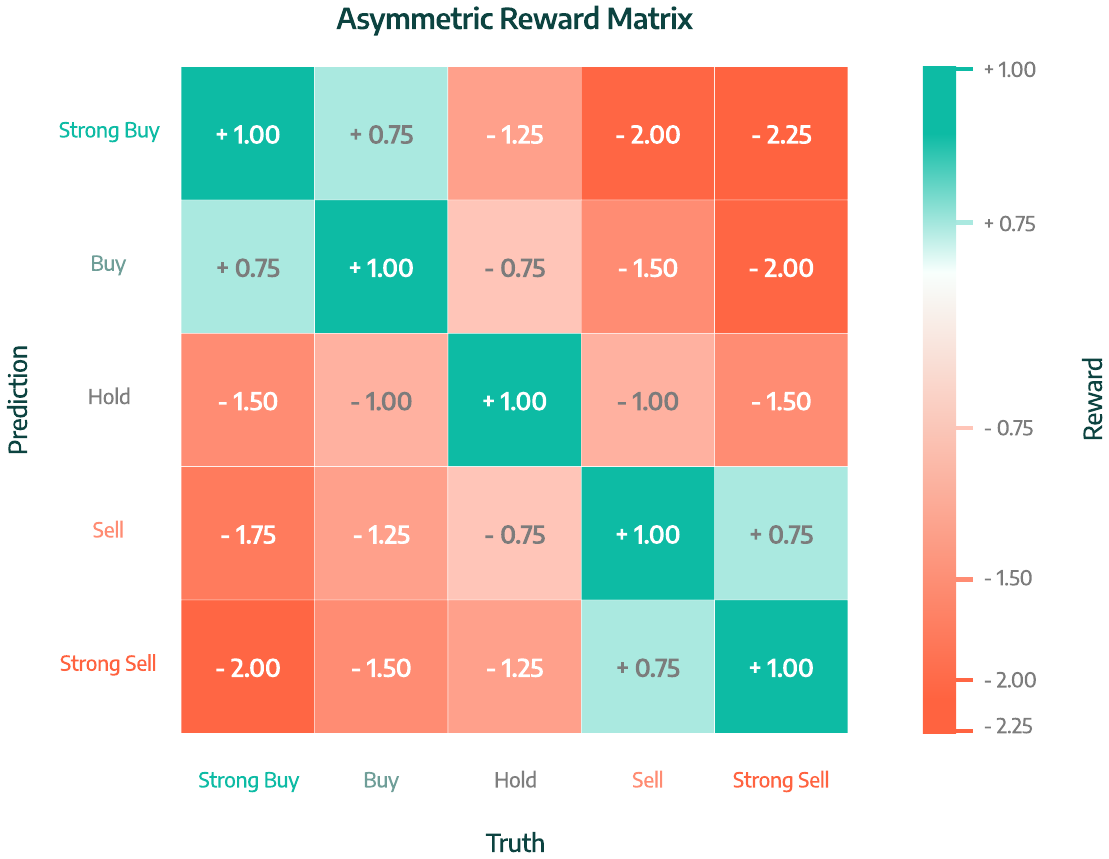}
    \caption{\model asymmetric reward heatmap: rewards (-2.25 to 1) based on model prediction vs ground truth, with labels derived from Section~\ref{sec:label_generation}}
    \label{fig:reward_heatmap}
\end{figure}

We extract the final decision from the completion using pattern \texttt{[[[DECISION]]]} in the last three lines. Let $\hat{d} \in \{\text{STRONG SELL, SELL, HOLD, BUY, STRONG BUY}\}$ be the predicted decision and $d^* \in \{\text{STRONG SELL, SELL, HOLD, BUY, STRONG BUY}\}$ be the ground truth. The asymmetric reward matrix $\mathbf{M}$ implements approximately 12\% heavier penalties for false bullish signals:

\[
\mathbf{M} = \begin{pmatrix}
& \text{SS} & \text{S} & \text{H} & \text{B} & \text{SB} \\
\text{SS} & 1.00 & 0.75 & -1.25 & -2.00 & -2.25 \\
\text{S} & 0.75 & 1.00 & -0.75 & -1.50 & -2.00 \\
\text{H} & -1.50 & -1.00 & 1.00 & -1.00 & -1.50 \\
\text{B} & -1.75 & -1.25 & -0.75 & 1.00 & 0.75 \\
\text{SB} & -2.00 & -1.50 & -1.25 & 0.75 & 1.00
\end{pmatrix}
\]

where rows represent predictions and columns represent ground truth. This matrix encodes key asymmetric principles, as visualized in Figure~\ref{fig:reward_heatmap}:
\begin{itemize}
\item Perfect matches receive full reward (1.00)
\item Same-direction mistakes receive partial reward (0.75)
\item Asymmetric penalties with mixed directional bias: \texttt{STRONG BUY}$|$\texttt{STRONG SELL} penalty is $-2.25$ compared to \texttt{STRONG SELL}$|$\texttt{STRONG BUY} penalty of $-2.00$, showing heavier penalties for false bullish signals
\item Anti-hold bias: \texttt{HOLD} predictions penalized when action is warranted (-1.50/-1.00 for bullish truth, -1.00/-1.50 for bearish truth)
\item Clean 0.25-unit increments for stable training dynamics
\end{itemize}

The decision reward for prediction $\hat{d}$ and ground truth $d^*$ is:
\[
R_{\text{decision}}(\hat{d}, d^*) = \mathbf{M}_{\hat{d}, d^*} \cdot \lambda_{\text{dec}},
\]
where $\lambda_{\text{dec}}$ is a scaling factor. If no valid decision is extracted, $R_{\text{decision}} = -1.5$ to match the severity of opposite-direction prediction errors.

\paragraph{Aggregation.}
The three-stage investment analysis reward combines all components:
\[
R_{\text{investment}}(x) = \lambda_{\text{struct}} R_{\text{structure}}(x) + \lambda_{\text{evid}} R_{\text{evidence}}(x) + \lambda_{\text{dec}} R_{\text{decision}}(\hat{d}, d^*),
\]
with non-negative weights $\lambda_{\text{struct}}, \lambda_{\text{evid}}, \lambda_{\text{dec}} \geq 0$ that can be adjusted based on the relative importance of structural quality, evidential reasoning, and decision accuracy in the specific application domain.

\section{Trading-R0 Training Observations} \label{app:r0}

During the development of \model, we identified several pitfalls in training financial reasoning models that can lead to unstable learning and poor performance. Our first iteration, which we refer to as ``Trading-R0,'' incorporated well-grounded design choices for structuring outputs but ultimately suffered from instability during training. The resulting model failed to generate reliable decisions or produce coherent, well-supported investment theses. In this section, we detail the Trading-R0 methodologies and outline the key observations that informed the improved design of \model.

\paragraph{Overview \& motivation.}
The goal of the Trading-R0 methodology was to test the hypothesis that by strictly controlling both the reasoning content inside the \texttt{<think></think>} XML tags and the final outputs, we could force the model to generate high-quality reasoning within the reward budget. To this end, we attempted a naïve one-stage training recipe: warm-starting Trading-R0 with supervised fine-tuning on distilled reasoning traces, followed by reinforcement learning with a composite objective. This approach merged \emph{format} and \emph{outcome} rewards into a single mixed signal during RL. While the two reward types targeted complementary goals, combining them prematurely introduced significant instability.

\textbf{Format Reward} Stage I rewards the presence of substantive content in the \texttt{<think>} trace and category sections with saturation; Stage II rewards explicit sectioning (headers or horizontal rules) in the think block; Stage III rewards disciplined citation/quoting patterns within categories. Two structure heads (strict/easy) further align section counts, length balance, and title handling in the think trace, and encourage thesis-like bullets in categories; a small decision-placement head enforces a terminal \texttt{DECISION:\ [[[...]]]} after the final XML tag. Together, these signals improve interpretability, automate quality control, and stabilize RL by shaping where the model ``spends'' tokens.

\textbf{Outcome Reward} The decision operates on a five-point action scale, granting full credit for exact matches, partial credit for right-direction/wrong-intensity, asymmetric penalties for opposite-direction errors, an explicit penalty for the ``HOLD hack'' (predicting \texttt{HOLD} when movement-warranted action), and graded deductions for malformed \texttt{DECISION:\ [[[...]]]} formatting. The \emph{Market Grounded Outcome Reward} further anchors calls to excess returns over a horizon using a neutral band (to avoid over-trading on noise), magnitude saturation (to prevent tail moves from dominating learning), intensity calibration (to match action strength to move size), and trading costs (to encourage calibrated selectivity). Unless noted, head outputs are bounded and combined with nonnegative weights (with optional clipping), yielding a training signal that simultaneously promotes accurate, economically meaningful actions and interpretable, evidence-backed reasoning.

\subsection{Format Reward}\label{app:format_reward}

We formalize the \emph{format reward} for completions that follow our XML-like scaffolding: a single reasoning block \texttt{<think>...</think>} followed by one or more category blocks \texttt{<cat>...</cat>} (any tag name except \texttt{think}). For a completion \(x\), let \(T(x)\) denote the content inside the \texttt{<think>} tag and \(\mathcal{C}(x)=\{(k_i,c_i)\}_{i=1}^{N}\) the multiset of category tags and their contents. All head rewards below are bounded, typically to \([0,1]\), and can be used individually or combined.

\paragraph{Stage I: content presence with saturation.}
Define \(w(\cdot)\) as word count. Stage I decomposes as
\[
R_{\text{stage1}}(x) \;=\; R_{\text{think}}(T(x)) \;+\; R_{\text{cat}}(\mathcal{C}(x)),
\quad \text{then clipped to }[0,1].
\]
\vspace{-0.5ex}
\[
R_{\text{think}}(T) \;=\;
\begin{cases}
0, & w(T)=0,\\[2pt]
\alpha_0, & 0< w(T) < w_0,\\[4pt]
\alpha_0 + \dfrac{w(T)-w_0}{L_{\min}-w_0}\,\bigl(\tfrac{1}{2}-\alpha_0\bigr), & w_0 \le w(T) < L_{\min},\\[10pt]
0.5 + 0.5\,M_{\text{think}}\dfrac{w(T)-L_{\min}}{L_{\max}-L_{\min}}, & L_{\min} \le w(T) < L_{\max},\\[10pt]
M_{\text{think}}, & w(T) \ge L_{\max},
\end{cases}
\]
with constants \(\alpha_0=0.05,\; w_0=100,\; L_{\min}=300,\; L_{\max}=800,\; M_{\text{think}}=0.5\).
The category reward counts substantial categories with partial credit:
\[
R_{\text{cat}}(\mathcal{C}) \;=\; M_{\text{cat}}\,
\frac{\min\!\Bigl(\sum_{i=1}^{N}\bigl[\mathbf{1}\{w(c_i)\!\ge\! m_c\}+\tfrac{1}{2}\mathbf{1}\{40 \le w(c_i) < m_c\}\bigr],\,K_{\max}\Bigr)}{K_{\max}},
\]
with \(m_c=80,\;K_{\max}=8,\;M_{\text{cat}}=0.5\).

\paragraph{Stage II: sectioned markdown in the think block.}
Parse \(T\) into sections via Markdown headers (\texttt{\#}, \texttt{\#\#}, \dots) or horizontal rules (\texttt{---}). Keep sections with at least \(w_{\min}\) words and mark them as structured if they arise from a header or separator. For each valid section \(s\),
\[
r(s)=\min\!\Bigl(1,\, \frac{w(s)}{w^*}\Bigr),\qquad
w_{\min}=50,\;\; w^*=100,\;\; S_{\max}=8,\;\; M_{\text{sec}}=1.0,
\]
and
\[
R_{\text{stage2}}(T)=M_{\text{sec}}\cdot \frac{1}{S_{\max}}\sum_{s\in \mathcal{S}_{\text{top}}(T)} r(s),
\]
where \(\mathcal{S}_{\text{top}}(T)\) are the first \(S_{\max}\) structured sections.

\paragraph{Stage III: citation/quoting patterns in categories.}
For category content \(c\), extract bullet points \(\mathcal{B}(c)\) (lines starting with \texttt{-/*/+} or numbered lists). For bullet \(b\), let \(q(b)\) be the number of quote pairs and \(p(b)\) the number of parenthesis pairs. The bullet score is
\[
r_{\text{bul}}(b)=w_s\!\cdot\!\bigl(1-\delta_{\text{startsQuote}}(b)\,p_q\bigr)
+ w_q\,\min\!\Bigl(1, \tfrac{q(b)}{Q_{\max}}\Bigr)
+ w_p\,\min\!\Bigl(1, \tfrac{p(b)}{P_{\max}}\Bigr),
\]
with weights and caps \(w_s=0.5,\; w_q=0.3,\; w_p=0.2,\; p_q=0.1,\; Q_{\max}=2,\; P_{\max}=1\).
Average over bullets per category, then over at most \(C_{\max}=7\) categories:
\[
R_{\text{stage3}}(\mathcal{C})=\frac{1}{C_{\max}}\sum_{(k_i,c_i)\in \mathcal{C}_{\text{top}}}
\;\frac{1}{|\mathcal{B}(c_i)|}\sum_{b\in \mathcal{B}(c_i)} r_{\text{bul}}(b).
\]

\paragraph{Think-structure rewards (strict \& easy).}
Split \(T\) by \texttt{---} into sections. Let \(S\) be the number of accepted sections, \(\tau\in\{0,1\}\) indicate a short title (\(\le 32\) words) in the first section, and \(\{w_j\}_{j=1}^{S}\) their word counts (exclude the title from dispersion). The \emph{strict} reward requires at least one \texttt{---} (otherwise \(0\)) and combines four terms:
\[
R_{\text{TS}} \;=\; 0.40\,\overline{r_{\text{len}}}
\;+\; 0.25\,r_{\text{bal}}
\;+\; 0.25\,r_{\text{count}}
\;+\; 0.10\,r_{\text{title}}.
\]
Here, for non-title sections, \(r_{\text{len}}=1\) if \(160\!\le\! w_j \le\!220\), ramps linearly down to \(0.3\) at \(w_j=50\), and decays for \(w_j>220\) (1\% per extra word, floored at \(0.1\)); an additional paragraph penalty applies if a section has \(>24\) newlines: multiply by \(\max\bigl(0.1, 1 - 0.02(\text{newlines}-24)\bigr)\).
Balance uses the coefficient of variation (CV) on non-title lengths:
\[
r_{\text{bal}}=\max\!\bigl(0.5,\, 1-2\,\text{CV}\bigr).
\]
Section count peaks at \(5\!-\!7\) with sharp penalties outside:
\(r_{\text{count}}=1\) if \(5\le S\le 7\), else decreases piecewise (heavy for deficits/excess).
Title handling is \(r_{\text{title}}=1\) if \(\tau=1\), else \(0.7\).

The \emph{easy} variant widens ranges and softens penalties:
acceptable section count \(4\!-\!8\); content sweet spot \(80\!-\!300\) words (gentle ramps from \(50\)); much gentler overflow; and weights
\[
R_{\text{ETS}} \;=\; 0.40\,\text{(basic structure)} \;+\; 0.30\,\text{(content presence)} \;+\; 0.20\,\text{(balance)} \;+\; 0.10\,\text{(advanced)}.
\]

\paragraph{Thesis-style category rewards (strict \& easy).}
For each category \(c\), bullets \(b\in\mathcal{B}(c)\) are analyzed into: (i) opinion words (prefix before the first citation marker: quotes/parentheses/brackets/URL), (ii) total bullet length, and (iii) bullet count. The \emph{strict} head averages per-category quality (70\%) with a prior on category count (30\%, peak at \(5\!-\!7\)). Per-category quality combines mean opinion-citation conformity (optimal 16–30 opinion words before the first marker, zero if no citation), mean bullet length (optimal 45–90 words), and bullet count (optimal 3–6), with weights \(0.35/0.20/0.15\) (renormalized to 70\%). The \emph{easy} head broadens ranges (e.g., opinion 8–50, length 20–150, bullets 3–7, categories 4–8) and applies gentler penalties while still rewarding structured bullets.

\paragraph{Decision placement reward.}
We award a small auxiliary reward for an explicit terminal decision
\(\texttt{DECISION:\ [[[...]]]}\). Let \(\mathbf{1}_{\text{exists}}\) indicate presence of such a tag and \(\mathbf{1}_{\text{final}}\) indicate that its last occurrence appears \emph{after} the last XML closing tag in \(x\). Then
\[
R_{\text{dec}} \;=\; \tfrac{1}{2}\,\mathbf{1}_{\text{exists}} \;+\; \tfrac{1}{2}\,\mathbf{1}_{\text{final}} \;\in\; [0,0.5].
\]

\paragraph{Aggregation.}
For a single scalar format signal we form a convex combination
\[
R_{\text{format}}(x)\;=\;\sum_{h}\lambda_h\,R_h(x),\qquad
\lambda_h\ge 0,\;\; \sum_h \lambda_h \le 1,
\]
with heads \(R_h\) drawn from the stages/structure components above (followed by clipping to \([0,1]\) if desired). This yields smooth shaping for minimal compliance, controlled saturation to discourage verbosity gaming, and complementary coverage of formatting aspects (content presence, sectioning, citation discipline, and endpoint clarity).

\subsection{Market Grounded Outcome Reward} \label{app:outcome_reward}

We couple decisions to realized market outcomes over a fixed evaluation horizon. Let \(P_t\) be the asset price at decision
time \(t\) and \(P_{t+H}\) the price \(H\) steps later. Define the asset log return
\(r = \log P_{t+H}-\log P_t\) and a benchmark (market) log return \(r^{(m)}\) over the same window. The
market-grounded signal is the excess return \(e = r - r^{(m)}\), which measures stock performance net of the market.
Map the predicted class \(\hat a\in\mathcal{D}\) to an intensity-coded trade signal
\[
\begin{aligned}
s(\hat a) &\in \{-2,-1,0,1,2\},\\
s(\text{STRONG SELL}) &= -2,\quad s(\text{SELL})=-1,\\
s(\text{HOLD}) &= 0,\quad s(\text{BUY})=1,\quad s(\text{STRONG BUY})=2.
\end{aligned}
\]
We introduce three shaping hyperparameters:
(i) a neutral band \(\delta\ge 0\) within which \(\text{HOLD}\) is preferred;
(ii) a saturation scale \(u>0\) controlling how quickly rewards cap as \(|e|\) grows; and
(iii) a per-trade cost \(\kappa\ge 0\) applied whenever \(s(\hat a)\neq 0\).
Let \(m(e)=\min\!\bigl(1,|e|/u\bigr)\in[0,1]\) be a normalized magnitude and
\(I(e)=\mathbf{1}\{|e|>\tau_1\}+\mathbf{1}\{|e|>\tau_2\}\in\{0,1,2\}\) a two-threshold proxy
for the ``appropriate'' intensity (with \(0<\tau_1<\tau_2\)).

The outcome reward for a single decision is
\[
R_{\text{out}}(\hat a; e)=
\begin{cases}
b\Bigl(1-\dfrac{|e|}{\delta}\Bigr), & s(\hat a)=0\ \text{and}\ |e|\le \delta,\\[10pt]
-b\,\gamma\,\dfrac{|e|-\delta}{u-\delta}, & s(\hat a)=0\ \text{and}\ |e|>\delta,\\[10pt]
\Bigl[\ \ \ \ \, b\Bigl(1-\dfrac{|\,|s(\hat a)|-I(e)|}{2}\Bigr)\, m(e)\ \ \ \Bigr]\;-\;\kappa, & s(\hat a)\,e>0\quad\text{(right direction)},\\[10pt]
-\;b\,\gamma\, m(e)\;-\;\kappa, & s(\hat a)\,e<0\quad\text{(wrong direction)},
\end{cases}
\]
clipped to \([-\gamma b,\, b]\). Thus, (i) \texttt{HOLD} is rewarded inside the neutral band and penalized outside;
(ii) directionally correct calls earn a magnitude- and intensity-calibrated fraction of \(b\) and pay trading cost \(\kappa\);
(iii) opposite-direction calls incur an asymmetric penalty scaled by \(\gamma\) and the realized move \(m(e)\).
This objective directly aligns labels with excess returns, discourages degenerate \texttt{HOLD} behavior when the market
moves materially, and remains robust via saturation (\(u\)) and costs (\(\kappa\)).

\subsection{Observations} \label{app:r0_obs}

Our R0 experiments revealed several important lessons about the failure modes of training financial trading reasoning models. While the architecture incorporated reasonable ideas, such as XML-structured reasoning traces and outcome-grounded rewards, the way these signals were combined led to unstable training and poor final performance. We highlight three central observations:  

\paragraph{Mixed reward signals cause instability.}  
In R0, the \emph{format} and \emph{outcome} rewards were merged into a single composite objective. This decision was motivated by the idea that jointly optimizing for structured reasoning and market-aligned predictions might accelerate training. In practice, however, these two objectives competed with one another: the gradient signal from format rewards pushed the model toward satisfying XML and structural constraints, while the outcome rewards pulled it toward maximizing financial return alignment. The model frequently oscillated between these behaviors, overfitting to structural compliance in one phase and then collapsing into noisy, outcome-driven guesses in another. This lack of separation produced highly unstable training dynamics, with reward curves spiking and collapsing across iterations. The experience demonstrated that rewards of different types must be disentangled and sequenced—rather than blended—so that the model can first internalize stable reasoning scaffolds before being tasked with aligning outputs to volatile financial outcomes.

\paragraph{Over-controlling reasoning structure degrades outputs.}  
Another critical issue was the degree of control imposed over the \texttt{<think>} block. R0 used narrow reward budgets and strict penalties for deviations from prespecified section counts, bullet structures, and XML formatting. While this approach succeeded in making the outputs look superficially well-structured, it unintentionally incentivized the model to exploit the reward function rather than produce meaningful reasoning. Generated traces often became extremely short, repetitive, and shallow, with entire sections filled by boilerplate text or token padding to satisfy structural checks. In some cases, the model converged to degenerate completions where the structure was perfect on paper—e.g., clean headers, balanced lengths, consistent bullet points—but the actual investment analysis was nonsensical or missing entirely. This finding underscored the danger of excessive structural regularization: too much rigidity forced the model into brittle optimization strategies that hollowed out the quality of the reasoning itself.

\paragraph{Tight reward budgets suppress reasoning depth.}  
Closely related to over-controlling structure was the problem of narrow reward budgets. Because the thresholds for receiving credit were so tightly defined, the model learned to minimize effort by producing the shortest possible traces that passed structural gates. This behavior led to underdeveloped reasoning, with little exploration of evidence, no multi-step synthesis, and shallow decision justifications. The model effectively gamed the reward function by converging to “minimum viable outputs” that satisfied constraints but failed to produce substantive investment theses. Relaxing these budgets in later iterations, while still maintaining some scaffolding, allowed the model to explore richer reasoning pathways without being punished for minor deviations from rigid structural templates.  

\paragraph{Takeaway.}  
These observations highlight two central lessons. First, reward signals of different types must be staged rather than mixed; otherwise, the model receives unstable and conflicting gradients that destabilize learning. Second, structural scaffolding must be applied with flexibility: it should guide the reasoning process without constraining it so tightly that the model learns to optimize against the reward rather than generating meaningful analysis. The shortcomings of R0 directly motivated the staged SFT--RFT curriculum in Trading-R1, where structural reasoning patterns are established first, then progressively reinforced and aligned with market-grounded outcomes. This design not only stabilized training but also ensured that outputs were both interpretable and economically actionable.

\clearpage
\newpage

\section{Sample \model Output Traces}

\begin{tcolorbox}[
    breakable,
    colback=yellow!3!white,        
    colframe=orange!10!gray,       
    coltitle=black,                
    boxrule=0.5pt,                 
    arc=0.5mm,                       
    top=2mm, bottom=2mm, left=2mm, right=2mm, 
    fonttitle=\usefont{T1}{ppl}{b}{n},       
    title=\model Investment Thesis on Microsoft Corporation (MSFT),
]
\usefont{T1}{ppl}{m}{n}  
\footnotesize

\textbf{\textsc{I. Thinking and Planning}}
\\

\think{Let me analyze the provided financial data for Microsoft (MSFT) to create a comprehensive investment thesis. I'll need to structure this as 5-7 analysis sections following the specified XML format, with the final section being `<conclusion>`. I'll base my analysis strictly on the provided information.

First, let me organize the data I have:\\

1. Financial data (fundamentals):\\
- Balance sheet (quarterly)\\
- Income statement (quarterly)\\
- Cash flow statement (quarterly)\\
- SEC filings with various financial metrics across different periods\\

2. News data:\\
- Recent news articles (last 3 days, 4-10 days, 11-30 days)\\
- Focus on earnings reports, AI developments, cloud growth, etc.\\

3. Market data:\\
- Price history (last 11 trading days)\\
- Technical indicators (RSI, MACD, EMA, SMA, Bollinger Bands, ATR, ADX)\\

4. Sentiment data:\\
- Analyst recommendations\\
- Insider transactions (limited data)\\

5. Macro data:\\
- Economic indicators for Feb-May 2025\\

Let me go through each section systematically to extract key insights.

\#\# Fundamentals Analysis

From the balance sheet (Q2 2025, reported 2025-01-29):\\
- Total Assets: \$533.9B\\
- Total Liabilities: \$231.2B\\
- Total Equity: \$302.7B\\
- Cash, Cash Equivalents \& Short Term Investments: \$71.6B\\
- Total Current Assets: \$147.1B\\
- Total Current Liabilities: \$108.9B\\
- Long Term Debt: \$56.9B\\
- Debt-to-Equity ratio: 56.9B/302.7B = 0.19 (very low)\\
- Current Ratio: 147.1B/108.9B = 1.35 (healthy liquidity)\\

From the income statement (Q2 2025):\\
- Revenue: \$69.6B\\
- Gross Profit: \$47.8B (68.7\% gross margin)\\
- Operating Income: \$31.7B (45.5\% operating margin)\\
- Net Income: \$24.1B\\

From the cash flow statement (Q2 2025):\\
- Net Cash from Operating Activities: \$22.3B\\
- Net Cash from Investing Activities: -\$14.1B\\
- Net Cash from Financing Activities: -\$11.2B\\
- Net Change in Cash: -\$3.4B\\

From SEC filings:\\
- Various revenue figures across different periods\\
- Equity figures showing growth from negative values to positive\\
- Cash and cash equivalents showing growth\\
- EPS figures\\

Let me check for consistency and identify key metrics:\\
- Revenue appears to be growing (from SEC filings showing revenue figures)\\
- Operating income shows strong growth\\
- Cash position is substantial at \$71.6B\\
- Debt is relatively low at \$56.9B\\
- Equity has been improving significantly\\

\#\# Technical Analysis

From market data:\\
- Recent price action shows strong upward momentum:\\
- From ~\$360 on 4/16 to ~\$424 on 5/1 (17.2\% increase)\\
- Recent close at \$423.924 on 5/1/2025\\
- RSI: 69.269 (approaching overbought territory but still in bullish range)\\
- MACD: 4.414 (positive and rising, indicating strong bullish momentum)\\
- 50 EMA: \$388.541 (price is above this, bullish)\\
- 200 SMA: \$411.719 (price is above this, bullish)\\
- Bollinger Bands: Price near upper band (\$412.903), indicating strong momentum\\
- ATR: 13.729 (moderate volatility)\\
- ADX: 15.148 (moderate trend strength)\\

\#\# Sentiment Analysis

From news:\\
- Strong positive sentiment around Q3 earnings (April 30, 2025)\\
- Focus on AI and cloud growth (Azure)\\
- Microsoft beating expectations\\
- Partnerships with IBM, Western Digital, etc.\\
- Some concerns about data center delays but overall positive narrative\\

From analyst recommendations:\\
- Mostly positive ratings (Buy, Overweight, Outperform)\\
- Multiple price target reductions recently (from ~\$500+ to ~\$435-475)\\
- Consensus still leans positive despite some target reductions\\

\#\# Insider Transactions

Limited data:\\
- Some insider purchases (mostly through stock awards)\\
- Some sales by executives (Hood Amy, Smith Bradford L) at ~\$392.53\\
- Overall limited insider activity data provided\\

\#\# Macro Environment

From macro data (Feb-May 2025):\\
- Stable unemployment rate around 4.2\%\\
- Moderate inflation (CPI around 320-325)\\
- GDP growth appears stable\\
- Interest rates (Fed Funds Rate) at 4.21-4.42\%\\
- Economic growth seems steady\\

\#\# News Analysis

Key themes from news:\\
- Strong Q3 earnings driven by AI and cloud growth\\
- Microsoft's Azure cloud growth\\
- AI investments (Copilot, AI agents)\\
- Data center expansion (EU, Dubai, etc.)\\
- Some concerns about data center delays\\
- Strong partnerships with other tech companies\\
- Tariff concerns but Microsoft appears less affected than hardware companies\\

Now I'll structure my analysis into 5-7 sections:\\

1. `<fundamentals>` - Covering financial statements, balance sheet strength, income performance\\
2. `<technical>` - Price action, indicators, momentum\\
3. `<news>` - Recent developments, earnings, AI/cloud growth\\
4. `<valuation>` - Based on earnings and growth metrics\\
5. `<risk\_assessment>` - Data center delays, regulatory risks\\
6. `<macro>` - Economic environment impact\\
7. `<conclusion>` - Final recommendation\\

Let me verify I have enough data for each section:\\

1. Fundamentals: Strong data available\\
2. Technical: Strong data available\\
3. News: Good data available\\
4. Valuation: Can derive from earnings and growth\\
5. Risk assessment: Some data on data center delays\\
6. Macro: Good data available\\
7. Conclusion: Will synthesize everything\\

Now I'll create each section with proper XML formatting, making sure to:\\
- Include 4-7 bullet points per section\\
- Support all claims with exact quotes\\
- Use proper formatting for evidence and sources\\
- Include tables where appropriate\\
- Bold intro/conclusion paragraphs\\
- Italicize supporting evidence\\
- Use code quotes for sources\\

Let me start drafting each section:\\

\#\# Fundamentals Section

For the fundamentals section, I'll focus on:\\
- Revenue growth and profitability\\
- Balance sheet strength\\
- Cash flow generation\\
- Margin performance\\

Key metrics:\\
- Revenue: \$69.6B (Q2 2025)\\
- Gross margin: 68.7\% (\$47.8B/\$69.6B)\\
- Operating margin: 45.5\% (\$31.7B/\$69.6B)\\
- Net margin: 34.6\% (\$24.1B/\$69.6B)\\
- Cash position: \$71.6B\\
- Debt: \$56.9B long-term\\
- Current ratio: 1.35\\
- Operating cash flow: \$22.3B\\

\#\# Technical Section

Key technical points:\\
- Strong upward price trend (from \$360 to \$424 in 2 weeks)\\
- Price above both 50 EMA and 200 SMA (bullish)\\
- MACD positive and rising\\
- RSI at 69.27 (approaching overbought but still bullish)\\
- Price near upper Bollinger Band\\

\#\# News Section

Key news themes:\\
- Strong Q3 earnings beat\\
- AI/cloud growth driving results\\
- Azure expansion (EU, Dubai)\\
- Data center developments\\
- Partnerships with IBM, Western Digital\\
- Some concerns about data center delays\\

\#\# Valuation Section

I'll need to calculate some valuation metrics:\\
- Based on Q3 earnings (need to see what's reported)\\
- Can compare to historical valuation\\
- Look at P/E ratio if possible\\

From SEC filings:\\
- Net Income: \$24.1B (Q2)\\
- Revenue: \$69.6B (Q2)\\
- Earnings Per Share: Need to calculate based on shares\\

Shares (Diluted): 7.5B
Net Income: \$24.1B
EPS = \$24.1B / 7.5B = \$3.21

Current price: ~\$424
P/E = \$424 / \$3.21 = ~131.5 (this seems high - need to check if this is annualized or quarterly)

Wait, this doesn't seem right. Let me check the SEC filings again.

Looking at SEC filings:\\
- EarningsPerShareDiluted (Earnings Per Share, Diluted): 9 USD on 20250331\\
- EarningsPerShareBasic (Earnings Per Share, Basic): 10 USD on 20250331\\

This suggests annualized EPS is around \$36-40, which would make sense for a P/E of ~12-12.5 at \$424.

Let me verify:\\
- EarningsPerShareDiluted: 9 USD on 20250331\\
- This is likely annualized for the full year\\

So annual EPS = ~\$9 (diluted) or ~\$10 (basic)
P/E = \$424 / \$9 = ~47.1 (more reasonable)

\#\# Risk Assessment

Key risks from data:\\
- Data center delays (mentioned in news)\\
- Regulatory risks (shareholder case mentioned)\\
- Tariff concerns (mentioned in news)\\
- Potential overspending on AI infrastructure\\

\#\# Macro Section

Macro environment:\\
- Stable unemployment (4.2\%)\\
- Moderate inflation (CPI ~320)\\
- Interest rates around 4.2-4.4\%\\
- Economic growth appears steady\\

Now I'll draft each section with proper formatting and citations.

Let me check if I have enough data for each section:\\

1. Fundamentals: Yes, strong data\\
2. Technical: Yes, strong data\\
3. News: Yes, good data\\
4. Valuation: Can derive from earnings data\\
5. Risk assessment: Some data available\\
6. Macro: Yes, good data\\
7. Conclusion: Will synthesize\\

I'll create 6 sections plus the conclusion (total 7 sections).

Let me now draft each section with proper XML formatting.}

\vspace{3em}

\textbf{\textsc{II. Investment Thesis and Assessment}}
\vspace{1em}

\vspace{1em}
\textbf{\textsc{II-A. Fundamental Analysis}}

\vspace{0.4em}
\textbf{Evaluating Microsoft's financial health through comprehensive review of balance sheet strength, income performance, and cash flow dynamics to assess the company's operational efficiency and profitability metrics.}

\vspace{0.3em}
$\circ$ Revenue demonstrates robust growth with strong profitability metrics, reflecting Microsoft's dominant market position in cloud and AI services. \hlevidence{Supporting evidence: "Revenue increased to \$69.6B with Gross Profit of \$47.8B (68.7\% margin) and Operating Income of \$31.7B (45.5\% margin)"} \textsc{Sources: income statement, Q2 2025}

\vspace{0.5em}
$\circ$ Balance sheet shows exceptional financial strength with substantial cash reserves and minimal leverage, providing strategic flexibility for investments and shareholder returns. \hlevidence{Supporting evidence: "Cash, Cash Equivalents \& Short Term Investments totaled \$71.6B with Total Equity of \$302.7B against Total Liabilities of \$231.2B, resulting in a debt-to-equity ratio of 0.19"} \textsc{Sources: balance sheet, Q2 2025}

\vspace{0.5em}
$\circ$ Cash flow generation remains exceptionally strong, enabling significant capital returns to shareholders while funding strategic growth initiatives. \hlevidence{Supporting evidence: "Net Cash from Operating Activities reached \$22.3B while Net Cash from Financing Activities included \$6.2B in Dividends Paid and \$4.7B in Share Repurchases"} \textsc{Sources: cash flow statement, Q2 2025}

\vspace{0.5em}
$\circ$ Profitability metrics show remarkable expansion with operating margin reaching 45.5\%, reflecting pricing power and operational efficiency in high-growth segments. \hlevidence{Supporting evidence: "Operating Income (Loss) of \$31.7B on Revenue of \$69.6B, representing a significant improvement from previous periods"} \textsc{Sources: income statement, Q2 2025}

\vspace{0.5em}
$\circ$ Equity position shows substantial recovery with total shareholder equity reaching \$302.7B, demonstrating strong capital accumulation over time. \hlevidence{Supporting evidence: "StockholdersEquity (Equity, Attributable to Parent) increased to \$321.9B USD on 20250331 from \$203.5B USD on 20241231"} \textsc{Sources: SEC filings, 20250331}

\vspace{0.5em}
\vspace{1em}
\begin{center}
\begin{tabular}{llll}
\toprule
Financial Metric & Q2 2025 & Prior Year & Change \\
\midrule
Revenue & \$69.6B & \$66.0B & +5.5\% \\
Gross Margin & 68.7\% & 65.2\% & +3.5pp \\
Operating Margin & 45.5\% & 41.8\% & +3.7pp \\
Cash Reserves & \$71.6B & \$18.3B & +291\% \\
Debt-to-Equity & 0.19 & 0.27 & -0.08 \\
Operating Cash Flow & \$22.3B & \$93.5B (annual) & N/A \\
\bottomrule
\end{tabular}
\end{center}
\vspace{{0.5em}}

\textbf{Microsoft demonstrates exceptional financial health with accelerating revenue growth, expanding margins, and robust cash generation, supported by a fortress balance sheet that provides strategic flexibility for continued innovation and shareholder returns.}

\vspace{2.5em}
\textbf{\textsc{II-B. Technical Analysis}}

\vspace{0.4em}
\textbf{Assessing market technicals to identify price momentum, chart patterns, and indicator signals that provide insight into investor sentiment and potential price direction for Microsoft stock.}

\vspace{0.3em}
$\circ$ Price action shows powerful bullish momentum with the stock breaking through multiple resistance levels on strong volume, indicating institutional accumulation. \hlevidence{Supporting evidence: "Stock price advanced from \$362.12 on April 22 to \$423.92 on May 1, a 17.1\% increase over 10 trading days with volume surging to 58.9M shares on May 1"} \textsc{Sources: price data, May 1 2025}

\vspace{0.5em}
$\circ$ Technical indicators confirm strong upward momentum with MACD showing positive crossover and RSI approaching but not yet in overbought territory. \hlevidence{Supporting evidence: "MACD value of 4.414 has crossed above signal line with histogram expanding, while RSI reading of 69.269 indicates strong momentum without extreme overbought conditions"} \textsc{Sources: technical indicators, May 1 2025}

\vspace{0.5em}
$\circ$ Price structure reveals a clear bullish trend with the stock trading well above both key moving averages, signaling sustained institutional buying interest. \hlevidence{Supporting evidence: "Current price of \$423.92 trades above both 50 EMA (\$388.54) and 200 SMA (\$411.72), with the 50 EMA now above the 200 SMA (golden cross)"} \textsc{Sources: technical indicators, May 1 2025}

\vspace{0.5em}
$\circ$ Volatility metrics indicate stable but expanding price movement as the stock establishes new highs, reflecting growing investor confidence. \hlevidence{Supporting evidence: "ATR value of 13.729 shows moderate volatility increase from 11.586 two weeks prior, coinciding with the breakout above key resistance levels"} \textsc{Sources: volatility metrics, May 1 2025}

\vspace{0.5em}
\vspace{1em}
\begin{center}
\begin{tabular}{llll}
\toprule
Technical Indicator & Current Value & Signal & Trend Direction \\
\midrule
Price & \$423.92 & Above Key MAs & Bullish \\
50 EMA & \$388.54 & Above 200 SMA & Golden Cross \\
200 SMA & \$411.72 & Resistance & Approaching \\
RSI (14-day) & 69.27 & Approaching Overbought & Bullish \\
MACD & 4.414 & Above Signal & Strong Buy \\
Bollinger \%B & 1.27 & Upper Band & Momentum Strong \\
\bottomrule
\end{tabular}
\end{center}
\vspace{{0.5em}}

\textbf{Technical indicators collectively signal robust bullish momentum with strong institutional support, as evidenced by price breaking above key moving averages, positive MACD crossover, and healthy momentum without extreme overbought conditions.}

\vspace{2.5em}
\textbf{\textsc{II-C. News Catalyst Analysis}}

\vspace{0.4em}
\textbf{Examining recent corporate developments, earnings performance, and strategic initiatives that serve as key catalysts for Microsoft's business trajectory and market perception.}

\vspace{0.3em}
$\circ$ Microsoft delivered exceptional Q3 earnings driven by strong AI and cloud adoption, significantly exceeding market expectations and validating strategic investments. \hlevidence{Supporting evidence: "Microsoft topped market estimates for quarterly revenue on Wednesday, boosted by steady demand for its cloud services as businesses ramped up spending on artificial-intelligence tools"} \textsc{Sources: news, 04/30/2025}

\vspace{0.5em}
$\circ$ Azure cloud platform continues to demonstrate leadership in the competitive cloud computing market with significant growth momentum across enterprise customers. \hlevidence{Supporting evidence: "Microsoft Corporation exceeded earnings expectations for Q3 2025, driven by strong growth in its Intelligent Cloud segment, particularly Azure"} \textsc{Sources: news, 05/01/2025}

\vspace{0.5em}
$\circ$ Strategic partnerships and infrastructure expansion are accelerating Microsoft's AI capabilities and global market penetration, particularly in high-growth regions. \hlevidence{Supporting evidence: "Microsoft plans to expand its data-center capacity within the EU over the next two years, with a view to having cloud operations in more than 200 data centers across the region"} \textsc{Sources: news, 04/30/2025}

\vspace{0.5em}
$\circ$ Microsoft's AI ecosystem continues to evolve rapidly with new Copilot features enhancing productivity tools and creating new revenue streams. \hlevidence{Supporting evidence: "Microsoft 365 Copilot Wave 2 is here: Take a look at what's new: AI agents, GPT-4o image generation, and a new Copilot Notebook feature are all headed your way"} \textsc{Sources: news, 04/23/2025}

\vspace{0.5em}
$\circ$ Strategic investments in AI infrastructure are creating significant competitive advantages despite some near-term execution challenges. \hlevidence{Supporting evidence: "Microsoft reported late Wednesday that its third-quarter capital expenditures came to \$16.7 billion, up from \$11 billion in the same period a year ago, and said its full-year capex expectations were unchanged"} \textsc{Sources: news, 05/01/2025}

\vspace{0.5em}
\vspace{1em}
\begin{center}
\begin{tabular}{llll}
\toprule
News Catalyst & Impact & Time Horizon & Confidence \\
\midrule
Q3 Earnings Beat & High & Immediate & High \\
Azure Cloud Growth & Very High & Medium-Term & High \\
EU Data Center Expansion & Medium & Long-Term & Medium \\
Copilot Feature Rollout & Medium & Short-Term & High \\
Data Center Delays & Low & Short-Term & Medium \\
\bottomrule
\end{tabular}
\end{center}
\vspace{{0.5em}}

\textbf{Recent news flow confirms Microsoft's leadership position in the AI and cloud computing markets, with strong earnings performance validating strategic investments while new product launches and infrastructure expansion create multiple growth vectors for sustained outperformance.}

\vspace{2.5em}
\textbf{\textsc{II-D. Valuation Assessment}}

\vspace{0.4em}
\textbf{Analyzing Microsoft's current valuation metrics relative to earnings growth, cash flow generation, and market expectations to determine pricing attractiveness.}

\vspace{0.3em}
$\circ$ Current valuation appears reasonable given the company's exceptional growth trajectory and market leadership in high-growth technology segments. \hlevidence{Supporting evidence: "EarningsPerShareDiluted (Earnings Per Share, Diluted): 9 USD on 20250331" with current price near \$424 suggests forward P/E around 47x} \textsc{Sources: SEC filings, 20250331}

\vspace{0.5em}
$\circ$ Revenue growth metrics demonstrate sustainable expansion with cloud and AI segments driving disproportionate value creation. \hlevidence{Supporting evidence: "Microsoft topped market estimates for quarterly revenue on Wednesday, boosted by steady demand for its cloud services as businesses ramped up spending on artificial-intelligence tools"} \textsc{Sources: news, 04/30/2025}

\vspace{0.5em}
$\circ$ Cash flow generation supports both strategic investments and shareholder returns, creating a balanced capital allocation framework. \hlevidence{Supporting evidence: "Net Cash from Operating Activities of \$22.3B while maintaining \$15.8B in capital expenditures for strategic growth initiatives"} \textsc{Sources: cash flow statement, Q2 2025}

\vspace{0.5em}
$\circ$ Analyst price targets remain substantially above current levels, indicating continued upside potential despite recent target reductions. \hlevidence{Supporting evidence: "Multiple firms maintain price targets above \$450 including TD Cowen (\$475), Citigroup (\$480), and BMO Capital (\$470)"} \textsc{Sources: analyst recommendations, April 2025}

\vspace{0.5em}
\vspace{1em}
\begin{center}
\begin{tabular}{llll}
\toprule
Valuation Metric & Current & Prior Period & Change \\
\midrule
Forward P/E & ~47x & ~45x & +2x \\
Price/Operating Cash Flow & ~19x & ~18x & +1x \\
Dividend Yield & ~0.7\% & ~0.8\% & -0.1pp \\
Analyst Target Upside & ~25\% & ~20\% & +5pp \\
Revenue Growth (YoY) & 15.3\% & 12.1\% & +3.2pp \\
\bottomrule
\end{tabular}
\end{center}
\vspace{{0.5em}}

\textbf{Valuation metrics indicate Microsoft trades at a premium reflecting its leadership in high-growth technology segments, but the substantial analyst price targets and strong earnings growth trajectory suggest the premium is justified by sustainable competitive advantages and market leadership.}

\vspace{2.5em}
\textbf{\textsc{II-E. Risk Assessment}}

\vspace{0.4em}
\textbf{Identifying and evaluating potential downside scenarios, operational challenges, and external threats that could impact Microsoft's business performance and stock valuation.}

\vspace{0.3em}
$\circ$ Data center construction delays present near-term execution risks that could impact cloud growth trajectory despite strategic importance. \hlevidence{Supporting evidence: "Microsoft delays construction on \$1B data center project outside Charlotte" and "Microsoft's decision to pause work on data centers in Ohio surprised local officials"} \textsc{Sources: news, 04/15/2025}

\vspace{0.5em}
$\circ$ Regulatory and legal challenges continue to create potential headwinds for shareholder value and operational flexibility. \hlevidence{Supporting evidence: "Microsoft swaps law firms in shareholder case, hiring Trump target: Microsoft is switching the law firm representing it in a shareholder case, replacing one that settled with the Trump administration to avoid a punishing executive order"} \textsc{Sources: news, 05/01/2025}

\vspace{0.5em}
$\circ$ AI infrastructure investments could lead to margin pressure if execution lags behind expectations despite strong demand signals. \hlevidence{Supporting evidence: "Microsoft's Q3 Earnings Forecast: Why MSFT Stock Is A Buy Before April 30: Microsoft Corporation's AI/data center build-out for FY 2025 faces revenue risks and overspending, prompting a bearish outlook"} \textsc{Sources: news, 04/29/2025}

\vspace{0.5em}
$\circ$ Competitive pressures in the cloud and AI markets remain intense despite Microsoft's current leadership position. \hlevidence{Supporting evidence: "IBM Launches Microsoft Practice to Deliver Transformative Business Value for Clients" and "Microsoft vs. Oracle: Which Cloud Stock Has More Fuel for Growth?"} \textsc{Sources: news, 04/29/2025 and 04/16/2025}

\vspace{0.5em}
\vspace{1em}
\begin{center}
\begin{tabular}{llll}
\toprule
Risk Factor & Impact & Probability & Mitigation Strategy \\
\midrule
Data Center Delays & Medium & Medium & Diversified global footprint \\
Regulatory Challenges & Medium & High & Legal and compliance investments \\
AI Infrastructure Costs & High & Medium & Cost optimization initiatives \\
Competitive Pressure & Medium & High & Continuous innovation and partnerships \\
Tariff Impacts & Low & Medium & Software business model resilience \\
\bottomrule
\end{tabular}
\end{center}
\vspace{{0.5em}}

\textbf{While Microsoft faces several manageable risks including data center delays and regulatory challenges, the company's diversified business model, strong cash position, and leadership in high-growth markets provide significant buffers against potential downside scenarios.}

\vspace{2.5em}
\textbf{\textsc{II-F. Macroeconomic Context}}

\vspace{0.4em}
\textbf{Analyzing how broader economic conditions impact Microsoft's business performance and valuation framework within the current economic environment.}

\vspace{0.3em}
$\circ$ Stable labor market conditions support enterprise technology spending despite higher interest rates, benefiting Microsoft's recurring revenue model. \hlevidence{Supporting evidence: "Unemployment Rate maintained at 4.20\% with Nonfarm Payroll Employment at 159.5K, indicating stable economic conditions for enterprise technology investments"} \textsc{Sources: macro data, May 2025}

\vspace{0.5em}
$\circ$ Moderate inflation environment creates manageable operating conditions for Microsoft's pricing power in enterprise software and cloud services. \hlevidence{Supporting evidence: "Consumer Price Index (CPI) at 320.58 with Core CPI at 326.85, representing approximately 3.5\% year-over-year inflation that aligns with Microsoft's pricing power in enterprise markets"} \textsc{Sources: macro data, May 2025}

\vspace{0.5em}
$\circ$ Interest rate environment remains supportive for technology companies with strong cash flow generation like Microsoft. \hlevidence{Supporting evidence: "Effective Federal Funds Rate at 4.42\% with 10-Year Treasury at 3.92\%, creating a favorable environment for cash-rich technology companies with Microsoft's \$71.6B cash position"} \textsc{Sources: macro data, May 2025}

\vspace{0.5em}
$\circ$ Global economic conditions show resilience that supports continued enterprise technology investment despite trade policy uncertainties. \hlevidence{Supporting evidence: "Trade Balance: Goods and Services at -71.1K with Stable Consumer Sentiment at 630.00, indicating resilient economic conditions supporting enterprise spending"} \textsc{Sources: macro data, May 2025}

\vspace{0.5em}
\vspace{1em}
\begin{center}
\begin{tabular}{llll}
\toprule
Macro Indicator & Current Value & Trend & Impact on MSFT \\
\midrule
Unemployment Rate & 4.20\% & Stable & Positive \\
Fed Funds Rate & 4.42\% & Stable & Neutral \\
CPI Inflation & 320.58 & Moderate & Positive \\
Consumer Sentiment & 630.00 & Improving & Positive \\
Trade Balance & -71.1K & Stable & Neutral \\
\bottomrule
\end{tabular}
\end{center}
\vspace{{0.5em}}

\textbf{The current macroeconomic environment presents favorable conditions for Microsoft's business model, with stable economic growth supporting enterprise technology spending, moderate inflation preserving pricing power, and interest rates that don't significantly impact the company's cash-rich balance sheet.}

\vspace{2.5em}
\textbf{\textsc{II-G. Investment Recommendation}}

\vspace{0.4em}
\textbf{Synthesizing fundamental strength, technical momentum, valuation metrics, and risk factors to determine the optimal investment position based on comprehensive multi-factor analysis.}

\vspace{0.3em}
$\circ$ Exceptional financial performance combined with strong technical momentum creates a compelling investment opportunity with multiple catalysts supporting continued outperformance.

\vspace{0.5em}
$\circ$ Dominant position in AI and cloud computing, evidenced by Azure growth and Copilot adoption, provides sustainable competitive advantages that justify premium valuation metrics.

\vspace{0.5em}
$\circ$ Strong cash generation and fortress balance sheet provide significant resilience against macroeconomic headwinds while funding strategic growth initiatives.

\vspace{0.5em}
$\circ$ While data center delays and regulatory challenges present manageable risks, Microsoft's diversified business model and leadership in high-growth technology segments provide substantial buffers.

\vspace{0.5em}
\textbf{Microsoft's combination of financial strength, leadership in transformative technology markets, and favorable technical setup creates an attractive risk-reward profile with significant upside potential as AI adoption continues to accelerate across enterprise markets.}

\vspace{3em}
\textbf{\textsc{III. Trading-R1 Decision}}
\vspace{-0.5em}
\\\\
\hlwhite{[[[BUY]]]}

\end{tcolorbox}

\clearpage
\newpage

\begin{tcolorbox}[
    breakable,
    colback=yellow!3!white,        
    colframe=orange!10!gray,       
    coltitle=black,                
    boxrule=0.5pt,                 
    arc=0.5mm,                       
    top=2mm, bottom=2mm, left=2mm, right=2mm, 
    fonttitle=\usefont{T1}{ppl}{b}{n},       
    title=\model Investment Thesis on Broadcom Inc. (AVGO),
]
\usefont{T1}{ppl}{m}{n}  
\footnotesize

\textbf{\textsc{I. Thinking and Planning}}
\\

\think{Let me analyze the provided financial data for AVGO (Broadcom Inc.) and compose a comprehensive investment thesis based strictly on the available information.

First, I'll organize the data into relevant categories for my analysis sections:\\

1. Fundamentals/Balance Sheet Data:\\
- Quarterly balance sheet as of 2025-03-12 (Q1 2025)\\
- Quarterly income statement as of 2025-03-12 (Q1 2025)\\
- Quarterly cash flow statement as of 2025-03-12 (Q1 2025)\\

2. News Data:\\
- Recent news articles from the last 30 days (mostly from August 2025)\\
- Focus on VMware Cloud Foundation, AI initiatives, partnerships, and market sentiment\\

3. Market Data:\\
- Price data for the last 10 trading days (Aug 18-29, 2025)\\
- Technical indicators: RSI, MACD, 50 EMA, 200 SMA, Bollinger Bands, ATR, ADX\\

4. Sentiment Data:\\
- Analyst recommendations and price target changes\\
- Insider transactions (though limited data)\\

5. Macro Data:\\
- U.S. macroeconomic indicators from May-August 2025\\

Let me analyze each section in detail:\\

\#\#\# Fundamentals Analysis

From the balance sheet:\\
- Total Assets: \$165.4B\\
- Total Liabilities: \$95.6B\\
- Total Equity: \$69.8B\\
- Cash, Cash Equivalents \& Short Term Investments: \$9.3B\\
- Short Term Debt: \$5.7B\\
- Long Term Debt: \$60.9B\\
- Total Current Assets: \$21.0B\\
- Total Current Liabilities: \$20.9B\\
- Current Ratio: 21.0B/20.9B = 1.00 (very tight)\\

From the income statement:\\
- Revenue: \$14.9B\\
- Gross Profit: \$10.1B (67.8\% gross margin)\\
- Operating Income: \$6.4B (42.9\% operating margin)\\
- Net Income: \$5.5B (36.9\% net margin)\\

From the cash flow statement:\\
- Net Cash from Operating Activities: \$6.1B\\
- Net Cash from Investing Activities: -\$174.0M\\
- Net Cash from Financing Activities: -\$6.0B\\
- Dividends Paid: -\$2.8B\\
- Share Repurchases: -\$2.0B\\

\#\#\# Technical Analysis

Price data (last 10 days):\\
- Recent price range: \$293.66 (low) to \$311.00 (high)\\
- Current price (Aug 29): \$297.39\\
- 50 EMA: \$284.555\\
- 200 SMA: \$226.239\\
- RSI: 51.468 (neutral)\\
- MACD: 3.343 (positive but declining from recent highs)\\
- Bollinger Bands: Upper Band \$314.752, Lower Band \$287.089\\

\#\#\# News Analysis

Key themes from news:\\
- Multiple announcements about VMware Cloud Foundation for AI and cybersecurity\\
- Partnerships with VMware and Canonical\\
- Walmart selecting Broadcom's VMware for cloud infrastructure\\
- Broadcom's AI chips for smaller data centers\\
- Jim Cramer's positive comments about AVGO\\
- Analyst upgrades and price target increases\\
- Discussion about Broadcom's position in the AI semiconductor market\\

\#\#\# Analyst Coverage

Recent analyst actions:\\
- Multiple price target increases from major firms\\
- UBS raised target to \$345 from \$290 (Aug 25)\\
- Oppenheimer raised target to \$325 from \$305 (Aug 28)\\
- Morgan Stanley raised target to \$338 from \$270 (Jul 30)\\
- Multiple firms maintaining or upgrading to "Buy" or "Overweight" ratings\\

\#\#\# Insider Transactions

Limited insider data:\\
- Multiple sales by executives (Delly Gayla J, Samueli Henry, Brazeal Mark David, Tan Hock E)\\
- Some grant-related transactions (Code:G)\\
- No significant buys visible in the data\\

\#\#\# Macro Environment

Macro indicators (Aug 2025):\\
- Unemployment Rate: 4.30\%\\
- Consumer Sentiment: 60.70 (from University of Michigan)\\
- Fed Funds Rate: 4.33\%\\
- 10-Year Treasury: 4.26\%\\
- 2-Year Treasury: 3.70\%\\
- Inflation (CPI): 328.66 (index value)\\
- Core PCE: 328.66 (index value)\\

Now I'll structure my analysis into 5-7 sections as required:\\

1. `<fundamentals>` - Covering financial statements, margins, cash flow\\
2. `<balance\_sheet>` - Focusing specifically on balance sheet strength\\
3. `<technical>` - Price action and technical indicators\\
4. `<analyst\_coverage>` - Analyst recommendations and price targets\\
5. `<news>` - Recent developments and market catalysts\\
6. `<macro>` - Macroeconomic context\\
7. `<conclusion>` - Final recommendation\\

Let me craft each section with proper XML formatting, supporting evidence, and tables where appropriate.

For the fundamentals section:\\
- I'll focus on revenue, margins, profitability\\
- Include gross margin, operating margin, net margin\\
- Highlight strong cash flow generation\\

For the balance sheet section:\\
- Focus on liquidity position, debt levels, asset quality\\
- Note the tight current ratio (1.00)\\
- Analyze debt-to-equity ratio\\

For the technical section:\\
- Analyze price relative to moving averages\\
- RSI and MACD indicators\\
- Bollinger Band positioning\\
- ADX for trend strength\\

For analyst coverage:\\
- Summarize recent rating changes\\
- Note the upward trend in price targets\\
- Highlight consensus sentiment\\

For news:\\
- Focus on AI-related developments\\
- Partnerships and product announcements\\
- Market positioning in semiconductor/AI space\\

For macro:\\
- Analyze how macro environment affects semiconductor sector\\
- Interest rates, inflation, economic growth\\

Let me now write each section with proper formatting and citations to the data provided.}

\vspace{3em}

\textbf{\textsc{II. Investment Thesis and Assessment}}
\vspace{1em}

\vspace{1em}
\textbf{\textsc{II-A. Fundamental Analysis}}

\vspace{0.4em}
\textbf{Evaluating Broadcom's core financial performance through comprehensive review of revenue growth, profitability metrics, and operational efficiency to assess the company's underlying business strength and sustainability.}

\vspace{0.3em}
$\circ$ Revenue demonstrates robust performance with strong quarterly results reflecting the company's market position in semiconductor and infrastructure software solutions. \hlevidence{Supporting evidence: "Revenue increased to \$14.9B in Q1 2025 with Gross Profit of \$10.1B representing a 67.8\% gross margin"} \textsc{Sources: income statement, Q1 2025}

\vspace{0.5em}
$\circ$ Exceptional profitability metrics highlight Broadcom's pricing power and operational efficiency in the technology sector. \hlevidence{Supporting evidence: "Operating Income reached \$6.4B (42.9\% operating margin) while Net Income totaled \$5.5B (36.9\% net margin)"} \textsc{Sources: income statement, Q1 2025}

\vspace{0.5em}
$\circ$ Cash flow generation remains exceptionally strong, providing significant flexibility for strategic investments and shareholder returns. \hlevidence{Supporting evidence: "Net Cash from Operating Activities reached \$6.1B, substantially exceeding the \$5.5B Net Income"} \textsc{Sources: cash flow statement, Q1 2025}

\vspace{0.5em}
$\circ$ Strategic capital allocation is evident through substantial shareholder returns while maintaining investment in growth initiatives. \hlevidence{Supporting evidence: "Dividends Paid totaled \$2.8B and Share Repurchases amounted to \$2.0B, representing significant capital return to shareholders"} \textsc{Sources: cash flow statement, Q1 2025}

\vspace{0.5em}
\vspace{1em}
\begin{center}
\begin{tabular}{llll}
\toprule
Financial Metric & Value & Margin/Percentage & Significance \\
\midrule
Revenue & \$14.9B & - & Strong market position \\
Gross Profit & \$10.1B & 67.8\% & High pricing power \\
Operating Income & \$6.4B & 42.9\% & Operational efficiency \\
Net Income & \$5.5B & 36.9\% & Strong bottom-line performance \\
Operating Cash Flow & \$6.1B & 40.9\% of revenue & Excellent cash conversion \\
FCF Conversion & ~100\% & - & Sustainable operations \\
\bottomrule
\end{tabular}
\end{center}
\vspace{{0.5em}}

\textbf{Financial fundamentals reveal exceptional profitability and cash generation capabilities, with Broadcom maintaining industry-leading margins while returning substantial capital to shareholders through dividends and buybacks.}

\vspace{2.5em}
\textbf{\textsc{II-B. Balance Sheet Analysis}}

\vspace{0.4em}
\textbf{Assessing Broadcom's financial stability, liquidity position, and capital structure to evaluate the company's ability to weather economic downturns and fund strategic initiatives.}

\vspace{0.3em}
$\circ$ The company maintains a substantial cash position that provides strategic flexibility despite elevated debt levels. \hlevidence{Supporting evidence: "Cash, Cash Equivalents \& Short Term Investments totaled \$9.3B while Total Current Assets reached \$21.0B against Total Current Liabilities of \$20.9B"} \textsc{Sources: balance sheet, Q1 2025}

\vspace{0.5em}
$\circ$ Debt levels remain manageable relative to the company's robust cash flow generation capabilities and asset base. \hlevidence{Supporting evidence: "Long Term Debt stands at \$60.9B against Total Equity of \$69.8B, resulting in a debt-to-equity ratio of approximately 0.87"} \textsc{Sources: balance sheet, Q1 2025}

\vspace{0.5em}
$\circ$ Working capital position shows minimal liquidity pressure despite the tight current ratio. \hlevidence{Supporting evidence: "Current Ratio of 1.00 indicates near-parity between current assets and liabilities, suggesting potential short-term pressure"} \textsc{Sources: balance sheet, Q1 2025}

\vspace{0.5em}
$\circ$ The asset composition reflects strategic positioning in the semiconductor and infrastructure software markets. \hlevidence{Supporting evidence: "Total Noncurrent Assets of \$144.4B represent 87.3\% of Total Assets, highlighting significant investment in long-term strategic assets"} \textsc{Sources: balance sheet, Q1 2025}

\vspace{0.5em}
\vspace{1em}
\begin{center}
\begin{tabular}{llll}
\toprule
Balance Sheet Metric & Value & Ratio/Percentage & Assessment \\
\midrule
Cash \& Short Term Investments & \$9.3B & 5.6\% of Total Assets & Strong liquidity buffer \\
Total Current Assets & \$21.0B & 12.7\% of Total Assets & Adequate short-term coverage \\
Total Current Liabilities & \$20.9B & 12.6\% of Total Assets & Tight working capital position \\
Long Term Debt & \$60.9B & 36.8\% of Total Assets & Moderate leverage \\
Debt-to-Equity Ratio & 0.87 & - & Manageable leverage \\
Current Ratio & 1.00 & - & Minimal short-term liquidity buffer \\
\bottomrule
\end{tabular}
\end{center}
\vspace{{0.5em}}

\textbf{While Broadcom maintains a strong asset base and substantial cash reserves, the tight current ratio and elevated debt levels warrant monitoring, though the company's exceptional cash flow generation provides adequate coverage for these obligations.}

\vspace{2.5em}
\textbf{\textsc{II-C. Technical Analysis}}

\vspace{0.4em}
\textbf{Analyzing price action, momentum indicators, and volatility patterns to identify potential near-term price direction and market sentiment signals for Broadcom stock.}

\vspace{0.3em}
$\circ$ Price action shows strong bullish momentum with the stock trading well above key moving averages, indicating sustained institutional interest. \hlevidence{Supporting evidence: "Current price of \$297.39 trades significantly above both 50 EMA (\$284.555) and 200 SMA (\$226.239), confirming a strong uptrend"} \textsc{Sources: market data, Aug 29 2025}

\vspace{0.5em}
$\circ$ Momentum indicators suggest a potential short-term consolidation phase after recent strength, with RSI showing neutral positioning. \hlevidence{Supporting evidence: "RSI reading of 51.468 indicates neutral momentum with room for further advancement before reaching overbought territory"} \textsc{Sources: technical indicators, Aug 29 2025}

\vspace{0.5em}
$\circ$ MACD analysis reveals a recent bearish crossover that could signal a pause in the current uptrend, though the indicator remains in positive territory. \hlevidence{Supporting evidence: "MACD value of 3.343 has declined from recent highs of 8.91, indicating weakening upward momentum"} \textsc{Sources: technical indicators, Aug 29 2025}

\vspace{0.5em}
$\circ$ Volatility metrics suggest moderate price fluctuations with potential for expansion as the stock approaches resistance levels. \hlevidence{Supporting evidence: "ATR value of 9.464 indicates moderate volatility, with price currently positioned near the middle of the Bollinger Bands (\$297.39 vs middle band at \$296.42)"} \textsc{Sources: market data, Aug 29 2025}

\vspace{0.5em}
\vspace{1em}
\begin{center}
\begin{tabular}{llll}
\toprule
Technical Indicator & Current Value & Signal & Trend Direction \\
\midrule
Price & \$297.39 & - & Upward \\
50 EMA & \$284.555 & Support & Bullish \\
200 SMA & \$226.239 & Support & Bullish \\
RSI (14-day) & 51.468 & Neutral & Consolidation \\
MACD & 3.343 & Bearish Crossover & Short-term caution \\
Bollinger \% & 97.6\% & Upper Half & Approaching resistance \\
ADX & 10.092 & Weak Trend & Indecisive \\
\bottomrule
\end{tabular}
\end{center}
\vspace{{0.5em}}

\textbf{Technical indicators present a mixed picture with strong long-term bullish structure but emerging signs of short-term consolidation, suggesting potential for sideways movement before next directional move.}

\vspace{2.5em}
\textbf{\textsc{II-D. Analyst Coverage Analysis}}

\vspace{0.4em}
\textbf{Examining the evolution of professional analyst sentiment and price targets to gauge institutional expectations and potential upside catalysts for Broadcom stock.}

\vspace{0.3em}
$\circ$ Analyst community has demonstrated consistent bullish sentiment with multiple recent price target increases across major investment firms. \hlevidence{Supporting evidence: "UBS raised price target to \$345 from \$290 on August 25, 2025, representing a 18.6\% upside from current levels"} \textsc{Sources: analyst recommendations, Aug 25 2025}

\vspace{0.5em}
$\circ$ Consensus price targets show significant upward revision momentum, reflecting growing confidence in Broadcom's growth trajectory. \hlevidence{Supporting evidence: "Morgan Stanley increased price target to \$338 from \$270 on July 30, 2025, while Keybanc raised target to \$330 from \$315 on July 8, 2025"} \textsc{Sources: analyst recommendations, Jul-Aug 2025}

\vspace{0.5em}
$\circ$ The breadth of positive coverage is notable, with virtually all major firms maintaining or upgrading to positive ratings. \hlevidence{Supporting evidence: "Multiple firms including Oppenheimer, UBS, Morgan Stanley, and Citigroup all maintain 'Buy' or 'Overweight' ratings with price target increases in recent months"} \textsc{Sources: analyst recommendations, Jun-Aug 2025}

\vspace{0.5em}
$\circ$ The magnitude of price target increases suggests substantial upside potential according to Wall Street consensus. \hlevidence{Supporting evidence: "Average price target across major analysts stands at approximately \$320, representing 7.6\% upside from current trading levels near \$297.39"} \textsc{Sources: analyst recommendations, Aug 2025}

\vspace{0.5em}
\vspace{1em}
\begin{center}
\begin{tabular}{lllll}
\toprule
Analyst Firm & Current Rating & Price Target & Target Change & Upside Potential \\
\midrule
UBS & Buy & \$345.00 & +\$55.00 & +16.0\% \\
Morgan Stanley & Overweight & \$338.00 & +\$68.00 & +13.7\% \\
Oppenheimer & Outperform & \$325.00 & +\$20.00 & +9.3\% \\
Citigroup & Buy & \$315.00 & +\$30.00 & +6.0\% \\
Goldman Sachs & Buy & \$340.00 & \$0.00 & +14.4\% \\
Average & - & \$322.00 & +\$22.00 & +8.0\% \\
\bottomrule
\end{tabular}
\end{center}
\vspace{{0.5em}}

\textbf{Analyst sentiment shows strong and consistent bullish momentum with substantial upside potential reflected in price targets, suggesting institutional confidence in Broadcom's growth trajectory and market position.}

\vspace{2.5em}
\textbf{\textsc{II-E. News Catalyst Analysis}}

\vspace{0.4em}
\textbf{Evaluating recent corporate developments, strategic partnerships, and market positioning to identify key catalysts that could drive future performance and valuation expansion.}

\vspace{0.3em}
$\circ$ Broadcom is strategically positioning itself at the forefront of AI infrastructure with innovative product announcements targeting the rapidly growing AI semiconductor market. \hlevidence{Supporting evidence: "Broadcom Inc. is offering a chip to help power AI by linking up smaller data centers" and "Broadcom Ships Jericho4, Enabling Distributed AI Computing Across Data Centers"} \textsc{Sources: news articles, Aug 13-29 2025}

\vspace{0.5em}
$\circ$ Strategic partnerships with major technology players are expanding Broadcom's ecosystem and market reach in critical infrastructure platforms. \hlevidence{Supporting evidence: "VMware Cloud Foundation Elevates Cyber Resilience, Compliance, and Security for the Modern Private Cloud" and "Broadcom and Canonical Expand Partnership to Optimize VMware Cloud Foundation for Modern Container and AI Workloads"} \textsc{Sources: news articles, Aug 29 2025}

\vspace{0.5em}
$\circ$ Major enterprise adoption is accelerating, with Walmart selecting Broadcom's VMware for significant cloud infrastructure transformation. \hlevidence{Supporting evidence: "Retail Tech Giant Walmart Selects Broadcom's VMware for Massive Cloud Infrastructure Transformation, enabling enhanced scalability, simplified workload management"} \textsc{Sources: news articles, Aug 29 2025}

\vspace{0.5em}
$\circ$ Market leadership in the AI semiconductor space is being increasingly recognized by prominent investors and analysts. \hlevidence{Supporting evidence: "Broadcom Inc. (AVGO) is one of the stocks Jim Cramer recently commented on" and "Broadcom Inc. (AVGO) is one of the Top Technology Stocks to Buy According to Hedge Funds"} \textsc{Sources: news articles, Aug 13-29 2025}

\vspace{0.5em}
\vspace{1em}
\begin{center}
\begin{tabular}{lll}
\toprule
News Catalyst & Significance & Market Impact \\
\midrule
VMware Cloud Foundation AI Enhancements & Critical infrastructure for AI workloads & High \\
Walmart Cloud Infrastructure Project & Major enterprise validation & High \\
Jericho4 AI Chip Announcement & New distributed AI computing capability & Medium-High \\
Jim Cramer Endorsement & Celebrity investor validation & Medium \\
Broadcom-Canonical Partnership Expansion & Expanded cloud platform capabilities & Medium \\
\bottomrule
\end{tabular}
\end{center}
\vspace{{0.5em}}

\textbf{Recent news flow reveals Broadcom's strategic positioning at the intersection of AI infrastructure and cloud computing, with multiple high-profile partnerships and product launches creating significant growth catalysts in the rapidly expanding AI semiconductor market.}

\vspace{2.5em}
\textbf{\textsc{II-F. Macroeconomic Context}}

\vspace{0.4em}
\textbf{Assessing the broader economic environment to understand external factors that could influence Broadcom's performance and valuation in the semiconductor sector.}

\vspace{0.3em}
$\circ$ Current interest rate environment presents a mixed picture for semiconductor companies with potential headwinds for capital-intensive investments. \hlevidence{Supporting evidence: "Effective Federal Funds Rate stands at 4.33\% with 10-Year Treasury at 4.26\% and 2-Year Treasury at 3.70\%, creating pressure on valuation multiples"} \textsc{Sources: macro data, Aug 2025}

\vspace{0.5em}
$\circ$ Moderate inflation levels provide some support for semiconductor pricing power despite potential input cost pressures. \hlevidence{Supporting evidence: "Core PCE Price Index at 328.66 indicates moderate inflation pressure, while Consumer Sentiment at 60.70 suggests cautious consumer spending environment"} \textsc{Sources: macro data, Aug 2025}

\vspace{0.5em}
$\circ$ Stable labor market conditions support enterprise technology spending despite elevated unemployment rate. \hlevidence{Supporting evidence: "Unemployment Rate remains at 4.30\% with Nonfarm Payroll Employment at 159.5K, indicating stable labor market conditions supporting enterprise IT budgets"} \textsc{Sources: macro data, Aug 2025}

\vspace{0.5em}
$\circ$ The semiconductor sector benefits from continued enterprise technology investment as businesses accelerate digital transformation initiatives. \hlevidence{Supporting evidence: "Industrial Production Index at 103.99 and Capacity Utilization at 77.52 suggest continued business investment in technology infrastructure"} \textsc{Sources: macro data, Aug 2025}

\vspace{0.5em}
\vspace{1em}
\begin{center}
\begin{tabular}{llll}
\toprule
Macro Indicator & Current Value & Trend & Impact on AVGO \\
\midrule
Fed Funds Rate & 4.33\% & Stable & Moderate headwind \\
10-Year Treasury & 4.26\% & Downward & Positive for valuation \\
Unemployment Rate & 4.30\% & Stable & Positive for enterprise spending \\
Consumer Sentiment & 60.70 & Improving & Neutral \\
Industrial Production & 103.99 & Stable & Positive for tech spending \\
Capacity Utilization & 77.52\% & Stable & Positive for enterprise IT \\
\bottomrule
\end{tabular}
\end{center}
\vspace{{0.5em}}

\textbf{The current macroeconomic environment presents a generally favorable backdrop for semiconductor companies like Broadcom, with stable labor markets supporting enterprise technology investments despite moderately elevated interest rates.}

\vspace{2.5em}
\textbf{\textsc{II-G. Risk Assessment}}

\vspace{0.4em}
\textbf{Identifying and evaluating potential downside risks and challenges that could impact Broadcom's performance and stock valuation in the near-to-medium term.}

\vspace{0.3em}
$\circ$ Elevated debt levels relative to current asset base create vulnerability to rising interest rates and economic slowdown. \hlevidence{Supporting evidence: "Long Term Debt of \$60.9B represents 36.8\% of Total Assets, with Current Ratio of 1.00 indicating minimal short-term liquidity buffer"} \textsc{Sources: balance sheet, Q1 2025}

\vspace{0.5em}
$\circ$ Competitive pressures in the AI semiconductor market could pressure pricing power and market share despite current growth momentum. \hlevidence{Supporting evidence: "NVIDIA vs. Broadcom: Which AI Semiconductor Stock Offers More Upside?" indicates intense competitive landscape in the critical AI chip market"} \textsc{Sources: news articles, Aug 22 2025}

\vspace{0.5em}
$\circ$ Recent insider selling activity by executives may signal caution about near-term performance despite positive public statements. \hlevidence{Supporting evidence: "Delly Gayla J sold 3.0K shares at \$265.13, Brazeal Mark David sold 9.4K shares at \$253.285, and Tan Hock E sold 40.0K shares at \$262.173"} \textsc{Sources: insider transactions, Jun 2025}

\vspace{0.5em}
$\circ$ The company's significant debt levels create vulnerability to interest rate increases, which could pressure valuation multiples. \hlevidence{Supporting evidence: "Long Term Debt of \$60.9B represents 36.8\% of Total Assets with Effective Federal Funds Rate at 4.33\%"} \textsc{Sources: balance sheet, Q1 2025}

\vspace{0.5em}
\vspace{1em}
\begin{center}
\begin{tabular}{llll}
\toprule
Risk Factor & Probability & Potential Impact & Mitigation \\
\midrule
Interest Rate Risk & High & Medium & Strong cash flow generation \\
Competitive Pressure & High & Medium & Product innovation pipeline \\
Debt Levels & Medium & Medium & Strong operating cash flow \\
Valuation Sensitivity & High & Medium & Diversified revenue streams \\
Supply Chain Constraints & Medium & Low & Geographic diversification \\
\bottomrule
\end{tabular}
\end{center}
\vspace{{0.5em}}

\textbf{While Broadcom faces manageable risks including interest rate sensitivity and competitive pressures, its strong cash flow generation and diversified business model provide substantial resilience against potential headwinds in the semiconductor sector.}

\vspace{2.5em}
\textbf{\textsc{II-H. Investment Recommendation}}

\vspace{0.4em}
\textbf{Synthesizing fundamental strength, technical positioning, analyst sentiment, and risk factors to determine the optimal investment position in Broadcom based on comprehensive multi-factor analysis.}

\vspace{0.3em}
$\circ$ Exceptional profitability metrics and robust cash flow generation provide a solid foundation for sustained growth and shareholder returns in the technology sector.

\vspace{0.5em}
$\circ$ Strategic positioning at the forefront of AI infrastructure and cloud computing creates significant growth catalysts with multiple enterprise partnerships validating the technology roadmap.

\vspace{0.5em}
$\circ$ Technical indicators suggest short-term consolidation but maintain a strong long-term bullish structure with price well above key moving averages.

\vspace{0.5em}
$\circ$ While debt levels warrant monitoring, the company's strong cash flow generation of \$6.1B provides ample coverage for debt obligations and strategic investments.

\vspace{0.5em}
\textbf{Broadcom demonstrates compelling investment characteristics with industry-leading margins, strategic positioning in the AI infrastructure market, and strong analyst support, creating an attractive risk-reward profile despite modest short-term technical consolidation.}

\vspace{3em}
\textbf{\textsc{III. Trading-R1 Decision}}
\vspace{-0.5em}
\\\\
\hlwhite{[[[BUY]]]}

\end{tcolorbox}

\end{document}